\documentclass[oneside,11pt]{article}

\usepackage[utf8]{inputenc}
\usepackage[T1]{fontenc}

\usepackage{amsthm}
\usepackage{amsmath}
\usepackage{amsfonts}
\usepackage{amssymb}

\usepackage{graphicx}
\usepackage{float}
\usepackage{booktabs}
\usepackage{multirow}
\usepackage{rotating}
\usepackage{array}
\usepackage{xcolor}
\usepackage{subfig}
\usepackage[ruled]{algorithm2e}

\usepackage{breakcites}

\usepackage{enumitem}

\usepackage[top=1.5cm,bottom=2cm,right=2.5cm,left=2.5cm]{geometry}
\usepackage{titling}

\usepackage{natbib}

\usepackage{ifplatform}

\usepackage{textcomp}

\ifwindows
\fi

\ifwindows
	\usepackage{bbm}
\else
	\iflinux
		\usepackage{bbm}
	\else
		\usepackage{bbm}
	\fi
\fi

\usepackage{bm}

\newcommand{\Arg}{\mathrm{Arg}}
\newcommand{\E}{\mathrm{E}}
\newcommand{\ic}{\mathrm{i}}
\newcommand{\e}{\mathrm{e}}

\usepackage[pdftex,final,bookmarksnumbered,bookmarksopen=false,breaklinks,colorlinks]{hyperref}
\hypersetup{
	final,
	bookmarksnumbered=true,
	bookmarksopen=true,
	bookmarksopenlevel=0,
	unicode=false,          
	pdftoolbar=true,        
	pdfmenubar=true,        
	pdffitwindow=false,     
	pdftitle={Recent advances in directional statistics},    
	pdfdisplaydoctitle=true, 
	pdftoolbar=true,
	pdfmenubar=true,
	pdflang={English},
	pdfauthor={Arthur Pewsey, Eduardo Garcia-Portugues},     
	pdfsubject={arXiv paper},   
	pdfcreator={Eduardo Garcia-Portugues},   
	pdfproducer={Eduardo Garcia-Portugues}, 
	pdfkeywords={Classification}{Clustering}{Dimension reduction}{Distributional models}{Exploratory data analysis}{Hypothesis tests}{Nonparametric methods}{Regression}{Serial dependence}{Software}{Spatial statistics}, 
	pdfnewwindow=true,      
	breaklinks=true,
	hidelinks,
	linkcolor=black,          
	citecolor=black,        
	filecolor=magenta,      
	urlcolor=cyan           
}

\allowdisplaybreaks

\newif\ifmain
\maintrue
\ifmain

\else

\fi
\newif\ifsupplement
\supplementtrue
\newif\iffigstabs
\figstabstrue

\begin{document}

\ifmain

\title{Recent advances in directional statistics}
\setlength{\droptitle}{-1cm}
\predate{}%
\postdate{}%

\date{}

\author{Arthur Pewsey$^{1,3}$ and Eduardo Garc\'ia-Portugu\'es$^{2}$}

\footnotetext[1]{Department of Mathematics, University of Extremadura (Spain).}
\footnotetext[2]{Department of Statistics, Carlos III University of Madrid (Spain).}
\footnotetext[3]{Corresponding author. e-mail: \href{mailto:apewsey@unex.es}{apewsey@unex.es}.}

\maketitle


\begin{abstract}
	Mainstream statistical methodology is generally applicable to data observed in Euclidean space. There are, however, numerous contexts of considerable scientific interest in which the natural supports for the data under consideration are Riemannian manifolds like the unit circle, torus, sphere and their extensions. Typically, such data can be represented using one or more directions, and directional statistics is the branch of statistics that deals with their analysis. In this paper we provide a review of the many recent developments in the field since the publication of \cite{Mardia1999a}, still the most comprehensive text on directional statistics. Many of those developments have been stimulated by interesting applications in fields as diverse as astronomy, medicine, genetics, neurology, aeronautics, acoustics, image analysis, text mining, environmetrics, and machine learning. We begin by considering developments for the exploratory analysis of directional data before progressing to distributional models, general approaches to inference, hypothesis testing, regression, nonparametric curve estimation, methods for dimension reduction, classification and clustering, and the modelling of time series, spatial and spatio-temporal data. An overview of currently available software for analysing directional data is also provided, and  potential future developments discussed.
\end{abstract}
\begin{flushleft}
\end{flushleft}

\section{Introduction}
\label{sec:Intro}

Directional statistics is that branch of statistical methodology specifically designed for use with observations that are directions. A direction observed in the plane $\mathbb{R}^2$, like wind direction, can be represented by an angle, $\theta$, typically in $[0,2\pi)$ or $[-\pi,\pi)$, measured in a specified direction from a specified origin, or by the unit vector $\bm{x}=(\cos\theta,\sin\theta)'$ for which $||\bm{x}||=\sqrt{\bm{x}'\bm{x}}=1$. The natural support for such directions is the circumference of the unit circle, $\mathbb{S}^1$; data on them being referred to as \textit{circular}. The term ``circular data'' is also used to distinguish them from data with the real line $\mathbb{R}$ (or some subset of it) as their support, which henceforth we will refer to as \textit{linear} data. Certain calculations can be performed more efficiently using the complex representation $z=\e^{\ic \theta}$, where $\ic =\sqrt{-1}$, for which $|z|=1$ and $\Arg(z)=\theta\in [-\pi,\pi)$. Closely related to circular data are \textit{axial data}, which arise when axes, for which the angles $\theta$ and $\theta+\pi$ are indistinguishable, are observed. Observations made on directions in $\mathbb{R}^3$, like the positions of stars on the celestial sphere, can be represented by pairs of angles or  $3\times 1$ unit column vectors, have natural support the unit sphere, $\mathbb{S}^2$, and are referred to as being \textit{spherical}. Circular and spherical data are the most commonly occurring forms of directional data. Since their supports are compact manifolds, it is (generally) inappropriate, and can prove thoroughly misleading, to apply standard statistical methods, designed for observations with more familiar supports like $\mathbb{R}^{d}$, $d\geq1$, to them.\\

Other data types that fall within the remit of directional statistics include \textit{toroidal} and \textit{cylindrical} data: toroidal data, with support the unit torus, $\mathbb{T}^2=\mathbb{S}^1\times\mathbb{S}^1$, arising when observations on a pair of circular variables are made, and cylindrical data, with support the cylinder $\mathbb{S}^1\times\mathbb{R}$ or some subset of it, when observations are made on a pair consisting of one circular and one linear variable. For example, toroidal data are obtained when wind direction is recorded at two different meteorological stations, and cylindrical data if, instead, wind direction and velocity are jointly measured at the same station.\\

In applications, data on these various manifolds, or their generalisations, such as the unit $d$-dimensional sphere, $\mathbb{S}^d$, and the $d$-torus, $\mathbb{T}^d=(\mathbb{S}^1)^d$, $d \geq 1$, might be observed and analysed as regression, time series, spatial or spatio-temporal data. Henceforth, we will use the term ``spherical data'' to refer to data on any $\mathbb{S}^d$ with $d\geq1$, not just $\mathbb{S}^2$, unless specifically mentioned otherwise.\\

Directional statistics can also be applied to data that are not originally directions but which can be represented on, or transformed to, one of the manifolds referred to previously. For instance, times on the 24 hour clock can be analysed as circular data after transferring them to the unit circle \citep[see, e.g.,][]{Gill2010}. More generally, methods for spherical data can be applied to data originally observed in Euclidean space, $\bm{x}_1,\ldots,\bm{x}_n\in\mathbb{R}^{d+1}$, after their Euclidean normalisation to $\bm{x}_1/||\bm{x}_1||,\ldots,\bm{x}_n/||\bm{x}_n|| \in\mathbb{S}^d$, a form of transformation often encountered \citep[see, e.g.,][]{Banerjee2005}.\\

Rotation groups, Stiefel and Grassmann manifolds, the elements of which are orthonormal frames and subspaces of $\mathbb{R}^d$, respectively, and other sample spaces such as hyperboloids, complex projective spaces, and general manifolds, are also important to the field but here, because of length restrictions, we refer only tangentially to certain developments related to them. Specifically, we do not consider models for rotations in $\mathbb{R}^3$ despite the fact that a $3\times 3$ rotation matrix can be represented as a $4\times 1$ unit vector called a quaternion, and modelling such rotations is equivalent to modelling axial data on $\mathbb{S}^4$. We would direct the reader interested in these topics to \citet[][Chapter 13]{Mardia1999a}, \cite{Mardia2005}, \cite{Chirikjian2001}, \cite{Chikuse2012}, \cite{Arnold2018a}, and \cite{Rivest2018}. An important related field is shape analysis \citep{Kendall1999, Dryden2016}, where a \textit{preshape} corresponding to a configuration of $k$ landmarks in $\mathbb{R}^{d}$ can be regarded as a point on $\mathbb{S}^{d(k-1)-1}$.\\

The last article-length review of directional statistics was \cite{Jupp1989}. It contained an extensive bibliography which included virtually all publications on directional statistics between 1975 and 1988. In their review, the authors sought to unify the theory of directional statistics from a mathematical perspective. In attempting to doing so, they referred to five key underpinning ideas: (i) exponential families; (ii) transformation structure; (iii) tangent-normal decomposition; (iv) transformation (of a directional problem) to a multivariate one; (v) the central limit theorem (CLT), and three basic approaches to directional statistics, termed the embedding, wrapping, and intrinsic approaches. All of these underlying principles have been fundamental to the ongoing development of the field, apart perhaps from the first. Whilst exponential models have certain appealing mathematical and inferential properties, insistence on them has largely been abandoned in recent years, primarily because of an increasing awareness of the need to model distributional features beyond location and concentration, such as the varying levels of skewness and peakedness frequently exhibited by real data. Moreover, directional data are often multimodal and finite mixture distributions, which do not belong to the exponential family, are natural choices with which to model them.\\

Books covering numerous facets of directional statistics published prior to the review of \cite{Jupp1989} include \cite{Mardia1972}, \cite{Batschelet1981}, \cite{Watson1983}, \cite{Fisher1987}, and \cite{Fisher1993}. Those published after that review include \cite{Mardia1999a}, \cite{Jammalamadaka2001}, \cite{Pewsey2013}, \cite{Ley2017a}, and \cite{Ley2018}, the latter being an excellent overview of interesting and important modern applications of directional statistics. We take as our definition of ``recent developments'' those that have appeared in the literature since the publication of \cite{Mardia1999a}, still the most comprehensive book-length treatment of the field. Whilst many, but certainly not all, of the themes we discuss are addressed in the books of Ley and Verdebout, our aim has been to provide a concise review of the most important developments since the publication of \cite{Mardia1999a} which is as exhaustive as possible, subject to length constraints. Given the latter, we have concentrated on describing key ideas and directing the interested reader to relevant original sources where more detailed information can be found. With the increasing pace of advances in the field, it is perhaps inevitable that we will have overlooked some developments. We hope that the number of such omissions is minimal, and apologise in advance for any that might have arisen.\\

Important areas of application that have stimulated much of the recent research activity in the field include bioinformatics \citep{Boomsma2008,Mardia2018}, astronomy \citep{Cabella2009,Marinucci2011}, medicine \citep{Vuollo2016,Pardo2017}, genetics \citep{Eisen1998,Dortet-Bernadet2008}, neurology \citep{Gu2004, Kaufman2005}, aeronautics \citep{Horwood2014}, acoustics \citep{McMillan2013,Traa2013}, image analysis \citep{Jung2011, Esteves2020}, text mining \citep{Dhillon2001,Banerjee2005}, machine learning \citep{Hamsici2007,Sra2018}, and the modelling of wildfires \citep{Garcia-Portugues2014,Ameijeiras-Alonso2018} and sea conditions \citep{Jona-Lasinio2012,Jona-Lasinio2018,Lagona2018}.\\

The remainder of the paper is structured as follows. In Section \ref{sec:Exploratory_data_analysis} we review advances in exploratory data analysis before proceeding to distributional models in Section \ref{sec:Dist_models}, general approaches to inference in Section \ref{sec:Inference}, and to hypothesis testing in Section \ref{sec:Testing}. Section \ref{sec:CorrRegr} discusses developments for correlation and regression. Section \ref{sec:Nonparam} focuses on advances in nonparametric curve estimation, Section \ref{sec:DimRedMeths} on methods for dimension reduction, and Section \ref{sec:ClassClust} on classification and clustering. Developments in modelling serial dependence, and spatial and spatio-temporal data, are reviewed in Sections \ref{sec:StochProcTSA} and \ref{sec:Spatial}, respectively. Advances in data depth, the design and analysis of experiments, order-restricted analysis, outlier detection, and compositional data analysis are considered more briefly in Section \ref{sec:FurtherTopics}. An overview of the software currently available for analysing directional data is provided in Section \ref{sec:Software}. The paper ends with the brief Section \ref{sec:Concl_Future} in which conclusions are drawn and potential future developments discussed.

\section{Exploratory data analysis}
\label{sec:Exploratory_data_analysis}

As for other types of data, the exploratory analysis of directional data usually begins with an inspection of some graphical summary of the data. Various adaptations of the popular rose diagram have been developed recently. \cite{Munro2012} introduced a moving rose diagram and applied it to circular datasets from the Earth sciences. \cite{Rodgers2014} proposed the wrap-around time series plot for displaying time series exhibiting periodic patterns. \cite{Morphet2010} proposed the circular dataimage, a graphical tool that uses a colour wheel to encode directions over a map. Rose diagrams, circular histograms, and other circular plots were adapted in \cite{Xu2020} so as to obtain area-proportional displays.\\

Circular boxplots have been investigated only relatively recently (but see \cite{Anderson1993}). For $\theta_1,\ldots,\theta_n\in[0,2\pi)$, \cite{Abuzaid2012} advocated one centred on the circular median, $M=\arg\min_{\phi\in[0,2\pi)}\sum_{i=1}^n d_c(\phi,\theta_i),$ where
\begin{align}
	\quad d_c(\phi,\theta)=\pi-|\pi-|\theta-\phi||\label{eqn:CircDistance}
\end{align}
is the shortest arc length distance between the two angles $\phi,\theta\in[0,2\pi)$ when represented as points on the circumference of the unit circle. In an attempt to mimic more closely Tukey's original construction, \cite{Buttarazzi2018} proposed a depth-based boxplot in which the observations are ranked from the antimedian to the median. For both proposals, the fences are calibrated assuming an underlying von Mises distribution (see Section \ref{sec:circ_models}).\\

The SiZer, an abbreviation for ``significant zero crossings of derivatives'' \citep{Chaudhuri1999}, is a handy tool used to identify statistically significant features at different scales, such as modes and antimodes, in univariate linear data. A circular adaptation of the SiZer, the CircSiZer, based on the kernel density estimator \eqref{eq:KDE1} and bootstrap confidence intervals to assess the significance of smoothed derivatives, was proposed in \cite{Oliveira2014}. It can also be employed to explore significant features in linear-circular regression. In both contexts, smoothing is based on a von Mises kernel with concentration parameter $\kappa$ (see \eqref{eq:KDE1} and \eqref{eq:NW1}). This kernel was shown not to be ``causal'' by \cite{Huckemann2016}, in the sense that its convolution with a circular density function is not guaranteed to maintain or reduce the number of modes as the level of smoothing, $1/\kappa$, increases. They proved that, among all the circular kernels satisfying certain mild assumptions, the wrapped normal (see Section \ref{sec:circ_models}) is the only one that yields circular causality. Employing such a kernel, they proposed the Wrapped SiZer (WiZer), with asymptotic confidence intervals used to assess the significance of smoothed derivatives. Extension of the SiZer methodology to spherical data led to the SphereSiZer of \cite{Vuollo2018}, itself inspired by the adaptation of the SiZer to bivariate linear data by \cite{Godtliebsen2002}. The SphereSizer uses a von Mises--Fisher kernel density estimator for data on $\mathbb{S}^2$ (see \eqref{eq:KDE}), and bootstrap confidence intervals to assess the significance of smoothed gradients. It produces a movie, indexed by the smoothing scale, that displays statistically significant density gradients as a vector field and highlights spherical regions with high density.

\section{Distributional models}
\label{sec:Dist_models}

\subsection{Circular models}
\label{sec:circ_models}

The probability density function (pdf) of an absolutely continuous circular random variable (rv) $\Theta$, $f_{\Theta}$, is such that $f_{\Theta}(\theta)\geq 0$ and $f_{\Theta}(\theta+2\pi)=f_{\Theta}(\theta)$ for almost all $\theta\in\mathbb{R}$. Also, $\int_{\theta}^{\theta+2\pi}f_{\Theta}(\omega)\,\text{d}\omega=1$. Thus, $f_{\Theta}$ is non-negative, $2\pi$-periodic, and integrates to $1$ over any interval of length $2\pi$. As a consequence of this latter property, it is usual to define a circular pdf through its values on $[0,2\pi)$ or $[-\pi,\pi)$. For instance, the circular uniform distribution, the most fundamental model for circular data corresponding to there being no preferred direction, has pdf $f_{\Theta}(\theta)=1/(2\pi)$, $\theta\in [0,2\pi)$. The circular cumulative distribution function (cdf) is defined as the non-periodic function $F_\Theta(\theta)=\int_{\theta_0}^{\theta}f(\omega)\,\mathrm{d}\omega$, with $\theta_0$ typically being $0$ or $-\pi$.\\

Six general approaches have often been used to generate models for circular data \citep [][Section 3.5]{Mardia1999a}: wrapping, projection, perturbation, conditioning, diffusion, and characterisations such as maximum likelihood or maximum entropy. The latter leads to distributions whose entropy is maximal under certain constraints, usually on their moments. The classical von Mises (vM) model with pdf
\begin{align}\label{eqn:vMden}
	f_{\Theta}(\theta; \mu,\kappa)=\frac{1}{2\pi \mathcal{I}_0(\kappa)}\exp\{\kappa\cos(\theta-\mu)\},
\end{align}
where $\mu\in[0,2\pi)$ is the mean direction, $\kappa>0$ its concentration parameter and $\mathcal{I}_{\nu}$ denotes the modified Bessel function of the first kind and order $\nu$, can be derived using no less than five of these constructions \citep [][Section 3.5.4]{Mardia1999a}. Due to their relevance in the sequel, below we give brief descriptions of wrapping, projection, and perturbation.\\

If $X$ is a linear rv then $\Theta = X~(\text{mod}~2\pi)\in [0,2\pi)$ is its wrapped circular counterpart. Alternatively, using a complex representation, $\Theta = \Arg\{\exp(\ic X)\}\allowbreak \in [-\pi,\pi)$. If $\phi_X$ is the characteristic function (cf) of $X$ then the cf of $\Theta$ is the set $\{\phi_{k}: k=0,\pm 1,\ldots\}$ where $\phi_k=\E(\e^{\ic k\Theta})=\phi_X(k)$, the $\phi_k$ being referred to as the Fourier coefficients or trigonometric moments \citep[Section 4.2.2]{Pewsey2013} of $\Theta$. Thus, $\Theta$ inherits the cf of $X$. If $f_X$ is the pdf of $X$ then the pdf of $\Theta$ is $f_{\Theta}(\theta)=\sum_{k=-\infty}^{\infty}f_X(\theta+2\pi k)$, the infinite sum generally not reducing to a closed-form expression. An important exception is the pdf of the wrapped Cauchy (WC) distribution,
\begin{align*}
	f_{\Theta}(\theta; \mu,\rho)=\frac{1}{2\pi}\frac{1-\rho^2}{1+\rho^2-2\rho\cos(\theta-\mu)},
\end{align*}
where $\mu =\Arg\{\E(\e^{\ic \Theta})\}\in[-\pi,\pi)$ is the mean direction and $\rho=|\E(\e^{\ic \Theta})|\in[0,1]$ the mean resultant length. More generally, the trigonometric moments of the WC model are given by
\begin{align}\label{eqn:WCTrigmom}
	\phi_k=(\rho\e^{\ic \mu})^k,~~k=1,2,\ldots~.
\end{align}
Perhaps the best known wrapped model is the wrapped normal distribution, obtained when $X\sim\mathrm{N}(\mu,\sigma^2)$. It can be used to closely approximate the vM distribution, and vice versa \citep{Pewsey2005}. Appealing wrapped circular models investigated recently include the wrapped: skew-normal \citep{Pewsey2000, Pewsey2006}, exponential and Laplace \citep{Jammalamadaka2004a}, $t$ \citep{Pewsey2007}, stable \citep{Pewsey2008}, and generalized normal-Laplace \citep{Reed2009}.\\

Projection involves projecting univariate or bivariate linear random variables onto $\mathbb{S}^1$. For example, stereographic projection of the linear random variable $X$ produces the circular rv $\Theta=2\tan^{-1}(X)$ \citep{Abe2010}. Radial projection of a bivariate linear random vector $\bm{X}=(X_1,X_2)$ onto $\mathbb{S}^1$ results in the circular rv $\Theta=\Arg(X_1+\ic X_2)$ or, equivalently, the random point $\bm{X}/||\bm{X}||$ on $\mathbb{S}^1$. Perhaps the best-known distribution of this latter type is the projected normal, also known as offset normal or angular Gaussian, the pdf of which can be symmetric or asymmetric, unimodal or bimodal in shape \citep[Section 3.5.6]{Mardia1999a}. Projection is a natural construction when modelling measurements relative to the position of an observer.\\

Perturbation involves multiplying a pdf by a suitable function so as to modulate its shape in some desired way. The cardioid distribution, with pdf
\begin{align*}
	f_{\Theta}(\theta; \mu,\rho)=\frac{1}{2\pi}\{1+2\rho\cos(\theta-\mu)\},
\end{align*}
where $|\rho|< 1/2$, is an example of perturbation of the circular uniform model. \cite{Umbach2009} adapted the perturbation approach of \cite{Azzalini1985} to the circular context, a special case of which is the sine-skewed family of distributions studied by \cite{Abe2011a}. If $g_{\Theta}$ denotes a base symmetric unimodal circular pdf with mean direction $\mu$ then the pdf of its sine-skewed extension is
\begin{align*}
	f_{\Theta}(\theta;\mu,\lambda)=g_{\Theta}(\theta-\mu)\{1+\lambda\sin(\theta-\mu)\},
\end{align*}
where $\lambda\in[-1,1]$ is a skewing parameter. The symmetric base pdf is unperturbed when $\lambda=0$, otherwise it is skewed in the anticlockwise direction ($\lambda>0$) or the clockwise direction ($\lambda<0$). Sine-skewed densities have the same normalising constants as their base symmetric densities, but can model only moderate departures from symmetry and are not necessarily unimodal.\\

An overarching family of symmetric unimodal circular distributions containing, amongst others, the circular uniform, cardioid, vM, and WC distributions, was proposed by \cite{Jones2005}. Its pdf is
\begin{align*}
	f_{\Theta}(\theta;\mu,\rho,\psi)\propto \{1+\tanh(\kappa\psi)\cos(\theta-\mu)\}^{1/\psi},
\end{align*}
where $\mu\in[0,2\pi)$ is the mean direction, $\kappa \geq 0$ is a concentration parameter, and $\psi\in\mathbb{R}$ is a shape index.\\

Recently, \cite{Kato2015} proposed a highly flexible extension of the WC model obtained by broadening the trigonometric moments in \eqref{eqn:WCTrigmom} to $\gamma(\rho\e^{\ic \lambda})^{-1}\{\rho\e^{\ic (\mu+\lambda)}\}^k$. The resulting family is unimodal and has pdf
\begin{align}
	f_{\Theta}(\theta; \mu,\rho,\gamma,\lambda)=\frac{1}{2\pi} \left\{1+2\gamma\frac{\cos(\theta-\mu)-\rho\cos\lambda}{1+\rho^2-2\rho\cos(\theta-\mu-\lambda)}\right\},\label{eq:fourparam}
\end{align}
where $\mu\in[0,2\pi)$, $\rho\in[0,1)$, $\gamma\in[0,(1+\rho)/2]$, and $\lambda\in[-\pi,\pi)$ satisfies $\rho\gamma\cos\lambda \geq (\rho^2+2\gamma-1)/2$. Its cdf also has a closed form. Its reparametrisation in terms of standard trigonometric moments \citep{Pewsey2004a} has parameters with clear interpretations and is the one generally used to perform inference.\\
	
Constructions based on M\"{o}bius transformation, Brownian motion, and transformation of argument have also been used recently to generate more flexible families of circular models. A M\"{o}bius transformation preserving the unit circle maps a point on the unit circle, $\Theta$, to another, $\Theta^*$, via
\begin{align*}
	\e^{\ic \Theta^*}=\e^{\ic \phi}\frac{\e^{\ic \Theta}+r\e^{\ic \omega}}{r \e^{\ic (\Theta-\omega)}+1},
\end{align*}
where $\phi, \omega \in [-\pi,\pi)$ and $r\in[0,1)$, or equivalently via
\begin{align}\label{eqn:Mobius2}
	\Theta^*=\phi+\omega+2\tan^{-1}[w_r\tan\{(\Theta-\omega)/2\}],
\end{align}
where $w_r=(1-r)/(1+r)$. Applying this M\"{o}bius transformation to a circular uniform rv results in a WC rv \citep{McCullagh1996}. \cite{Kato2010a} and \cite{Wang2012} studied families obtained by applying the same transformation to vM and cardioid random variables, respectively. \cite{Jacimovic2017} related the former family to the dynamics of coupled oscillators. \cite{Kato2013} varied the Brownian motion specification leading to the WC distribution so as to generate a four-parameter extension of it. The families of \cite{Kato2010a}, \cite{Wang2012}, and \cite{Kato2013} have pdfs that can be symmetric or asymmetric and unimodal or bimodal in shape.\\

Transformation of argument involves replacing the argument of an existing circular pdf, $f_\Theta$, by some function of $\theta$. \cite{Jones2012} used this approach to derive inverse Batschelet distributions. The resulting four-parameter distributions are unimodal and highly flexible in shape. Unlike the smooth unimodal models of \cite{Kato2015}, inverse Batschelet distributions can adopt Laplace-like shapes. The most flexible unimodal circular models currently available are those of \cite{Jones2012} and \cite{Kato2015}.\\

Of the modelling approaches available for multimodal circular data, finite mixtures have proven the most popular. Mixture models with vM components have recently received renewed attention (\citet{Mooney2003,Fu2008}; see, also, Section \ref{sec:spherical_models}). The pdf of an $m$ component vM mixture is
\begin{align}\label{eqn:vMMixden}
	f_{\Theta}(\theta;\bm{p}, \bm{\mu},\bm{\kappa})=\sum_{j=1}^m p_jf_{\Theta}(\theta;\mu_j,\kappa_j),
\end{align}
where $\bm{p}=(p_1,\ldots,p_m)'$ is a vector of mixing probabilities satisfying $\sum_{j=1}^m p_j=1$, $\bm{\mu}=(\mu_1,\ldots,\mu_m)'$, $\bm{\kappa}=(\kappa_1,\ldots,\kappa_m)'$, and $f_{\Theta}(\theta;\mu_j,\kappa_j)$ is as in \eqref{eqn:vMden}. When the interpretation of the parameters of the component densities is straightforward, so is the interpretation of the parameters of a mixture.\\

More generally, \cite{Holzmann2004} established conditions for the identifiability of mixtures of location-scale extensions of wrapped circular models including the wrapped symmetric $\alpha$-stable, wrapped normal, and WC distributions. Mixtures with circular triangular \citep{McVinish2008}, skew-rotationally symmetric \citep{Miyata2019}, and power Batschelet \citep{Mulder2020} components have also been considered.\\

Finally, we consider three alternative approaches to modelling multimodal circular data. Generalized von Mises models \citep{Gatto2008,Gatto2009}, with the density of the generalized vM distribution of order $m$ being
\begin{align}
	f_{\Theta}(\theta;\bm{\mu},\bm{\kappa})=\exp\bigg\{\kappa_0+\sum_{j=1}^m \kappa_j\cos(j(\theta-\mu_j))\bigg\},\label{eq:gvm}
\end{align}
where $\bm{\mu}=(\mu_1,\ldots,\mu_m)'$, $\bm{\kappa}=(\kappa_1,\ldots,\kappa_m)'$, $\mu_j\in [0,2\pi/j)$, and $\kappa_j\geq 0$, have a long history dating back to \cite{Maksimov1967}. The normalising constant, $\e^{\kappa_0}$, must generally be computed numerically. \cite{Fernandez-Duran2004} revisited work by \cite{Fejer1916} when defining a family of circular distributions based on non-negative trigonometric (i.e.\ truncated Fourier) sums. Whilst they do not require the calculation of normalising constants, fitted densities of this type tend to have many parameters and display minor harmonic modes that need not be supported by the data. Recently, \cite{Taniguchi2020} proposed flexible models for circular data obtained by normalising the spectra of stochastic processes, the residue theorem being used to calculate their normalising constants. The interpretation of the parameters of all three of these types of model is, however, generally difficult.

\subsection{Models for toroidal data}
\label{sec:toroidal_models}

Let $(\Theta_1, \Theta_2)$ denote the angular coordinates of a random vector distributed on the torus $\mathbb{T}^2=\mathbb{S}^1\times\mathbb{S}^1$. Some of the approaches used to generate models for toroidal data are extensions of those introduced in Section \ref{sec:circ_models}. These include maximum entropy characterisation, projection, and wrapping \citep{Johnson1977,Baba1981,Mardia2008}. Models for univariate and bivariate axial data were proposed by \cite{Arnold2006}.\\

The bivariate von Mises model of \cite{Mardia1975b} is a maximum entropy (equivalently, an exponential family) distribution with pdf
\begin{align}\label{eqn:MardiaBvMden}
	f_{\Theta_1, \Theta_2}(\theta_1, \theta_2) \propto &\; \exp\{\kappa_1\cos(\theta_1-\mu_1)+\kappa_2\cos(\theta_2-\mu_2)\\
	& + (\cos(\theta_1-\mu_1),\sin(\theta_1-\mu_1))\bm{A}(\cos(\theta_2-\mu_2),\sin(\theta_2-\mu_2))'\}\nonumber
\end{align}
where ~$\mu_1,\mu_2\in[-\pi,\pi)$, $\kappa_1,\kappa_2 \geq 0$, and $\bm{A}$ is a $2\times 2$ matrix. The most compact form for its normalising constant involves a doubly infinite sum \citep{Mardia2010}. The model has eight parameters, three more than the minimum of five required to control the locations and concentrations of the two marginal variables and the dependence between them. Moreover, their interpretation is difficult \citep{Mardia2007}. In the search for five-parameter analogous of the bivariate normal distribution, \cite{Singh2002}, \cite{Mardia2007}, and \cite{Kent2008} proposed the sine, cosine, and hybrid submodels of \eqref{eqn:MardiaBvMden}, respectively. The properties of these three submodels were compared in \cite{Kent2008} and \cite{Mardia2012}. Their normalising constants are available as infinite sums. Their conditional distributions are vM, but their marginal pdfs are generally not and, for some parameter values, can be bimodal.\\

Extensions of \eqref{eqn:MardiaBvMden} and its cosine submodel, for use with data on $\mathbb{T}^d$, $d\geq2$, were proposed by \cite{Mardia2008} and \cite{Mardia2005}, respectively. No simple closed analytic form is generally available for the normalising constant of the sine multivariate von Mises model of \cite{Mardia2008}, but its conditional distributions are vM and thus its parameters can be estimated by maximising the pseudo-likelihood. Conditions on its parameters to ensure unimodality were established in \cite{Mardia2014} and pseudo-likelihood regularised approaches were given in \cite{RodriguezLujan2015,RodriguezLujan2017}. A multivariate extension of the second-order generalized vM (GvM$_2$) distribution (with $m=2$ in \eqref{eq:gvm}), obtained by conditioning a multivariate Gaussian distribution on $\mathbb{R}^{2d}$ to $\mathbb{T}^d$, was proposed by \cite{Navarro2017}. Its one-dimensional conditional distributions are GvM$_2$, and the sine multivariate vM model is a special case of it. \cite{Hassanzadeh2018} recently proposed a model for data on $\mathbb{T}^2$ obtained using a conditional specification construction involving GvM$_2$ pdfs.\\

The range of available models can be expanded beyond toroidal analogues of the bivariate normal distribution using the projection approach of \cite{Saw1983} to construct models with more flexible specified marginal distributions. A simpler marginal specification approach can be traced back to \cite{Wehrly1980}. They proposed toroidal pdfs of the form
\begin{align}\label{eqn:JWBivariateDensity}
f_{\Theta_1, \Theta_2}(\theta_1, \theta_2)=2\pi
f_{\Theta_1}(\theta_1)f_{\Theta_2}(\theta_2)f_{\Omega}(2\pi[
F_{\Theta_2}(\theta_2)-q F_{\Theta_1}(\theta_1)]),
\end{align}
where $f_{\Theta_j}$ and $F_{\Theta_j}$ are the marginal pdf and cdf of $\Theta_j$, $j=1,2$, $f_{\Omega}$ is a circular \textit{binding} pdf, and $q=\pm1$ determines whether the dependence is positive or negative. Various models obtained using \eqref{eqn:JWBivariateDensity} with different choices for $f_{\Theta_1}$, $f_{\Theta_2}$, and $f_{\Omega}$ are referred to in \cite{Jones2015}.\\

\cite{Kato2015a} considered a case of \eqref{eqn:JWBivariateDensity} having a closed-form pdf that is unimodal and pointwise symmetric, and marginal and conditional distributions that are all WC. This bivariate WC model can also be obtained by applying a M\"{o}bius transformation to a tractable toroidal model with circular uniform marginal distributions derived by \cite{Kato2009} using a Brownian motion construction.\\

\cite{Shieh2005} were the first to note the relationship between \eqref{eqn:JWBivariateDensity} and copulas \citep{Sklar1959}, toroidal pdfs being generated through
\begin{align}\label{eqn:copula}
f_{\Theta_1, \Theta_2}(\theta_1,
\theta_2)=f_{\Theta_1}(\theta_1)f_{\Theta_2}(\theta_2)c(
F_{\Theta_1}(\theta_1),F_{\Theta_2}(\theta_2)),
\end{align}
where $c$ is a \textit{copula} pdf. \cite{Garcia-Portugues2013} imposed periodic restrictions on $c$ to construct alternatives to \eqref{eqn:JWBivariateDensity}. \cite{Jones2015} revisited \eqref{eqn:JWBivariateDensity} and considered, instead of \eqref{eqn:copula},
\begin{align*}
f_{\Theta_1, \Theta_2}(\theta_1,
\theta_2)=4\pi^2f_{\Theta_1}(\theta_1)f_{\Theta_2}(\theta_2)c_{\circ}(2\pi
F_{\Theta_1}(\theta_1),2\pi F_{\Theta_2}(\theta_2)),
\end{align*}
where now $c_{\circ}$ is what they coined a \textit{circula} density, with arguments that are circular uniform. For any circula density, $c_{\circ}(\theta_1,\theta_2)=c_{\circ}(\theta_1\pm 2\pi k,\theta_2\pm 2\pi l), k,l\in \mathbb{Z}^+$. \cite{Jones2015} showed that \eqref{eqn:JWBivariateDensity} corresponds to $c_{\circ}(\phi_1,\phi_2)=\frac{1}{2\pi} f_{\Omega}(\phi_2-q\phi_1),$ the pdf of $(\Phi_1,\Phi_2)$, where $\Phi_1$ and $\Phi_2=\Phi_1-q\Omega~(\textrm{mod}~2\pi)$ are circular uniform random variables and $\Omega$ follows the circular pdf $f_{\Omega}$ independently of $\Phi_1$. This circula density is a particular case of the (infinite) Fourier series approach to obtaining circula densities proposed recently by \cite{Kato2018}. They considered six cases of their general construction, all having simple closed-form expressions for their densities. \cite{Jupp2015} extended the idea of circulas to compact Riemannian manifolds.\\

Recently, \cite{Ameijeiras-Alonso2019a} used a sine-skewing approach (see Section \ref{sec:circ_models}) to generate models for asymmetric data on $\mathbb{T}^d$, $d\geq 2$. Alternative approaches to modelling such data make use of pdfs obtained from truncated Fourier series \citep{Pertsemlidis2005, Fernandez-Duran2014} or normalised spectra of stochastic processes \citep{Taniguchi2020}. These models have properties analogous to those mentioned in Section \ref{sec:circ_models} for their circular counterparts.

\subsection{Models for cylindrical data}
\label{sec:cylindrical_models}

Let $(\Theta, X)$ denote the coordinates of a random vector distributed on the cylinder $\mathbb{S}^1\times\mathbb{R}$. Approaches used to generate models for cylindrical data have included wrapping \citep{Johnson1977}, conditioning, marginal specification, and maximum entropy characterisation.\\

\cite{Mardia1978a} conditioned a trivariate normal distribution to obtain a six-parameter cylindrical model for which the marginal distribution of $\Theta$ is vM and the conditional distribution of $X|\Theta=\theta$ is normal. More recently, \cite{Kato2008} proposed an eight-parameter extension of it having generalized vM distributions for $\Theta$ and $\Theta|X=x$.\\

An analogous marginal specification approach to that used to derive pdf \eqref{eqn:JWBivariateDensity} can be employed to obtain a cylindrical model with pdf
\begin{align}\label{eqn:JWMargSpecDen}
	f_{\Theta, X}(\theta, x)=2\pi
	f_{\Theta}(\theta)f_{X}(x)f_{\Omega}(2\pi[F_{\Theta}(\theta)-q
	F_{X}(x)]),
\end{align}
where $f_{\Theta}$ and $f_X$ are the marginal pdfs of $\Theta$ and $X$, respectively, and $F_{\Theta}$ and $F_X$ their cdfs. \cite{Johnson1978} considered cases of \eqref{eqn:JWMargSpecDen} with $X$ normally distributed,
and circular uniform or vM distributions for $\Theta$. Recently, other cases have been applied to model cylindrical data from disciplines such as wind energy analysis \citep{Carta2008,Zhang2018}, ocean engineering \citep{Soukissian2014}, and image analysis \citep{Roy2017}.\\

\cite{Johnson1978} also proposed three maximum entropy cylindrical models, with conditional distributions that are vM and normal or exponential. The dependence structures of all three models are, however, severely constrained. Their model having vM and exponential marginal distributions when $\Theta$ and $X$ are independent was recently extended by \cite{Abe2017}, \cite{Imoto2019}, and \cite{Abe2018} so as to admit skew and more flexible models for $\Theta$ and $X$, respectively.\\

Recently, \cite{Mastrantonio2018} proposed the joint projected normal and skew-normal distribution, the first model for multivariate cylindrical data. It is highly flexible and closed under marginalisation.

\subsection{Models for spherical data}
\label{sec:spherical_models}

Suppose $\bm{X}$ is a unit random vector on $\mathbb{S}^{d}$. Perhaps the best-known model for spherical data is the von Mises--Fisher (vMF) distribution, with pdf
\begin{align}\label{eqn:vMFDen}
	f_{\bm{X}}(\bm{x}; \bm{\mu}, \kappa)=\frac{\kappa^{(d-1)/2}}{(2 \pi)^{(d+1)/2} \mathcal{I}_{(d-1)/2}(\kappa)}\exp\{\kappa\bm{x}'\bm{\mu}\},
\end{align}
where $\bm{\mu}\in \mathbb{S}^{d}$ is the mean direction vector and $\kappa\geq0$ is a concentration parameter. Other classical models are the Kent, Fisher--Watson, Bingham--Mardia, Bingham, and Watson distributions \citep[Chapter 9]{Mardia1999a}, the last two being models for axial data. They are all submodels of the Fisher--Bingham exponential family of distributions with pdf
\begin{align}\label{eq:FB}
	f_{\bm{X}}(\bm{x}; \bm{\mu}, \kappa,\bm{A})\propto \exp\{\kappa\bm{x}'\bm{\mu}+\bm{x}'\bm{A}\bm{x}\},
\end{align}
where $\bm{A}$ is a symmetric $(d+1)\times (d+1)$ matrix, and $\bm{\mu}$ and $\kappa$ play the same roles as in \eqref{eqn:vMFDen}. Evaluation of the distribution's normalising constant is challenging but \cite{Kume2018} showed how it can be achieved using the holonomic gradient method. \cite{Kent2018} developed an efficient acceptance-rejection method of simulating variates from Fisher--Bingham distributions on spheres and related manifolds. \cite{Kent2016} introduced a five-parameter special case of the Fisher--Bingham model for use with data patterns that are unimodal and concentrated near a great circle. More recently, \cite{Kim2019} proposed two kinds of small-sphere distributions, one of which is a member of the Fisher–Bingham family. Previously, \cite{Oualkacha2009} had developed an alternative to the Bingham distribution for modelling symmetric axial data, with a simple closed-form normalising constant.\\

Rotationally symmetric (RS) spherical pdfs depend on $\bm{x}$ only through $\bm{x}'\bm{\mu}$ and, as a consequence, have contours that are circular when $\bm{x}\in \mathbb{S}^{2}$. Historically, the vMF has been the most important such model. In recent years, numerous other RS families have been proposed in the literature: Section 2.3.2 of \cite{Ley2017a} summarises many of them. The spherical logistic distribution of \cite{Moghimbeygi2020} provides a multimodal and RS extension of the vMF, with a closed-form normalising constant when $d=2$. Another recent addition is the highly tractable 
spherical Cauchy distribution of \cite{Kato2020}, which extends the WC to $\mathbb{S}^d$ and has a very simple normalising constant.\\

The pdf of the Kent distribution, i.e.\ \eqref{eq:FB} constrained to have $\bm{A}\bm{\mu}=\bm{0}$, has elliptical contours and hence can be used to model certain departures from isotropy. Recently, \cite{Paine2018} proposed the elliptically symmetric angular Gaussian (ESAG) as an alternative. As simulation from it and the computation of its pdf are far quicker than for the Kent model, this model is a particularly appealing alternative when the use of computer intensive methods is being contemplated. Other Kent-like alternatives with the advantages of the ESAG model are the scaled vMF family of \cite{Scealy2019}, which has an additional parameter controlling tail-weight, and the tangent models of \cite{Garcia-Portugues2020}.\\

The sine-skewed circular distributions of Section \ref{sec:circ_models} are special cases of the skew-rotationally symmetric (SRS) distributions proposed as models for asymmetric spherical data by \cite{Ley2017b}. In turn, SRS models are spherical analogues of the skew-symmetric linear models of \cite{Wang2004}. The perturbation of spherical distributions was studied in greater generality by \cite{Jupp2016}.\\

Asymmetric or bimodal spherical data can be modelled using the general projected normal family of distributions referred to in \citet[][Section 9.3.3]{Mardia1999a} and advocated more recently from a Bayesian perspective in \citet{Hernandez-Stumpfhauser2017}. \cite{Nunez-Antonio2020} proposed a projected gamma distribution to model data on the positive orthant of $\mathbb{S}^d$.\\

In Section \ref{sec:ClassClust} we summarise recent developments in the use of mixture distributions with spherical component pdfs as a means of modelling multimodal spherical data. The flexible directional log-spline pdfs of \cite{Ferreira2008}, based on thin-plate splines on $\mathbb{S}^d$ \citep{Taijeron1994}, provide an alternative means of modelling multimodality and skewness. They are given by
\begin{align}\label{eqn:DLSDen}
	f_{\bm{X}}(\bm{x}; \bm{c}, \bm{\mathcal{K}},m) = \exp\bigg\{c_0+\sum_{j=1}^mc_j R_d(\bm{x};\bm{k}_j)\bigg\},
\end{align}
where $m\geq 1$, $(c_0,c_1,\ldots,c_m)'\in \mathbb{R}^{m+1}$, $\bm{\mathcal{K}}=\{\bm{k}_1,\ldots,\bm{k}_m\}$ is a set of knot vectors in $\mathbb{S}^d$, and $R_d(\cdot;\bm{k}_j)$ are real-valued spline basis functions on $\mathbb{S}^d$ that, when evaluated at $\bm{x}\in\mathbb{S}^d$, are functions of $\bm{x}'\bm{k}_j$, for $j=1,\ldots,m$. \cite{Ferreira2008} proposed a Bayesian inferential approach for \eqref{eqn:DLSDen}. \cite{Fernandez-Duran2014a} constructed pdfs on $\mathbb{S}^2$ through non-negative trigonometric sum expansions in terms of spherical angles.

\section{General approaches to inference}
\label{sec:Inference}

Historically, inference for the models in Section \ref{sec:Dist_models} has generally been frequentist: sometimes using the method of (trigonometric) moments but more generally being likelihood-based. The maximum likelihood (ML) estimators of full exponential family models are moment estimators \citep[][Chapter 4]{VanderVaart2000} and, as a consequence, closed-form expressions exist for the ML estimators of, for example, the vM distribution and the cylindrical model of \cite{Mardia1978a}. Exact ML inference for the highly challenging Fisher--Bingham model and its submodels identified in Section \ref{sec:spherical_models} was developed recently in \cite{Kume2018}. More generally, maximisation of the log-likelihood has to be performed numerically. When available, method of moments estimates can be used as starting values for that process. For some models, statistical inference based on the full log-likelihood is intractable and pseudo-likelihood methods have been employed \citep{Kent2008}. Score matching estimators, inspired by the \cite{Hyvarinen2005} scoring rule, circumvent the need to calculate normalising constants for directional distributions \citep{Mardia2016,Mardia2018a,Takasu2018}. \\

Large-sample results for ML-based inference generally assume standard regularity conditions to apply and the asymptotic normality of ML estimators \citep[][Chapter 5]{VanderVaart2000}. \cite{Pewsey2004a} employed the delta method to obtain the asymptotic distribution of the fundamental measures of central location, concentration, skewness, and kurtosis used in the analysis of circular data. For some models, large-sample ML-based inference for parameter values on the boundary of the parameter space will be of interest, and the results of \cite{Self1987} can be employed \citep[see, e.g.,][]{Shieh2005}. For small-sized samples, bootstrap confidence interval constructions have become increasingly popular \citep[Chapter 5]{Pewsey2013}. Computer intensive resampling methods in hypothesis testing are mentioned in Section \ref{sec:Testing}.\\

Recently, Le Cam's local asymptotic normality approach to inference has been adapted to problems in directional statistics: see Section \ref{sec:Testing}, and Section 5 of \cite{Ley2017a}. The first such adaptation appeared in \cite{Ley2013}, where optimal rank-based estimators of the location parameter of rotationally symmetric spherical distributions were proposed. More recently, \cite{Paindaveine2020} considered inference under high concentration for the spherical location of a semi-parametric class of rotationally symmetric distributions.\\

Bayesian inferential techniques have become increasingly popular in recent years, often being implemented using Markov chain Monte Carlo (MCMC) methods. A general approach to MCMC simulation on embedded Riemannian manifolds was introduced by \cite{Byrne2013}, and illustrated for the Fisher--Bingham distribution. Bayesian approaches to inference have been developed for the: vM \citep{Damien1999} and mixtures thereof \citep{Mulder2020a}, WC \citep{Ghosh2019}, vMF \citep{NunezAntonio2005a, Hornik2013}, bivariate vM \citep{Mardia2010}, and projected normal \citep{NunezAntonio2005} distributions. \cite{Bhattacharya2009} considered Bayesian inference for circular distributions with unknown normalising constants. \cite{Fallaize2016} gave a Monte Carlo exact Bayesian method of inference for the Bingham distribution. Bayesian approaches based on projected and wrapped models have become popular for a range of applications: see Sections \ref{sec:spherical_models} and \ref{sec:Spatial}. Scoring rules provide an alternative to the traditional Bayesian formulation, and have been applied for the vMF distribution \citep{Giummole2019}.\\

Robust estimators have been proposed for the parameters of the vM and wrapped normal distributions \citep{Agostinelli2007}, the vMF distribution \citep{Kato2016}, and a range of other circular distributions in a series of papers referred to by \cite{Laha2019}.\\

Asymptotic results for extrinsic and intrinsic means on manifolds, including $\mathbb{S}^d$, were obtained in \cite{Bhattacharya2003,Bhattacharya2005}; see \cite{Bhattacharya2014} for a review on the topic. \cite{Hotz2013} gave a detailed comparison between extrinsic and intrinsic means on $\mathbb{S}^1$. Nonparametric inference on intrinsic means on circles and spheres, however,
can be fundamentally different from its Euclidean analogues due to the
effect of \emph{smeariness} (asymptotic rates are slower than $n^{-1/2}$) present on $\mathbb{S}^1$ \citep{Hotz2015} and $\mathbb{S}^d$ \citep{Eltzner2019}. The related effect of \emph{finite sample smeariness} has been studied on the circle by \cite{Hundrieser2020} and may affect all of the circular distributions mentioned previously. As a consequence, quantile-based tests may be inappropriate while suitable bootstrap tests remain valid.\\

The inferential impact of the reference systems used for circular distributions was explored recently in \cite{Mastrantonio2019}.

\section{Hypothesis testing}
\label{sec:Testing}

Here we consider hypothesis tests for uniformity, symmetry, location, concentration, goodness-of-fit, and other testing scenarios. Calibration of the tests has generally been based on asymptotic theory and, for small to moderate sized samples, the use of resampling methods.

\subsection{Uniformity}
\label{sec:TestingUniformity}

Uniformity (or isotropy), corresponding to there being no preferred direction, is the most important dividing hypothesis in directional statistics. \cite{Garcia-Portugues2020a} provide an extensive review of tests for it.\\

Sobolev tests \citep{Beran1968,Beran1969,Gine1975} form, by far, the most extensive class of tests for uniformity on $\mathbb{S}^d$. Given a sample $\bm{X}_1,\ldots,\bm{X}_n$ on $\mathbb{S}^d$, Sobolev statistics take the form
\begin{align}\label{eqn:Sobo}
	S_{n}(\{v_k^2\})=\frac{1}{n}\sum_{i,j=1}^n\sum_{k=1}^\infty v_k^2 h_k(\bm{X}_i,\bm{X}_j),
\end{align}
where
\begin{align*}
	h_k(\bm{u}, \bm{v}) = \begin{cases}
		2 \cos(k \cos^{-1}(\bm{u}'\bm{v})),~~~d=1, \\
		\big(1+ \frac{2k}{d}\big) C_k^{(d-1)/2} (\bm{u}' \bm{v}),~~~d>1,
	\end{cases}
\end{align*}
$C_k^{(d-1)/2}$ is the $k$-th Gegenbauer polynomial of index $(d-1)/2$, and the $v_k^2$ should decay fast enough to ensure convergence in (\ref{eqn:Sobo}). Different choices for $\{v_k^2\}$ give different local optimality properties, consistencies, and powers against specific kinds of alternatives. For example, the choices $v_k=\delta_{kj}$, $j=1,2,$ give, respectively, the test statistics of \cite{Rayleigh1919} and \cite{Bingham1974}. Both were modified by \cite{Jupp2001} to improve their convergence under the null hypothesis. The alternatives for which the Rayleigh and Bingham tests are inconsistent were identified by \cite{Ehler2011} as the minimisers of certain potentials over $\mathbb{S}^d$. The Rayleigh test plays a key role in the CUSUM-based test for circular uniformity developed by \cite{Lombard2012}. \cite{Pycke2007,Pycke2010} proposed uniformity tests on $\mathbb{S}^2$ and $\mathbb{S}^1$ based on the geometric mean of pairwise chordal distances, whilst \cite{Bakshaev2010} gave an analogous approach based on the arithmetic mean. The Bayesian optimality of Sobolev tests on $\mathbb{S}^1$ was studied by \cite{Sun2019}.\\

``Data-driven'' Sobolev tests are obtained by using an information criterion to truncate the infinite series in \eqref{eqn:Sobo}. This approach was used to obtain tests of uniformity on $\mathbb{S}^1$ by \cite{Bogdan2002}, and on compact Riemannian manifolds by \cite{Jupp2008, Jupp2009}. Such truncation simplifies the computation of \eqref{eqn:Sobo} and its asymptotic distribution, the latter effectively being chi-squared. A variation on this approach was pursued recently by \cite{Jammalamadaka2020}, who proposed increased levels of truncation of \eqref{eqn:Sobo} on $\mathbb{S}^1$ and $\mathbb{S}^2$, so as to obtain a normal limit rather than the usual weighted sum of chi-squared random variables appearing in the asymptotic null distribution of \eqref{eqn:Sobo}.\\

\cite{Su2011} considered spherical harmonics and exponential models as alternatives to uniformity, and derived score tests strongly related to Sobolev tests to test for uniformity against them. Also related to Sobolev tests, \cite{Garcia-Portugues2020b} proposed a class of tests based on the projected empirical cdf that yields extensions for data on $\mathbb{S}^d$ of the \cite{Watson1961} and \cite{Rothman1972} tests for circular uniformity, and a novel Anderson--Darling-like test for uniformity on $\mathbb{S}^d$.\\

Recent non-Sobolev tests for circular uniformity include the four-point Cram\'er--von Mises test of \cite{Feltz2001}, the likelihood-ratio test against a mixture with symmetric wrapped stable and circular uniform components of \cite{SenGupta2001}, the spacings-based Gini mean difference test of \cite{Tung2013}, and the Bayesian tests of \cite{Mulder2021} against the vM distribution and the kernel density estimator \eqref{eq:KDE1}. Tests for uniformity on $\mathbb{S}^d$ include that of \cite{Fay2013}, based on needlets (see Section \ref{sec:NP_DensityEst:series}), and those of \cite{Lacour2014} and \cite{Kim2016} for ``noisy'' data on $\mathbb{S}^2$, i.e. where the density of the observations is a convolution of an error pdf with a true underlying pdf. \cite{Cuesta-Albertos2009} proposed a projection-based test, and \cite{Ebner2018} one based on a sum of weighted nearest-neighbour distances. \cite{Cutting2020} investigated tests for uniformity on $\mathbb{S}^d$ against axial alternatives.\\

Tests for uniformity when $d\to\infty$ as $n\to\infty$ are scarcer. \cite{Cai2012} and \cite{Cai2013} proposed tests based on $\max_{i<j}|\bm{X}_i'\bm{X}_j|$. \cite{Paindaveine2016} and \cite{Cutting2017} studied a standardised Rayleigh statistic under uniformity and vMF alternatives, respectively.\\

Simulation studies comparing the performance of various tests for uniformity on $\mathbb{S}^1$ have been carried out by \cite{Landler2018}, on $\mathbb{S}^d$, $d\geq1$, by \cite{Garcia-Portugues2020b}, and on $\mathbb{S}^d$, $d\geq2$, by \cite{Figueiredo2003} and \cite{Figueiredo2007}. \cite{Humphreys2017} and \cite{Landler2019} performed simulation experiments to compare the performance of tests for circular uniformity when the data are grouped.

\subsection{Symmetry}
\label{sec:TestingSymmetry}

There are at least four forms of symmetry that might be of interest in the analysis of circular data: reflective symmetry about an unknown direction, reflective symmetry about a known median axis, reflective symmetry about a specified median axis, and $\ell$-fold symmetry. \cite{Pewsey2002a,Pewsey2004} described these various forms of symmetry and proposed simple, trigonometric moment-based, omnibus tests for the first two scenarios. More recently, \cite{Ameijeiras-Alonso2020}, \cite{Meintanis2019}, and \cite{Ley2014b} proposed tests for the same two setups that are optimal against the $k$-sine-skewed models of \cite{Umbach2009}.\\

As we saw in Section \ref{sec:spherical_models}, many of the models for spherical data are rotationally symmetric. Recently, \cite{Garcia-Portugues2020} developed semi-parametric tests for rotational symmetry when the axis of symmetry is known or unknown. Previously, \cite{Ley2017b} had developed two tests for rotational symmetry about a known centre within the class of skew-rotationally symmetric distributions.\\

For data on $\mathbb{S}^2$, \cite{Jammalamadaka2019a} proposed tests for various types of symmetry, as well as uniformity, based on spherical harmonics.

\subsection{Location and concentration}
\label{sec:Test:LocationConcentration}

Various tests for the parameters of vM and vMF models have been developed in recent years. \cite{Larsen2002} proposed improved likelihood-ratio tests for the: (i) mean direction of a vM distribution; (ii) equality of mean directions of two vM distributions; (iii) concentration of a vMF distribution; (iv) equality of concentrations of two vMF distributions. \cite{Watamori2005} introduced improved likelihood-ratio and score tests for homogeneity of concentration in vMF distributions. Their score tests were derived and studied from an alternative perspective in the review of \cite{Ley2014a}. \cite{Laha2015} investigated the robustness of tests for the locations of the vM and vMF models. \cite{Gatto2017} proposed a simultaneous test for the mean direction and concentration of a vMF distribution. For data from vMF distributions with a common unknown concentration, \cite{Rumcheva2017} proposed an improved version of the multi-sample likelihood-ratio test for the equality of mean directions.\\

Widening the scope to rotationally symmetric spherical distributions, \cite{Tsai2009} introduced asymptotically efficient rank tests for the equality of the modal direction vectors of two unimodal rotationally symmetric spherical distributions. \cite{Ley2015} investigated the high dimensional robustness of Watson's test for the mean direction. \cite{Paindaveine2015} proposed optimal rank-based tests for the mean direction, and \cite{Ley2017} used the invariance principle to construct rank-based semi-parametric tests for the homogeneity of mean directions. \cite{Paindaveine2017} investigated the problem of testing for a specified mean direction when the underlying distribution tends to uniformity. \cite{Cutting2017a} proposed tests for concentration in low and high dimensions, and \cite{Verdebout2015,Verdebout2017} tests for homogeneity of concentration.\\

A simultaneous saddlepoint test for the mean direction and dispersion of the wrapped symmetric stable model was proposed in \cite{Gatto2000}, and a nonparametric extension of it, for an assumed underlying unimodal circular distribution, in \cite{Gatto2006}. \cite{Amaral2007} proposed nonparametric bootstrap and permutation tests for the equality of the mean directions of directional distributions.\\

For axial data, \cite{Figueiredo2017} proposed and explored the performance of bootstrap and permutation counterparts of a high-concentration test for the homogeneity of principal axes of Watson distributions.

\subsection{Goodness-of-fit}
\label{sec:GOFTesting}

Given an independent and identically distributed circular sample, $\Theta_1,\ldots,\Theta_n$, with underlying cdf $F$, the goodness-of-fit testing problem of $H_0: F=F_0$ versus $H_1: F\neq F_0$, where $F_0$ is a fully specified cdf, can be addressed using tests for circular uniformity. Appealing to the probability integral transform, testing $H_0$ is equivalent to testing the sample $2\pi F_0(\Theta_1),\ldots,2\pi F_0(\Theta_n)$ on $[0,2\pi)$ for circular uniformity. When the parameters of a model are estimated, the sampling distributions of the test statistics are affected. However, those sampling distributions can be approximated using the parametric bootstrap \citep[][Chapter 6]{Pewsey2013}.\\

For data on other supports, the situation is more complicated as there is no canonical transformation to uniformity. Examples of bootstrap goodness-of-fit tests for models fitted to toroidal data appear in \cite{Jones2015}, \cite{Pewsey2016}, and \cite{Kato2018}: the approach used in the latter essentially being based on the multivariate probability integral transform. Almost-canonical transformations for other Riemannian manifolds have been proposed recently in \cite{Jupp2020}.\\

Using spherical harmonic expansions, \cite{Boulerice1997} developed goodness-of-fit tests for the vMF and Watson distributions. \cite{Jupp2005} considered weighted Sobolev goodness-of-fit tests for distributions on compact Riemannian manifolds. \cite{Deschepper2008} proposed a lack-of-fit test for linear-circular regression models based on the arcs generated by the circular observations. \cite{Wouters2009} proposed data-driven goodness-of-fit tests for the vM model based on orthonormal polynomials.\\

Smoothing-based approaches to testing the goodness-of-fit of parametric models to directional data have also been developed. \cite{Boente2014} and \cite{Garcia-Portugues2015} used kernel density estimators (see Section \ref{sec:NP_DensityEst}) to test the goodness-of-fit of spherical, and spherical-linear/spherical models, respectively. In the regression context, \cite{Garcia-Portugues2016} proposed a goodness-of-fit test for linear-spherical models based on an extension of \eqref{eq:NW}. 

\subsection{Other testing scenarios}

For bivariate circular data, the asymptotic sampling properties of likelihood-ratio tests of independence were considered by \cite{Shieh2005}, and their permutation analogues by \cite{Kato2015a} and \cite{Kato2018}. Nonparametric, kernel-based, tests for the independence of spherical-linear/spherical variables were proposed in \cite{Garcia-Portugues2014,Garcia-Portugues2015}.\\

Several tests for change-point detection in circular series have been introduced. Nonparametric tests include the permutation test of \cite{Byrne2009} and the CUSUM test of \cite{Lombard2017}. In vM-distributed series, likelihood \citep{Ghosh1999,Hawkins2015} and CUSUM \citep{Hawkins2017} procedures have been advocated. Bayesian approaches have also been considered \citep{Ghosh1999,SenGupta2008}.\\

Other tests for circular data include the kernel density estimate-based tests of \cite{Fisher2001} and \cite{Ameijeiras-Alonso2019b} for assessing the number of modes, and the test of \cite{Ducharme2012} for detecting vortices in two-dimensional vector fields. Recently, bootstrap-based tests using smoothing have been introduced in \cite{Zhang2019}, for comparing two samples, and in \cite{Alonso-Pena2019}, for testing in circular regression.

\section{Correlation and regression}
\label{sec:CorrRegr}

\subsection{Correlation}
\label{sec:Correlation}

\citet[Section 11.2]{Mardia1999a} provide details of some of the correlation coefficients for toroidal, cylindrical, and spherical data that have been proposed in the literature. Others are considered in \citet[Chapter 8]{Jammalamadaka2001}. Recently, \cite{Zhan2019} reviewed the correlation coefficients available for toroidal data and proposed two new ones. In the context of Bayesian network modelling, \cite{Leguey2019} and \cite{Leguey2019a} introduced mutual information measures of the dependence between circular and linear variables, and between two circular variables, respectively.

\subsection{Regression}
\label{sec:Regression}

Here we consider parametric regression models. Throughout, we use the generic notation $\cal{Y}$-$\cal{X}$ to denote that $\cal{Y}$ is the response variable and $\cal{X}$ is the explanatory variable. Recent developments in nonparametric regression are described in Section \ref{sec:NP_Regression}.

\subsubsection{Circular-circular regression}
\label{sec:Regr_circ-circ}

Circular-circular regression is used to model the relationship between a circular response variable, $\Psi$, and a circular explanatory variable, $\Theta$. \cite{Polsen2015} reviewed parametric circular-circular regression models and considered inference and diagnostic analysis for them. They focused on the general inverse tangent link-based regression model
\begin{align}\label{Eq:GeneralCCREgr}
	\Psi=\Arg\{g_1(\Theta;\bm{\eta})+\ic g_2(\Theta;\bm{\eta})\}+\varepsilon~~(\text{mod}~2\pi),
\end{align}
where the first term represents the conditional mean direction of $\Psi$ given $\Theta$, $g_1$ and $g_2$ are non-uniquely identifiable functions, $\bm{\eta}$ is a vector of parameters, and $\varepsilon$ is a circular error variable. The decentred model of \cite{Rivest1997} and the M\"{o}bius transformation-based models of \cite{Downs2002}, \cite{Kato2008a}, and \cite{Kato2010a} are all special cases of \eqref{Eq:GeneralCCREgr}. The first two have vM errors, whilst those of the models of \cite{Kato2008a} and \cite{Kato2010a} are WC and a four-parameter highly flexible unimodal extension of the WC different from \eqref{eq:fourparam}, respectively. \cite{Polsen2015} also related the bivariate regression model of \cite{Sarma1993} \citep[see also][Section 8.6]{Jammalamadaka2001}, having finite-order trigonometric polynomials for $g_1$ and $g_2$, to the model in \eqref{Eq:GeneralCCREgr}.\\

\cite{McMillan2013} proposed a hierarchical Bayesian approach for repeated measures angular data that are bimodal, based on a two-component circular-circular regression model with parameters that change according to a function expressed in a finite circular B-spline basis (see Section \ref{sec:NP_Regression:lin}).

\subsubsection{Circular-linear regression}
\label{sec:Regr_circ-lin}

Circular-linear regression is used to model the relationship between a circular response variable, $\Theta$, and a vector containing one or more covariates denoted here by $\bm{X}$. \citet[][Section 11.3.2]{Mardia1999a} refer to the use of the link-based models of \cite{Fisher1992} in this context. It would appear that the most popular such link function has been $2\tan^{-1}$. For this choice, the conditional mean is
\begin{align*}
	\E(\Theta|\bm{X}=\bm{x})=2\tan^{-1}(\bm{\beta}'\bm{x}),
\end{align*}
where $\bm{\beta}$ is a vector of regression coefficients. This link function maps the origin of $\mathbb{R}$ to the angle 0, and 
the two extremes of $\mathbb{R}$ to the angle furthest from $0$, namely $-\pi\equiv\pi$. \cite{Presnell1998} identified important practical difficulties with estimating the parameters of such models using ML methods. As a means of circumventing those inherent inferential problems, \cite{George2006} proposed a semi-parametric Bayesian approach. \cite{Artes2008} developed analysis of covariance tests for link-based models.\\

Instead of using link functions, \cite{Presnell1998} proposed an alternative modelling approach based on projecting (see Section \ref{sec:circ_models}) the unobserved responses from a multivariate linear model onto $\mathbb{S}^1$. This approach has become increasingly popular, particularly in Bayesian applications \citep[see, e.g.,][]{Nunez-Antonio2011,Wang2013a,Hernandez-Stumpfhauser2016}. The interpretation of predictor effects in projected normal regression models has been considered recently by \cite{Cremers2018}.\\

A different tack was taken by \cite{Lund2002}, who evaded the problem of devising a meaningful regression function through the use of a tree-based approach to predicting a circular response from a combination of circular and linear predictors.\\

Various approaches to modelling longitudinal data have been developed recently. \cite{Artes2000} considered the use of estimating equations when the angular response is assumed to follow a circular distribution parametrised by its mean direction and mean resultant length. \cite{DElia2001} proposed a variance components model with fixed and random effects. \cite{Lagona2016a} introduced a regression model for correlated circular data which assumes that the angular measurements arise from the sine multivariate vM distribution of \cite{Mardia2008}. All three of these proposals make use of the link function approach of \cite{Fisher1992}. The model of \cite{Lagona2016a} was extended by \cite{Mulder2017}, who employed weakly informative priors within a Bayesian framework to elude the problems with ML estimation for link-based models. Other researchers have adapted projected normal models. \cite{Nunez-Antonio2014} investigated one in which the components are specified as mixed linear models. \cite{Maruotti2016a} considered a mixed linear model with correlated random coefficients controlling dependence that can be represented as a finite mixture of projected normal distributions. \cite{Maruotti2016} proposed a time-dependent extension of the projected normal regression model with a hidden Markov heterogeneity structure.\\

Many of the above proposals have been used to model animal orientation data. Other models for such data include that of \cite{Rivest2016}, which features a consensus model for the angular response, based on circular and linear covariates, combined with vM errors. Recently, \cite{Rivest2019} proposed a random effects circular regression model for clustered circular data where both the cluster effects and the regression errors have vM distributions. Their model is based on the multivariate angular pdf with vM-distributed cluster-level random effects of \cite{Holmquist2017}. Other approaches to modelling animal orientation data are considered in Section \ref{sec:Spatial}.

\subsubsection{Linear-circular regression}
\label{sec:Regr_lin-circ}

Linear-circular regression is applied to model the relationship between a linear response variable and one or more circular covariates. The standard approach is to regress the linear variable on sums of trigonometric polynomials of the circular variables, using least squares to estimate the parameters \citep{Johnson1978}. \cite{Bhattacharya2009a} and \cite{SenGupta2015} have considered Bayesian approaches to linear-circular modelling.\\

Recently, \cite{Cremers2020a} proposed several regression models for a cylindrical response variable with linear and circular components.

\subsubsection{Spherical response}
\label{sec:Regr_spherical}

Spherical regression was first considered by \cite{Chang1986}. \cite{Downs2003} made use of M\"{o}bius transformation, stereographic projection and link functions to develop $\mathbb{S}^2$-$\mathbb{S}^2$ regression models with the conditional distribution between response and predictor being vMF. \cite{Hinkle2014} proposed polynomial models for manifold-linear regression. For $\mathbb{S}^d$-$\mathbb{S}^d$ regression, \cite{Rosenthal2014} employed projective linear transformations to model the conditional mean direction of the response, combined with a vMF error structure. \cite{Cornea2017} proposed a more general semi-parametric intrinsic manifold-manifold regression model that incorporates parametric link functions and a nonparametric error structure. Very recently, \cite{Paine2020} introduced a very general regression model for an $\mathbb{S}^2$-valued response with covariates that can be spherical, linear or categorical, and two kinds of anisotropic error distributions. In its most general formulation, a preliminary orthogonal transformation of the response is assumed to follow an anisotropic distribution with covariate-dependent parameters. For $\mathbb{S}^d$-$\mathbb{R}^q$ regression, \cite{Scealy2019} proposed a flexible heteroscedastic model with scaled vMF errors.\\

Related regression problems for a $\mathbb{S}^2$-valued response include the fitting of small circles to spherical data \citep{Rivest1999} and the analysis of rotational deformations through fitting small circles on the sphere \citep{Schulz2015}.

\section{Nonparametric curve estimation}
\label{sec:Nonparam}

Here we review advances in nonparametric curve estimation. See Section \ref{sec:Testing} for nonparametric tests, and later sections for other nonparametric methods.

\subsection{Density estimation}
\label{sec:NP_DensityEst}

\subsubsection{Smoothing-based}
\label{sec:NP_DensityEst:smooth}

Kernel density estimation (KDE) on $\mathbb{S}^d$ dates back to \cite{Beran1979}, \cite{Hall1987}, and \cite{Bai1988}. In the latter's formulation, the kernel estimator for a sample $\bm{X}_1,\ldots,\bm{X}_n$ from the target pdf $f$, is given by
\begin{align}
	\hat{f}(\bm{x};h)=\frac{c_L(h)}{n}\sum_{j=1}^n L\left(\frac{1-\bm{x}'\bm{X}_j}{h^2}\right),\quad \bm{x}\in\mathbb{S}^d,\label{eq:KDE}
\end{align}
where $h>0$ denotes the bandwidth, $L:[0,\infty)\rightarrow[0,\infty)$ a kernel, and $c_L(h)$ a normalising constant. For the vMF kernel $L(r)=e^{-r}$ and $d=1$, \eqref{eq:KDE} reduces to 
\eqref{eqn:vMMixden} with common concentration $\kappa=1/h^2$, namely
\begin{align}
	\hat{f}(\theta;\kappa)=\frac{1}{2\pi\mathcal{I}_0(\kappa)n}\sum_{j=1}^n \exp\{\kappa\cos(\theta-\Theta_j)\},\quad \theta\in[-\pi,\pi).\label{eq:KDE1}
\end{align}

Several extensions and modifications of \eqref{eq:KDE} and \eqref{eq:KDE1} have been proposed. For $d\geq2$, \cite{Klemela2000} used $L(\kappa\cos^{-1}(\bm{x}'\bm{X}_j))$ in \eqref{eq:KDE} to analyse estimators of $f$ and its derivatives. Extending \eqref{eq:KDE1} to $[-\pi,\pi)^d$, \cite{DiMarzio2011} introduced a class of Fourier-based sine-order circular kernels containing many well-known circular pdfs. \cite{Garcia-Portugues2013b} considered the extension of \eqref{eq:KDE} to $\mathbb{S}^d\times\mathbb{R}$. \cite{Amiri2017} transformed \eqref{eq:KDE} into a sequentially updating estimator. \cite{Tsuruta2017a} showed that using a WC kernel instead of a vM kernel in \eqref{eq:KDE1} worsens the optimal asymptotic mean integrated squared error (AMISE) rate from $n^{-4/5}$ to $n^{-2/3}$, despite both kernels being second sine-order. This motivated \cite{Tsuruta2017} to propose a class of $p$-th order kernels with an optimal AMISE rate of $n^{-2p/(2p+1)}$.\\

Bandwidth selection is crucial to KDE and hence was also addressed in most of the aforementioned contributions. Plug-in selectors as alternatives to cross-validation (CV) have received most attention. \cite{Taylor2008} proposed the first plug-in selector for \eqref{eq:KDE1} by deriving the AMISE under the assumption that $f$ is vM. The plug-in rule of \cite{Oliveira2012} employed the AMISE of \cite{DiMarzio2011}, but used a two-component vM mixture in its curvature term. \cite{Garcia-Portugues2013a} gave plug-in selectors for \eqref{eq:KDE} using the AMISE and MISE for mixtures of vMF pdfs. Recently, \cite{Tsuruta2020} studied the convergence rates of direct plug-in and CV selectors for KDE on $\mathbb{S}^1$. \cite{PhamNgoc2019} proposed a bandwidth selector for \eqref{eq:KDE} with a convergence rate of $n^{-2p/(2p+d)}$ for $p$-th order kernels.\\

Asymptotic results obtained for \eqref{eq:KDE} include: CLTs for the integrated squared error of KDEs on $\mathbb{S}^d$ \citep{Zhao2001}, and $\mathbb{S}^d\times\mathbb{R}$ and $\mathbb{S}^{d_1}\times\mathbb{S}^{d_2}$ \citep{Garcia-Portugues2015}; lower bounds for asymptotic minimax risks \citep{Klemela2003}; laws for the iterated logarithm \citep{Wang2001,Wang2003}; large and moderate deviations \citep{Gao2010,Li2014}.\\

Convolutions on $\mathbb{S}^{d}$ are intimately related with KDE and are key to fast computation. They have been studied in \cite{Egecioglu2000}, \cite{Dokmanic2010}, and \cite{LeBihan2016}.\\

As alternatives to \eqref{eq:KDE} and \eqref{eq:KDE1}, \cite{Wang2000b} introduced a nearest-neighbour estimator of $f$ and \cite{Park2012,Park2013} considered KDE via the tangent space of $\mathbb{S}^d$. \cite{DiMarzio2017} matched trigonometric moments of $f$ with their smoothed sample versions to derive pdf estimators. As in high-order KDE, such estimators lower the bias and retain the variance order of \eqref{eq:KDE1}, although negative values are possible. Using a different approach, \cite{DiMarzio2016b,DiMarzio2018a} investigated local likelihood \citep{Loader1996} for pdfs on $[-\pi,\pi)^d$ by using local approximation of $\log f$. KDE based on the heat kernel on $\mathbb{S}^d$ \citep[see][]{Hartman1974}, the $d=1$ case of which being the wrapped normal kernel, was applied in \cite{Zhang2019}.\\

Extensions of \eqref{eq:KDE1} enable the construction of smoothed estimators for circular cdfs \citep{DiMarzio2012} and conditional pdfs \citep{DiMarzio2016}. More generally, KDE has also been developed for compact Riemannian manifolds \citep{Pelletier2005,Henry2009}, with inherent reduced tractability.

\subsubsection{Series-based}
\label{sec:NP_DensityEst:series}

An alternative approach to estimating a circular pdf is to use sample trigonometric moments as estimates of coefficients in its Fourier series expansion. Such estimates generally exhibit harmonic peaks and troughs and can be negative, although the latter defect can be circumvented by imposing constraints \citep{Fernandez-Duran2004}. Instead, periodic Bernstein polynomials might be considered. However, as \cite{Carnicero2018} have shown, imposing periodicity on such polynomials increases the error rate from $n^{-4/5}$ to $n^{-2/3}$. An interesting connection between Fourier-based estimation and \eqref{eq:KDE1} arises through the use of the WC kernel \citep{Chaubey2018}.\\

Spherical harmonics \cite[see, e.g.,][]{Dai2013} extend Fourier orthogonal bases to $\mathbb{S}^d$ with increasing complexity as $d$ grows. Hence, pdf estimation through spherical harmonics inherits both the advantages and disadvantages of Fourier series estimation on the circle. A compelling alternative, are \textit{needlets} (\cite{Narcowich2006}; see also \citet[][]{Marinucci2008}), a class of spherical wavelets. Needlets build on spherical harmonics to form a \textit{tight frame} on $L^2(\mathbb{S}^d)$ that is not a basis, as redundancy is allowed, but has superior localisation properties. Needlet coefficients can be estimated from sample spherical harmonic coefficients. \cite{Baldi2009} approached adaptive pdf estimation on $\mathbb{S}^d$ by thresholding needlet coefficients, and \cite{Kueh2012} studied the latter estimator under varying local pdf regularity. Like Fourier-based estimates, needlet-based pdf estimates can take negative \nolinebreak[4]values.\\

Circular deconvolution, i.e.\ the estimation of a pdf on $\mathbb{S}^1$ from noisy observations (see Section \ref{sec:TestingUniformity}), has been tackled with increasing generality in \cite{Efromovich1997}, \cite{Comte2003}, and \cite{Johannes2013}. Spherical deconvolution has been studied through spherical harmonics \citep{Healy1998,Kim2002,Kim2004} and needlets \citep{Kerkyacharian2011}.

\subsubsection{Bayesian-based}

Density estimation using Dirichlet process mixtures (DPMs) is a popular nonparametric Bayesian approach and has been employed with directional variables too. \cite{Lennox2009} provided a DPM model having sine bivariate vM distributions to model pairs of dihedral angles on $\mathbb{T}^2$. \cite{Straub2015} proposed a DPM model of Gaussian distributions in distinct tangent spaces to $\mathbb{S}^d$. DPM models with projected normal distributions have been advocated by \cite{NunezAntonio2018} on $\mathbb{S}^1$ and by \cite{Abraham2019} on $\mathbb{T}^d$. Density estimation through DPM on manifolds was addressed in \citet[Chapter 13]{Bhattacharya2012}.

\subsection{Regression estimation}
\label{sec:NP_Regression}

\subsubsection{Linear response}
\label{sec:NP_Regression:lin}

The Nadaraya--Watson estimator for linear-spherical regression is
\begin{align}
	\hat{m}(\bm{x};h)=\frac{c_L(h)}{n\hat{f}(\bm{x};h)}\sum_{j=1}^n Y_jL\left(\frac{1-\bm{x}'\bm{X}_j}{h^2}\right),\quad \bm{x}\in\mathbb{S}^d,\label{eq:NW}
\end{align}
which, for the vMF kernel and $\mathbb{S}^1$, reduces to
\begin{align}
	\hat{m}(\theta;\kappa)=\frac{\sum_{j=1}^n Y_j\exp\{\kappa\cos(\theta-\Theta_j)\}}{\sum_{j=1}^n\exp\{\kappa\cos(\theta-\Theta_j)\}},\quad \theta\in[-\pi,\pi).\label{eq:NW1}
\end{align}

As an extension of \eqref{eq:NW1}, \cite{DiMarzio2009} introduced local polynomial regression for predictors on $\mathbb{T}^d$ through a sine term-based Taylor expansion. Their approach was extended further by \cite{Qin2011} to accommodate circular and multivariate predictors using product kernels, a broadly applicable approach to combine different predictors. \cite{Tsuruta2018} showed the different optimal error rates for the (second sine-order) WC and vM kernels.\\

On $\mathbb{S}^d$, \eqref{eq:NW} was considered by \cite{Wang2000a} and \cite{Wang2002} when deriving laws for iterated logarithm and exponential error bounds, respectively. \cite{DiMarzio2014} extended \eqref{eq:NW1} to local polynomial regression using a Taylor expansion within the tangent-normal decomposition. \cite{Garcia-Portugues2016} used a different Taylor expansion yielding a local linear estimator that, for $d=1$, coincides with the \cite{DiMarzio2009} proposal. \cite{DiMarzio2019} built on their construction in \cite{DiMarzio2014} to perform local polynomial logistic regression with a spherical predictor.\\

\cite{Monnier2011} proposed needlet-based regression for a uniformly distributed predictor on $\mathbb{S}^d$ and Gaussian noise, while \cite{Lin2019} weakened those assumptions and introduced regularisation on the needlet coefficients.\\

Thin-plate splines on $\mathbb{S}^d$ \citep{Taijeron1994} offer an alternative smoothing approach to kernel methods. Such splines have been considered for improving brain conformal mapping to $\mathbb{S}^2$ \citep{Zou2007}. \cite{Kaufman2005} introduced circular Bayesian adaptive regression splines for modelling the firing rates of neurons activated by movements of a monkey's wrist. Quadratic B-splines on the circle were constructed in \cite{McMillan2013}.\\

Related to regression for a $\mathbb{S}^1$ predictor, \cite{Hall2000}, \cite{Hall2003}, and \cite{Genton2007} studied the estimation of periodic functions over an (unwrapped) time domain. \cite{Klemela1999} considered the estimation of a function on $\mathbb{S}^d$ observed in Gaussian continuous time white \nolinebreak[4]noise.

\subsubsection{Circular or spherical response}
\label{sec:NP_Regression:cir}

\cite{Boente1991} considered estimators for $\mathbb{S}^d$-$\mathbb{R}^q$ regression based on locally-weighted spherical means, with nearest-neighbour or Nadaraya--Watson weights. Their construction was generalised to local polynomial $\mathbb{S}^1$-$\mathbb{S}^1$ and $\mathbb{S}^1$-$\mathbb{R}$ \citep{DiMarzio2013}, $\mathbb{S}^1$-$\mathbb{R}^q$ \citep{Meilan-Vila2020}, and $\mathbb{S}^d$-$\mathbb{S}^q$ \citep{DiMarzio2014} regression through local circular and spherical means. A novel approach to $\mathbb{S}^d$-$\mathbb{S}^d$ regression, based on local polynomial expansions of the \textit{rotation} function, was advocated by \cite{DiMarzio2019b}. \\

Quantile $\mathbb{S}^1$-$\mathbb{S}^1$ and $\mathbb{S}^1$-$\mathbb{R}$ regression was developed by \cite{DiMarzio2016a} through inversion of the conditional circular distribution and smoothing a circular check function.\\

From a Bayesian perspective, \cite{Scott2011} estimated the regression function on $\mathbb{S}^2$ by imposing shrinkage priors on its needlet coefficients. \cite{Navarro2017} proposed multivariate generalized vM circular processes as a replacement for Gaussian processes in circular regression.\\

More generally, \cite{Cheng2013} addressed linear-manifold regression through local linear regression on the tangent space, and \cite{Lin2017} gave an extrinsic Nadaraya--Watson estimator for manifold-linear regression.

\section{Dimension reduction methods}
\label{sec:DimRedMeths}

\subsection{Principal component analysis}

\subsubsection{General manifolds}
\label{sec:DimRedMeths:PCA:M}

Principal component analysis (PCA) for data on a Riemannian manifold $\mathcal{M}$ of dimension $d$, such as $\mathbb{S}^d$ or $\mathbb{T}^d$, has received considerable attention lately. Approaches to manifold PCA can be classified using two broad dichotomies: (i) \textit{extrinsic} (based on tangent space) versus \textit{intrinsic} (geodesic-based); (ii) \textit{forward} (sequential computation of the $j$-th principal component, $j=1,\ldots,d$) versus \textit{backward} (computation of a sequence of nested subspaces of decreasing dimension within $\mathcal{M}$ based on constraints; \cite{Damon2014}). \cite{Huckemann2010a} gave a detailed review of the topic, and \cite{Marron2014} and \cite{Pennec2018} more recent overviews.\\

\cite{Fletcher2004} introduced principal geodesic analysis (PGA) as an analogue of PCA in symmetric spaces such as $\mathbb{S}^d$ and $\mathbb{T}^d$. It is centred upon the intrinsic sample mean on $\mathcal{M}$, $\hat{\bm{\mu}}$, and defines the first principal geodesic as the one passing through $\hat{\bm{\mu}}$ that minimises the sum of squared intrinsic residuals. Other principal geodesics are obtained sequentially by imposing orthogonality at $\hat{\bm{\mu}}$. PGA involves a complex optimisation process, only solved later by \cite{Sommer2014}. This complexity led \cite{Fletcher2004} to propose tangent PCA (tPCA) as an approximation. tPCA performs PCA with the log-mapped data onto the tangent plane at $\hat{\bm{\mu}}$ and then obtains the principal geodesics on $\mathcal{M}$ spanned by the tangent principal directions. When $\mathcal{M}=\mathbb{S}^2$, the principal components of PGA and tPCA are great circles that pass through $\hat{\bm{\mu}}$. \\

Two limitations of PGA are exemplified on $\mathbb{S}^2$: (a) great circles are forced to cross at $\hat{\bm{\mu}}$; (b) great circles are unable to describe certain forms of variation (see Section \ref{sec:DimRedMeths:PCA:Sd}). \cite{Huckemann2006} tackled (a) by introducing geodesic PCA (GPCA) for Riemannian manifolds, a forward-type method with a backward shift that locates a data centre $\tilde{\bm{\mu}}$ \textit{after} finding the best fitting geodesic. The other components cross orthogonally at $\tilde{\bm{\mu}}$, a restriction circumvented by horizontal component analysis \citep{Sommer2013}. \cite{Curry2019} recently proposed principal symmetric space approximation (PSSA), which considers totally geodesic subspaces (great subspheres, on $\mathbb{S}^d$) and is computationally tractable on certain manifolds.\\

A non-geodesic approach to PCA on $\mathcal{M}$ is barycentric subspace analysis \citep{Pennec2018}. It considers $k$-dimensional \textit{affine spans} (great subspheres if $\mathcal{M}=\mathbb{S}^d$) spanned by $k+1$ $\mathcal{M}$-affinely independent points, whose successive addition/removal yields a forward/backward-type sequence of nested subspaces.\\

\cite{Zhang2013} proposed probabilistic PGA, in which the normal distribution used in probabilistic PCA \citep{Tipping1999} is replaced by what the authors refer to as the \textit{Riemannian normal distribution}, with pdf $f(\bm{x};\boldsymbol{\mu},\sigma^2)\propto\exp\{-d_g(\bm{x},\boldsymbol{\mu})^2/(2\sigma^2)\}$, where $\bm{x},\boldsymbol{\mu}\in\mathcal{M}$ and $d_g$ is the intrinsic distance on $\mathcal{M}$ ($d_g(\bm{x},\boldsymbol{\mu})=\cos^{-1}(\bm{x}'\boldsymbol{\mu})$ if $\mathcal{M}=\mathbb{S}^d$). \cite{Sommer2019} advocated an alternative to PGA based on an anisotropic normal distribution over $\mathcal{M}$, generated from the marginal distributions of a diffusion process on $\mathcal{M}$ with a constant infinitesimal covariance.\\

The previous approaches assume a parametric form for the first principal curve on $\mathcal{M}$. Instead, the \textit{principal flow} of \cite{Panaretos2014} is defined as the curve of maximal data variation on $\mathcal{M}$ that, starting at $\hat{\bm{\mu}}$, is tangential to the vector field formed by the first eigenvector of the local tangent covariance matrix. Higher-order principal flows, which are always curves, are defined analogously.\\

\cite{Dai2018} adapted tPCA for functional data on $\mathcal{M}$ (e.g., flight trajectories on $\mathbb{S}^2$) by replacing PCA by functional PCA on the tangent plane.\\

Nonparametric inference on backward nested principal component subspaces, generalising the result of \cite{Anderson1963} on asymptotic inference for classical PCA, has been provided by \cite{Huckemann2018}.

\subsubsection{Methods for spherical data}
\label{sec:DimRedMeths:PCA:Sd}

In relation to limitation (b) of Section \ref{sec:DimRedMeths:PCA:M}, and for the specific case of $\mathbb{S}^2$, \cite{Jung2011} advocated principal arc analysis (PAA), a non-geodesic approach designed to improve the flexibility of GPCA. PAA employs small circles on $\mathbb{S}^2$ as the primary modes of data variation, an idea generalised to $\mathbb{S}^d$ by \cite{Jung2012} as principal nested spheres (PNS). By iteratively performing a series of tangent-normal decompositions on $\mathbb{S}^d$, PNS is a backward-type approach that produces a sequence of subspheres isomorphic to $\mathbb{S}^j$, $j=d-1,\ldots,1$, that none of the methods in Section \ref{sec:DimRedMeths:PCA:M} are able to match in terms of flexibility.\\

Despite the generality of the approaches in Section \ref{sec:DimRedMeths:PCA:M}, the success of PNS highlights the advantages of focusing on specific manifolds, such as $\mathbb{S}^d$ or $\mathbb{T}^d$, and exploiting their peculiarities so as to obtain more informative methods. Other examples of the benefits of specificity include PAA on direct product manifolds such as $(\mathbb{R}^3\times\mathbb{R}^+\times\mathbb{S}^2\times\mathbb{S}^2)^m$, introduced by \cite{Jung2011}, and composite PNS for skeletal models, proposed by \cite{Pizer2013}.

\subsubsection{Methods for toroidal data}
\label{sec:DimRedMeths:PCA:Td}

Torus-specific PCA proposals have been stimulated by the need to analyse dihedral angles in bioinformatics, and the inapplicability of most of the methods in Section \ref{sec:DimRedMeths:PCA:M} due to the pathological behaviour of geodesics on $\mathbb{T}^d$.\\

The majority of toroidal PCA methods resort to some sort of transformation prior to applying classical PCA. \cite{Mu2005} proposed dihedral PCA (dPCA) by mapping angles from $[-\pi,\pi)^d$ to $\mathbb{T}^{d}$ and then performing PCA. Complex dPCA \citep{Altis2007} performs PCA on the complex representation of angles. \citet{Riccardi2009} proposed angular PCA (aPCA), based on applying PCA to toroidal data centred on their circular means. \cite{Kent2009} gave a trigonometric moment characterisation of the covariance matrix in a wrapped normal model on $\mathbb{T}^d$, facilitating PCA on it. \cite{Nodehi2015} applied PGA on $\mathbb{T}^d$ in what they called dPGA. The latter two approaches yield principal directions that almost surely wrap around infinitely. With regard to this issue, \cite{Kent2015} discussed desirable properties for principal component curves on $[-\pi,\pi)^2$. \cite{Sittel2017} introduced a variation on aPCA, called dPCA+, that shifts each variable so that $-\pi\equiv\pi$ is located at the lowest pdf region for minimising the distortion when PCA is applied. \cite{Sargsyan2012,Sargsyan2015} used a non-injective mapping from $\mathbb{T}^d$ to $\mathbb{S}^d$ that equates toroidal angles in $[-\pi,\pi)^d$ to hyperspherical coordinate angles, even though the latter are defined\nopagebreak[4] in $[0,\pi]^{d-1}\times [-\pi,\pi)$, then applied  PGA.\\

A better-grounded approach to torus PCA is T-PCA \citep{Eltzner2018}, which maps $\mathbb{T}^{d}$ to $\mathbb{S}^{d}$ with a deformation, which, for $d=2$, corresponds to cutting $\mathbb{T}^2$ at a data-driven point to form a cylinder, contracting the circles at its ends to single points, and reconnecting at those points. Principal nested deformed spheres \citep{Eltzner2015} is an extension of the T-PCA approach to data on a polysphere $\mathbb{S}^{d_1}\times\cdots\times\mathbb{S}^{d_m}$. PSSA can also be applied to toroidal data, with geodesics on $\mathbb{T}^d$ identified using model selection.

\subsection{Other dimension reduction methods}

Nonlinear dimension reduction methods for Euclidean data have also been adapted to the directional context. \cite{Lunga2013} modified $t$-stochastic neighbour embedding ($t$-SNE) of \cite{Maaten2008} to obtain a dimension reduction method, from $\mathbb{S}^d$ to $\mathbb{S}^q$, $q\ll d$, using neighbourhoods formed by the extension of the WC distribution to $\mathbb{S}^d$ in \cite{Kato2009}. \cite{Wang2016a} proposed a modification of $t$-SNE with vMF neighbourhoods.\\

\cite{Wilson2014} revisited multidimensional scaling on $\mathbb{S}^d$, proposing a new approach that obviated the minimisation of stress functions based on spherical distances inherent in former approaches. They used their approach to map textures of 3D objects onto spheres \citep{Elad2005}, and model normalised time-warping functions \citep{Veeraraghavan2009}. \cite{Lu2019} adapted $t$-SNE for dimension reduction from $\mathbb{R}^d$ to $\mathbb{S}^q$, $q\ll d$, highlighting the benefits of the clusterings obtained on the spherical geometry. Note that these transformations of multivariate data into spherical data, termed \textit{spherical embeddings}, call for the use of directional statistics with data which were not originally directions.

\section{Classification and clustering}
\label{sec:ClassClust}

\subsection{Classification}
\label{sec:Classification}

\cite{SenGupta2005} introduced a discrimination rule based on the chordal distance between a new circular observation and those from two known circular populations. More recently, \cite{DiMarzio2018} considered nonparametric circular classification based on KDE and local logistic regression. \cite{Pandolfo2018a} studied the depth-versus-depth classifier for circular data. \cite{Leguey2019a} proposed Bayesian classification algorithms for WC-distributed circular predictors.\\

\cite{SenGupta2011} proposed generalised likelihood-ratio tests for classifying toroidal and cylindrical data into two populations when one of the misclassification probabilities is assumed to be known. \cite{Fernandes2016} developed a logistic classifier for use with circular and linear predictors.\\

Classification rules for data on $\mathbb{S}^d$ from Watson and vMF populations were developed by \cite{Figueiredo2006} and \cite{Figueiredo2009}, respectively. \cite{Bhattacharya2012a} proposed a Dirichlet process mixture model classifier comprising vMF kernels. \cite{Lopez-Cruz2015} considered the naive Bayes classifier with vMF-related conditional distributions of directional predictors. Techniques for the classification of image textures, based on multi-resolution directional filters, were proposed by \cite{Kim2018}. \cite{DiMarzio2019a} considered KDE-based nonparametric classification. The cosine depth \textit{distribution} classifier was introduced in \cite{Demni2019}. A comparison of different classification rules on $\mathbb{S}^2$ was performed by \cite{Tsagris2019}.

\subsection{Clustering}
\label{sec:Clustering}

The development of clustering methods for directional data has been a major research theme lately, particularly amongst the machine learning community.\\

The two most popular approaches to clustering data on $\mathbb{S}^d$ are spherical $k$-means and the use of (finite) mixture models with vMF components, the vMF extension of \eqref{eqn:vMMixden}. Both have often been applied after projecting data in $\mathbb{R}^{d+1}$ to $\mathbb{S}^d$. The spherical $k$-means approach \citep{Dhillon2001} maximises the cosine similarity measure $\sum_{j=1}^n \bm{X}'_j\bm{c}_{(j)}$ between the sample $\bm{X}_1,\ldots,\bm{X}_n$ and $k$ centroids $\bm{c}_1,\ldots,\bm{c}_k\in\mathbb{S}^d$, where $\bm{c}_{(j)}$ is the centroid of the cluster containing~$\bm{X}_j$.\\

\cite{Dhillon2003a} and \cite{Banerjee2003, Banerjee2005} gave Expectation-Maximisation (EM) algorithms for fitting vMF mixtures. Such mixtures accommodate spherical $k$-means as a particular limiting case \citep{Banerjee2005}. Other approaches to fitting vMF mixtures include those of: \cite{Yang1997}, based on embedding fuzzy partitions in the mixtures; \cite{Taghia2014}, a Bayesian approach employing variational inference; \cite{Gopal2014}, who proposed Bayesian graphical modelling approaches based on variational inference and collapsed Gibbs sampling; \cite{Qiu2015}, which used a new information criterion to determine the number of clusters; \cite{Kasarapu2015}, based on the Bayesian minimum message length criterion to determine the optimal number of components; \cite{Mashal2015}, a $k$-means++ method for identifying favourable mixture starting values; \cite{Salah2019}, a co-clustering approach based on diagonal block mixtures of vMF distributions. Mixtures with vMF components have been employed to model data from photometry \citep{Hara2008}, text mining \citep{Banerjee2009}, speech recognition \citep{Tang2009}, radiation therapy \citep{Bangert2010}, pattern recognition \citep{Calderara2011}, multichannel array processing \citep{Costa2014}, and collaborative filtering \citep{Salah2017}. vMF mixtures have been used to cluster pdf objects based on their spherical-valued wavelet coefficients \citep{Montanari2013}.\\

Clustering approaches based on mixtures with other types of directional distributions have also been advocated. To increase cluster shape modelling flexibility on $\mathbb{S}^d$, \cite{Peel2001} used mixtures of Kent distributions, whereas \cite{Dortet-Bernadet2008} proposed ones with inverse stereographic projections of multivariate normal distributions. Mixture models with wrapped normal components were investigated by \cite{Agiomyrgiannakis2009}, and used to cluster X-ray position data in \cite{Abraham2013}. Bayesian approaches to fitting projected normal mixtures have been proposed by \cite{Wang2014} and \cite{Rodriguez2020}, and for general projected normal mixtures by \cite{Hernandez-Stumpfhauser2017}. \cite{Franke2016} developed an EM algorithm to fit the latter type of mixtures to data on $\mathbb{S}^2$. For data on $\mathbb{T}^d$, \cite{Mardia2012a} proposed mixtures of concentrated sine multivariate vM components, with approximated normalising constants, to cluster dihedral angles of an amino acid. For cylindrical data, \cite{Lagona2011,Lagona2012} developed latent-class mixture models to cluster incomplete environmental data.\\

Mixture model-based approaches for clustering axial data have also been proposed. \cite{Bijral2007}, \cite{Souden2013}, and \cite{Sra2013} developed EM-based algorithms to fit mixtures of Watson distributions. \cite{Hasnat2014} provided an alternative approach to fitting such models based on Bregman divergence. A clustering approach based on mixtures of Bingham distributions was developed by \cite{Yamaji2011}.\\

Variations of spherical $k$-means include the spherical fuzzy and possibilistic $c$-means proposed by \cite{Kesemen2016} and \cite{Benjamin2019}, respectively, and the adaptations by \cite{Maitra2010} for computational efficiency. On $\mathbb{S}^1$, \cite{Baragona2003} further investigated an alternative partitioning based on the statistic of \cite{Lund1999a}. A nonparametric alternative to $k$-means is kernel mean shift clustering, introduced on $\mathbb{S}^d$ by \cite{Oba2005}. It was then extended by \cite{Cetingul2009} to $\mathbb{S}^d$ and other specific manifolds, using intrinsic and extrinsic perspectives, and later reintroduced on $\mathbb{S}^d$ with minor variations \citep{ChangChien2010,Yang2014}. A modification of kernel mean shift that uses time-varying bandwidths was adapted to spherical data by \cite{Hung2015}.

\section{Modelling serial dependence}
\label{sec:StochProcTSA}

\subsection{Discrete-time processes}

Let $\{\Theta_t\}_{t=1,2,\ldots}$ denote a discrete-time circular process, and $\{\theta_t\}_{t=1,2,\ldots,n}$ a corresponding circular time series. \cite{Mardia1999a} provide a summary of the projected normal, wrapped, linked autoregressive moving average (ARMA), and circular autoregressive (CAR) models for circular time series considered in \cite{Fisher1994} and \citet[][Section 7.2]{Fisher1993}.\\

If $\{(X_t,Y_t)\}_{t=1,2,\ldots}$ is a stationary bivariate normal process then $\{\Theta_t\}_{t=1,2,\ldots}$, where $\Theta_t=\Arg(X_t+\ic Y_t)$, is a projected normal process. If $\{X_t\}_{t=1,2,\ldots}$ is a process on $\mathbb{R}$ then $\{\Theta_t\}_{t=1,2,\ldots}$, where $\Theta_t=X_t~(\text{mod}~2\pi)$, is the corresponding wrapped process. The wrapped AR processes of \cite{Breckling1989} provide an example. A linked process is defined through $\Theta_t=\mu+g(X_t)$ where $\mu\in[-\pi,\pi)$ and $g$ is a link function defined in \citet[][Section 7.2.4]{Fisher1993} as a mapping from $\mathbb{R}$ to $(-\pi,\pi)$. A linked ARMA($p,q$) process is obtained if $X_t=g^{-1}(\Theta_t)$ is an ARMA($p,q$) process. A CAR process is one for which $\Theta_t|\Theta_{t-1}=\theta_{t-1},\ldots,\Theta_{t-p}=\theta_{t-p}$ is vM-distributed with mean direction
\begin{align*}
	\mu_t=\mu+g[\alpha_1g^{-1}(\theta_{t-1}-\mu)+\cdots+\alpha_pg^{-1}(\theta_{t-p}-\mu)]
\end{align*}
and concentration parameter $\kappa$, where $\mu\in[-\pi,\pi)$ and $\alpha_1,\ldots,\alpha_p\in\mathbb{R}$. \cite{Artes2009} proposed an extension of the CAR model with covariates. Processes with distributions other than the vM can be defined analogously to CAR. \cite{Erdem2011} consider four ARMA-based approaches to the short-term forecasting of wind speed and direction.\\
 
Markov models can be constructed using a transition pdf, $f_{\Theta_t|\Theta_{t-1}=\theta_{t-1}}$, derived from a bivariate circular pdf. \cite{Hughes2007} used this approach, dating back to \cite{Wehrly1980}, to obtain stationary Markov processes from the sine and cosine bivariate vM models referred to in Section \ref{sec:cylindrical_models}, and the M\"{o}bius transformation-based regression model of \cite{Downs2002}. Similarly, \cite{Kato2010} employed the regression model of \cite{Kato2008a} to derive a stationary Markov process. \cite{Yeh2013} proposed a circular Markov process based on a transition pdf that belongs to the class of generalized vM distributions referred to in Section \ref{sec:circ_models}. \cite{LeBihan2016} studied Markov processes with rotationally symmetric transition pdfs on $\mathbb{S}^d$, specifically analysing the vMF case.\\

In a paper that has stimulated much research into the modelling of time series, spatial and spatio-temporal data (see Section \ref{sec:Spatial}), \cite{Holzmann2006} introduced hidden Markov models (HMMs) \citep{Zucchini2016} for circular as well as cylindrical time series. Such models offer considerable flexibility in their serial dependence properties and use mixtures of varying distributions to model different underlying regimes. More specifically, \cite{Holzmann2006} considered circular HMMs with state-dependent vM, wrapped normal or WC distributions, and marginal distributions that are mixtures of each. They also proposed a cylindrical HMM with state-dependent vM distributions for the circular component. \cite{Bulla2012} extended this approach to develop a multivariate hidden Markov model for bivariate circular and bivariate linear data with sine bivariate vM and bivariate skew-normal pdfs. An HMM for toroidal time series using sine bivariate vM pdfs and allowing for missing observations was proposed in \cite{Lagona2013}. \cite{Hokimoto2014} developed a model incorporating a non-homogeneous HMM with cylindrical covariates. \cite{Ailliot2015} proposed Markov-switching autoregressive models based on a non-homogeneous hidden Markov chain for circular time series with vM innovations. HMMs for use with cylindrical time series have been proposed by \cite{Lagona2015a}, \cite{Mastrantonio2015a}, and \cite{Mastrantonio2016b}. \cite{Mastrantonio2015a} considered a projected normal-based extension of the model of \cite{Bulla2012} that allows for conditional correlation between the circular and linear variables. The Dirichlet process mixture model of \cite{Mastrantonio2016b} is designed for use with discrete cylindrical variables.\\

HMMs with toroidal components have also been employed in protein structure modelling. \cite{Boomsma2008} proposed one with cosine bivariate vM distributions to model the pairs of dihedral angles describing protein backbones. \cite{Lennox2010} considered a Dirichlet process mixture of HMMs with sine bivariate vM distributions for the dihedral angles. \cite{Golden2017} developed an HMM to model the evolution of pairs of proteins, with bivariate wrapped normal diffusions \citep{Garcia-Portugues2019} used to describe dihedral angle evolution.\\

Recently, \cite{Mazumder2017} proposed a state-space model for circular time series, with a circular latent process, based on wrapped Gaussian processes \citep{Mazumder2016}. \cite{Beran2020} introduced a class of linked processes for circular time series, allowing for long-range dependence, obtained by transforming Gaussian processes. \cite{Hokimoto2008} extended the multiple regression model of \cite{Johnson1978} to develop a time series model for data on $\mathbb{T}^d\times\mathbb{R}^q$.\\

Nonparametric kernel-based trend estimation in circular time series was tackled in \cite{DiMarzio2012a}. \cite{Beran2016} proposed a class of nonparametric normalised symmetric linear estimators for the trend of $\mathbb{S}^d$-valued time series.

\subsection{Continuous-time processes}

Continuous-time processes involving directional data have received considerably less attention than discrete-time processes. \cite{Hill1997} introduced a random walk model whose reorientation process follows a vM diffusion \citep{Kent1975}. Several variations of such random walk models have been developed for biological purposes: see \cite{Codling2008} for a review. \cite{Garcia-Portugues2019} proposed various Langevin diffusions on the torus that can be viewed as analogues of Ornstein--Uhlenbeck processes, and studied likelihood-based estimation approaches for them. \cite{Sommer2019} considered anisotropic diffusion processes on Riemannian manifolds, and \cite{Jensen2019} simulated diffusion bridges on $\mathbb{T}^d$. \cite{Ball2008} introduced Brownian motion and Ornstein--Uhlenbeck processes on the shape space of $\mathbb{R}^2$.\\

\cite{Kurz2019} provided an overview of various recursive filtering algorithms involving a variety of circular, toroidal, and spherical distributions. Analogues of the Kalman filter based on the vMF and Bingham distributions on $\mathbb{S}^d$ were proposed by \cite{Chiuso1998} and \cite{Kurz2014}, respectively. Filtering using the wrapped normal distribution on $\mathbb{S}^1$ was considered by \cite{Traa2013}. \cite{Pitt1999} proposed auxiliary particle filter methods and applied them to a ship tracking problem modelled using a WC process.

\section{Spatial and spatio-temporal modelling}
\label{sec:Spatial}

The modelling of spatial and space-time directional data is one of the branches of directional statistics that has experienced particularly important advances in recent years. Many contributions involve fitting hierarchical Bayesian spatial models to meteorological data using MCMC methods. As an approach to modelling hurricane winds, \cite{Modlin2012} proposed a Bayesian hierarchical model for vector fields featuring a wrapped normal conditional autoregressive model. \cite{Jona-Lasinio2012} formulated a similar model incorporating, instead, a wrapped Gaussian spatial process to model wave directions at different sea locations. Using a different perspective, \cite{Wang2013}, \cite{Wang2014}, and \cite{Wang2015} considered models based on projected normal processes for modelling wave direction and height at different sites.\\

As \cite{Lagona2015} has pointed out, such Bayesian hierarchical models require specific assumptions on the prior distributions of the parameters of interest and ad-hoc MCMC for fitting. Instead, \cite{Lagona2015} developed an HMM to model the temporal evolution of the sea surface in terms of time-varying circular-linear patterns that arise through latent environmental conditions. Fitting is performed using a pseudo-likelihood approach.\\

Extending the wrapped normal-based Bayesian approach of \cite{Jona-Lasinio2012}, \cite{Mastrantonio2016} introduced a wrapped skew-normal process, for use with spatio-temporal circular data, which is capable of modelling asymmetric marginal distributions. \cite{Mastrantonio2016a} extended and compared the processes of \cite{Jona-Lasinio2012} and \cite{Wang2014} to the spatio-temporal setting by introducing space-time dependence and space- and time-varying covariate information.\\

In \cite{Lagona2016} and \cite{Ranalli2018}, hidden Markov random field models were proposed for the analysis of cylindrical spatial series, enabling segmentation of latent environmental conditions. \cite{Jona-Lasinio2018} and \cite{Lagona2018} provided overviews of many of the developments discussed above. \cite{Ameijeiras-Alonso2019} extended the approach of \cite{Ranalli2018} to develop a hidden Markov random field for the spatial segmentation of wildfires, using a mixture of \cite{Kato2015} pdfs with parameters varying according to a latent nonhomogeneous Potts model.\\

Next we consider models for animal orientation data based on random walks and HMMs that provide alternatives to those in Section \ref{sec:Regr_circ-lin}. \cite{Morales2004} proposed a Bayesian approach to fitting multiple random walks to animal movement data with paths composed of random step-lengths and turning angles. Each step and turn is assigned to a random walk characteristic of a hidden behavioural state. A similar approach was proposed by \cite{McClintock2012}, with movement paths considered to be movement strategies between which animals switch in response to environmental factors. The authors combined a variety of methodologies to develop a suite of models based on biased and correlated random walks that allow for different forms of movement. \cite{Nicosia2017} proposed a hidden-state random walk model in which a circular-linear process models the direction and distance between consecutive positions of an animal, and the hidden states describe the animal's behaviour.\\

Random fields on $\mathbb{S}^2$ are discussed in depth in \cite{Marinucci2011}. Amongst many other advances, their monograph analyses recent high-frequency limit results and tests for Gaussianity and isotropy of scalar-valued random fields, and considers applications in the analysis of the cosmic microwave background. Recent research into isotropic Gaussian random fields on $\mathbb{S}^2$ has developed CLTs for functionals of needlet coefficients \citep{Baldi2009a}, limit results for the first Minkowski functional \citep{Leonenko2017}, isotropy tests based on spherical harmonics \citep{Sahoo2019}, and tests for the detection of local maxima on isotropic fields \citep{Cheng2020}. The construction of valid covariance functions on $\mathbb{S}^d$, for use in geostatistics, has been summarised in the excellent overview of \cite{Gneiting2013}. New covariance functions on $\mathbb{S}^d$ include the spatio-temporal covariance functions of \cite{Porcu2016} and the matrix-valued covariance functions of \cite{Guella2018}; see also the review by \cite{Porcu2020}. A review of advances in the construction of\nopagebreak[4] covariance functions and process models on $\mathbb{S}^2$ was given in \cite{Jeong2017}. \\

\cite{Irwin2002} gave a review of spatio-temporal nonlinear filtering and illustrated the use of cylindrical filtering in the analysis of battlespace data.

\section{Other topics}
\label{sec:FurtherTopics}

\subsection{Statistical depth}
\label{sec:StatisticalDepth}

\cite{Agostinelli2013} studied, mainly for $\mathbb{S}^1$ and $\mathbb{S}^2$, the angular simplicial depth of \cite{Liu1992} and the angular Tukey depth of \cite{Small1987}. Within the class of rotationally symmetric distributions on $\mathbb{S}^d$, \cite{Ley2014} defined a depth based on the quantiles of the sample projections onto the mean direction. \cite{Pandolfo2018} introduced computationally tractable distance-based depths on $\mathbb{S}^d$, illustrating their use in location estimation and classification. A nonparametric approach to constructing tolerance regions for spherical data was proposed by \cite{Mushkudiani2002}.

\subsection{Design and analysis of experiments}
\label{sec:design}

\cite{Otieno2012} provided an overview of the design of experiments involving directional variables, and methods available for analysing the data obtained from them. Recently, optimal designs for linear-spherical regression, based on Fourier series and spherical harmonics, have been established for $\mathbb{S}^1$ by \cite{Dette2003}, for $\mathbb{S}^2$ by \cite{Dette2005} and \cite{Dette2009}, and for $\mathbb{S}^d$, with $d > 2$, by \cite{Dette2019}.

\subsection{Order-restricted analysis}
\label{sec:OrderRestrictedAnalysis}

The random-periods model (RPM) of \cite{Liu2004} is a nonlinear regression model used to estimate the phase angles of periodically expressed genes. \cite{Rueda2009} developed circular isotonic regression estimation to infer the relative order of phase angles from the unconstrained estimates of the RPM. \cite{Fernandez2012} proposed a test for a specified ordering of phase angles assuming the unconstrained estimators of the RPM to be vM-distributed. \cite{Barragan2015} developed methods for estimating and testing for a common ordering of phase angles across multiple experiments. A review of such developments was provided by \cite{Rueda2015}. Subsequently, \cite{Rueda2016} proposed a piecewise circular regression model for the relationship between the phase angles of cell-cycle genes in two species with differing periods, and \cite{Barragan2017} considered the problem of aggregating different circular orders for the peak expressions of genes coming from heterogeneous datasets. Recently, \cite{Larriba2020} proposed a circular signal plus error model for identifying components of systems displaying rhythmic temporal patterns. \\

Independently of these developments, \cite{Klugkist2012}, \cite{Baayen2012}, and \cite{Baayen2014} proposed ANOVA tests under order restrictions on the mean directions of vM distributions.

\subsection{Outlier detection}
\label{sec:OutlierDetection}

New tests for detecting outliers in circular data, based on the circular distance \eqref{eqn:CircDistance}, sums of such distances, and gaps, were introduced and compared with existing procedures in a series of papers referred to by \cite{Mahmood2017a}. \cite{Sau2018} developed a minimum distance approach to estimating the parameters of spherical models that provides an outlier detection tool. Outlier detection tests for cylindrical, simple circular regression, and circular time series data were\nopagebreak[4] proposed in \cite{Sadikon2019}, \cite{Abuzaid2013}, and \cite{Abuzaid2014}, respectively.\\

Eigenvalue, likelihood-ratio, and geodesic distance-based tests for detecting outliers in axial data from an assumed underlying Watson distribution were developed in \cite{Figueiredo2005}, \cite{Figueiredo2007}, and \cite{Barros2017}.

\subsection{Compositional data analysis}
\label{sec:CompDataAnal}

Compositional data analysis is used when the data under consideration are vectors of non-negative proportions summing to one. The most popular approach to analysing such data is that of \cite{Aitchison1986}. However, an alternative approach, based on the square-root transformation from a unit simplex to $\mathbb{S}^d$, was discussed in \cite{Stephens1982}. Recently, that approach has been further developed by \cite{Scealy2011, Scealy2014, Scealy2014a, Scealy2017}. The relationship between compositional and directional data was further exploited by \cite{Cuesta-Albertos2009} in the context of testing for uniformity.

\section{Software}
\label{sec:Software}

Historically, a major impediment to the application of directional statistics was a lack of software implementing the methodology particular to it. In recent years, the advent of the \textsf{R} statistical computing environment \citep{R2020} and its ecosystem of contributed packages has partially addressed that paucity. An overview of many such packages was given by \cite{Pewsey2018a}. Relevant packages written in other languages include \texttt{CircStat} \citep{Berens2009} and \texttt{PyCircStat} \citep{Berens2018} for data on $\mathbb{S}^1$, \texttt{libDirectional} \citep{Kurz2019} for data on $\mathbb{T}^d$ and $\mathbb{S}^d$, \texttt{Mocapy++} for constructing probabilistic models of biomolecular structure \citep{Paluszewski2010}, and the promising \texttt{geomstats} \citep{Miolane2020} for manifold-valued data.

\subsection{General-purpose packages}
\label{sec:SoftwareMainRPackages}

There are two main \textsf{R} packages designed for use with directional data: \texttt{circular} \citep{Agostinelli2017} and \texttt{Directional} \citep{Tsagris2020}. Both include functions for the analysis of data on $\mathbb{S}^1$, $\mathbb{T}^2$, and $\mathbb{S}^1\times\mathbb{R}$. \texttt{Directional} also has routines for data on $\mathbb{S}^d$. \\

For data on $\mathbb{S}^1$, the \texttt{circular} package has functions for: descriptive statistics; KDE; pdf evaluation, simulation, and estimation for a range of classical and more recently proposed circular distributions; tests for uniformity, homogeneity, goodness-of-fit, and change points; one-way ANOVA. It also has functions for $\mathbb{S}^1$-$\mathbb{S}^1$ and $\mathbb{S}^1$-$\mathbb{R}$ regression and includes a variety of datasets. Many of \texttt{circular}'s capabilities were illustrated in \cite{Pewsey2013}. The latter's companion workspace, \texttt{CircStatsInR}, includes over 150 routines for techniques not implemented in \texttt{circular}.\\

Amongst its more specific capabilities, \texttt{Directional} implements techniques on $\mathbb{S}^d$ for: descriptive statistics; spherical data visualisation; constructing convenient rotation matrices and transformations; KDE; pdf computation, simulation, and ML estimation for various spherical distributions; $\mathbb{S}^d$-$\mathbb{S}^d$ correlation and regression, ANOVA, classification, and clustering.

\subsection{More specific \textsf{R} packages}
\label{sec:SoftwareMoreSpecificRPackages}

Here we provide an overview of more specific \textsf{R} packages and their functionality, following the order used to present themes in the previous sections.\\

Various graphical representations for data on $\mathbb{S}^1$ are supported in \texttt{bpDir} \citep{Buttarazzi2020}, \texttt{season} \citep{Barnett2020}, and \texttt{bReeze} \citep{Graul2018}. Visualisation of data on $\mathbb{S}^2$ is\nopagebreak[4] facilitated by the outstanding \texttt{rgl} \citep{Adler2020} and \texttt{plot3D} \citep{Soetaert2019} packages.\\

Efficient modelling with vMF mixtures on $\mathbb{S}^d$ is implemented in \texttt{movMF} \citep{Hornik2014}. Several mixture models can be fitted using Bayesian methods to data on $\mathbb{S}^1$ and $\mathbb{T}^2$ with \texttt{BAMBI} \citep{Chakraborty2019}. Non-negative trigonometric sums can be fitted to data on $\mathbb{T}^d$ and $\mathbb{S}^2$ using \texttt{CircNNTSR} \citep{Fernandez-Duran2016}.\\

Tests for uniformity and rotational symmetry on $\mathbb{S}^d$ are available in \texttt{sphunif} \citep{Garcia-Portugues2020c} and \texttt{rotasym} \citep{Garcia-Portugues2020e}, respectively.\\

Bayesian projected normal regression models for data on $\mathbb{S}^1$ are implemented in \texttt{bpnreg} \citep{Cremers2020}. Also for data on $\mathbb{S}^1$, nonparametric kernel methods for density and regression estimation are available in \texttt{NPCirc} \citep{Oliveira2014}. KDE and bandwidth selection on $\mathbb{S}^d$ are supported in \texttt{DirStats} \citep{Garcia-Portugues2020f}. Nonparametric $\mathbb{S}^d$-$\mathbb{S}^d$ regression is implemented in \texttt{nprotreg} \citep{Taylor2018}. Smoothing splines on $\mathbb{S}^2$ are supported in \texttt{mgcv} \citep{Wood2017}.\\

Principal nested spheres and spherical $k$-means clustering can be performed with \texttt{shapes} \citep{Dryden2019} and \texttt{skmeans} \citep{Hornik2012}, respectively.\\

Markov switching autoregressive models with vM innovations are implemented in \texttt{NHMSAR} \citep{Monbet2020}. Animal orientation data can be analysed using \texttt{CircMLE} \citep{Fitak2017}, \texttt{FLightR} \citep{Rakhimberdiev2019}, \texttt{move} \citep{Kranstauber2020}, and \texttt{moveHMM} \citep{Michelot2019}. Tools for toroidal diffusions are provided in \texttt{sdetorus} \citep{Garcia-Portugues2020d}.\\

Bayesian methods for fitting spatial and spatio-temporal models to circular data are implemented in \texttt{CircSpaceTime} \citep{Jona-Lasinio2020}. Spherical random fields can be analysed using \texttt{RandomFields} \citep{Schlather2015}. Routines for the management and analysis of cosmic microwave background data on $\mathbb{S}^2$ are available in \texttt{rcosmo} \citep{Fryer2020}.\\

Methods for analysing data on $\mathbb{S}^1$ under order restrictions are supported by \texttt{isocir} \citep{Barragan2016}. Outlier detection methods for $\mathbb{S}^1$-$\mathbb{S}^1$ regression are available in \texttt{CircOutlier} \citep{Ghazanfarihesari2016}. Depths on $\mathbb{S}^1$ and $\mathbb{S}^2$ can be computed using \texttt{depth} \citep{Genest2019}.\\

Intrinsic means and fundamental geodesic tools for $\mathbb{S}^d$ and other manifolds are available in \texttt{RiemBase} \citep{You2020}.

\section{Conclusions and future developments}
\label{sec:Concl_Future}

We hope that the previous sections provide both seasoned and neophyte researchers with a concise, comprehensive and useful overview of the widespread developments in directional statistics that have taken place over the last two decades. As often happens in research, most of those developments evolved in an uncoordinated way through the efforts of individuals and research groups working independently of one another. Given this background, predicting how the field might develop over the next 20 years is essentially impossible. That said, the further development of models with greater flexibility, techniques for high-dimensional and complex directional data involving combinations of different data types, as well as Bayesian, nonparametric, and resampling methods, would appear highly probable in the short term as such developments would be consistent with current trends. More generally, progress in all the areas covered in the previous sections is certainly possible and will no doubt evolve through responses to interesting new applications and the exigencies of the Riemannian supports of directional data, often incorporating appropriate adaptations of methodologies from other fields of statistics. The development of software to implement new techniques will continue to be crucial to the wider and proper application of directional statistics.     

\section*{Supplementary materials}

The BibTeX file \texttt{DirectionalStats.bib} includes entries for over 1700 references related to directional statistics. The file is available at \url{https://github.com/egarpor/DirectionalStatsBib}. We hope researchers in the field will find this resource useful.

\section*{Acknowledgements}

We are most grateful to three anonymous referees and, in alphabetical order, to Davide Buttarazzi, Marco Di Marzio, Miguel Fern\'{a}ndez, Stephan Huckemann, Peter Jupp, Shogo Kato, Christophe Ley, Kanti Mardia, and Louis-Paul Rivest, for their enthusiastic feedback on our original submission and helpful suggestions as to how it might be further improved. This work was supported by grants PGC2018-097284-B-100, IJCI-2017-32005, and MTM2016-76969-P from the Spanish Ministry of Economy and Competitiveness, and GR18016 from the Junta de Extremadura. All four grants were co-funded with FEDER funds from the European Union.

\setlength\bibsep{0cm}
\setlength\bibhang{0.5cm}

\begin{thebibliography}{}
	
	\bibitem[Abe and Ley, 2017]{Abe2017}
	Abe, T. and Ley, C. (2017).
	\newblock A tractable, parsimonious and flexible model for cylindrical data,
	with applications.
	\newblock {\em Econometrics and Statistics}, 4:91--104.
	\newblock \href {http://dx.doi.org/10.1016/j.ecosta.2016.04.001}
	{\path{doi:10.1016/j.ecosta.2016.04.001}}.
	
	\bibitem[Abe and Pewsey, 2011]{Abe2011a}
	Abe, T. and Pewsey, A. (2011).
	\newblock Sine-skewed circular distributions.
	\newblock {\em Statistical Papers}, 52(3):683--707.
	\newblock \href {http://dx.doi.org/10.1007/s00362-009-0277-x}
	{\path{doi:10.1007/s00362-009-0277-x}}.
	
	\bibitem[Abe and Shimatani, 2018]{Abe2018}
	Abe, T. and Shimatani, I.~K. (2018).
	\newblock Cylindrical distributions and their applications to biological data.
	\newblock In Ley, C. and Verdebout, T. (Eds.), {\em Applied Directional
		Statistics}, Chapman \& Hall/CRC Interdisciplinary Statistics Series, pp.
	163--185. CRC Press, Boca Raton.
	
	\bibitem[Abe et~al., 2010]{Abe2010}
	Abe, T., Shimizu, K., and Pewsey, A. (2010).
	\newblock Symmetric unimodal models for directional data motivated by inverse
	stereographic projection.
	\newblock {\em Journal of the Japan Statistical Society}, 40(1):45--61.
	\newblock \href {http://dx.doi.org/10.14490/jjss.40.045}
	{\path{doi:10.14490/jjss.40.045}}.
	
	\bibitem[Abraham et~al., 2013]{Abraham2013}
	Abraham, C., Molinari, N., and Servien, R. (2013).
	\newblock Unsupervised clustering of multivariate circular data.
	\newblock {\em Statistics in Medicine}, 32(8):1376--1382.
	\newblock \href {http://dx.doi.org/10.1002/sim.5589}
	{\path{doi:10.1002/sim.5589}}.
	
	\bibitem[Abraham et~al., 2019]{Abraham2019}
	Abraham, C., Servien, R., and Molinari, N. (2019).
	\newblock A clustering {B}ayesian approach for multivariate non-ordered
	circular data.
	\newblock {\em Statistical Modelling}, 19(6):595--616.
	\newblock \href {http://dx.doi.org/10.1177/1471082X18790420}
	{\path{doi:10.1177/1471082X18790420}}.
	
	\bibitem[Abuzaid et~al., 2013]{Abuzaid2013}
	Abuzaid, A.~H., Hussin, A.~G., and Mohamed, I.~B. (2013).
	\newblock Detection of outliers in simple circular regression models using the
	mean circular error statistic.
	\newblock {\em Journal of Statistical Computation and Simulation},
	83(2):269--277.
	\newblock \href {http://dx.doi.org/10.1080/00949655.2011.602679}
	{\path{doi:10.1080/00949655.2011.602679}}.
	
	\bibitem[Abuzaid et~al., 2012]{Abuzaid2012}
	Abuzaid, A.~H., Mohamed, I.~B., and Hussin, A.~G. (2012).
	\newblock Boxplot for circular variables.
	\newblock {\em Computational Statistics}, 27(3):381--392.
	\newblock \href {http://dx.doi.org/10.1007/s00180-011-0261-5}
	{\path{doi:10.1007/s00180-011-0261-5}}.
	
	\bibitem[Abuzaid et~al., 2014]{Abuzaid2014}
	Abuzaid, A.~H., Mohamed, I.~B., and Hussin, A.~G. (2014).
	\newblock Procedures for outlier detection in circular time series models.
	\newblock {\em Environmental and Ecological Statistics}, 21(4):793--809.
	\newblock \href {http://dx.doi.org/10.1007/s10651-014-0281-8}
	{\path{doi:10.1007/s10651-014-0281-8}}.
	
	\bibitem[Adler et~al., 2020]{Adler2020}
	Adler, D., Murdoch, D., et~al. (2020).
	\newblock {\em {rgl}: {3D} Visualization Using {OpenGL}}.
	\newblock {R} package version 0.100.54.
	\newblock URL: \url{https://CRAN.R-project.org/package=rgl}.
	
	\bibitem[Agiomyrgiannakis and Stylianou, 2009]{Agiomyrgiannakis2009}
	Agiomyrgiannakis, Y. and Stylianou, Y. (2009).
	\newblock Wrapped {G}aussian mixture models for modeling and high-rate
	quantization of phase data of speech.
	\newblock {\em IEEE Transactions on Audio Speech and Language Processing},
	17(4):775--786.
	\newblock \href {http://dx.doi.org/10.1109/tasl.2008.2008229}
	{\path{doi:10.1109/tasl.2008.2008229}}.
	
	\bibitem[Agostinelli, 2007]{Agostinelli2007}
	Agostinelli, C. (2007).
	\newblock Robust estimation for circular data.
	\newblock {\em Computational Statistics \& Data Analysis}, 51(12):5867--5875.
	\newblock \href {http://dx.doi.org/10.1016/j.csda.2006.11.002}
	{\path{doi:10.1016/j.csda.2006.11.002}}.
	
	\bibitem[Agostinelli and Lund, 2017]{Agostinelli2017}
	Agostinelli, C. and Lund, U. (2017).
	\newblock {\em {R} package {circular}: Circular Statistics}.
	\newblock {R} package version 0.4-93.
	\newblock URL: \url{https://CRAN.R-project.org/package=circular}.
	
	\bibitem[Agostinelli and Romanazzi, 2013]{Agostinelli2013}
	Agostinelli, C. and Romanazzi, M. (2013).
	\newblock Nonparametric analysis of directional data based on data depth.
	\newblock {\em Environmental and Ecological Statistics}, 20(2):253--270.
	\newblock \href {http://dx.doi.org/10.1007/s10651-012-0218-z}
	{\path{doi:10.1007/s10651-012-0218-z}}.
	
	\bibitem[Ailliot et~al., 2015]{Ailliot2015}
	Ailliot, P., Bessac, J., Monbet, V., and P\`ene, F. (2015).
	\newblock Non-homogeneous hidden {M}arkov-switching models for wind time
	series.
	\newblock {\em Journal of Statistical Planning and Inference}, 160:75--88.
	\newblock \href {http://dx.doi.org/10.1016/j.jspi.2014.12.005}
	{\path{doi:10.1016/j.jspi.2014.12.005}}.
	
	\bibitem[Aitchison, 1986]{Aitchison1986}
	Aitchison, J. (1986).
	\newblock {\em The Statistical Analysis of Compositional Data}, volume~25 of
	{\em Monographs on Statistics and Applied Probability}.
	\newblock Chapman \& Hall, London.
	\newblock \href {http://dx.doi.org/10.1007/978-94-009-4109-0}
	{\path{doi:10.1007/978-94-009-4109-0}}.
	
	\bibitem[Alonso-Pena et~al., 2019]{Alonso-Pena2019}
	Alonso-Pena, M., Ameijeiras-Alonso, J., and Crujeiras, R.~M. (2019).
	\newblock Nonparametric tests for circular regression.
	\newblock {\em arXiv:1910.07825}.
	
	\bibitem[Altis et~al., 2007]{Altis2007}
	Altis, A., Nguyen, P.~H., Hegger, R., and Stock, G. (2007).
	\newblock Dihedral angle principal component analysis of molecular dynamics
	simulations.
	\newblock {\em The Journal of Chemical Physics}, 126(24):244111.
	\newblock \href {http://dx.doi.org/10.1063/1.2746330}
	{\path{doi:10.1063/1.2746330}}.
	
	\bibitem[Amaral et~al., 2007]{Amaral2007}
	Amaral, G. J.~A., Dryden, I.~L., and Wood, A. T.~A. (2007).
	\newblock Pivotal bootstrap methods for {$k$}-sample problems in directional
	statistics and shape analysis.
	\newblock {\em Journal of the American Statistical Association},
	102(478):695--707.
	\newblock \href {http://dx.doi.org/10.1198/016214506000001400}
	{\path{doi:10.1198/016214506000001400}}.
	
	\bibitem[Ameijeiras-Alonso et~al., 2019a]{Ameijeiras-Alonso2019b}
	Ameijeiras-Alonso, J., Benali, A., Crujeiras, R.~M., Rodr\'{i}guez-Casal, A.,
	and Pereira, J.~M. (2019a).
	\newblock Fire seasonality identification with multimodality tests.
	\newblock {\em Annals of Applied Statistics}, 13(4):2120--2139.
	\newblock \href {http://dx.doi.org/10.1214/19-AOAS1273}
	{\path{doi:10.1214/19-AOAS1273}}.
	
	\bibitem[Ameijeiras-Alonso et~al., 2018]{Ameijeiras-Alonso2018}
	Ameijeiras-Alonso, J., Crujeiras, R.~M., and Rodr\'iguez~Casal, A. (2018).
	\newblock Directional statistics for wildfires.
	\newblock In Ley, C. and Verdebout, T. (Eds.), {\em Applied Directional
		Statistics}, Chapman \& Hall/CRC Interdisciplinary Statistics Series, pp.
	187--210. CRC Press, Boca Raton.
	
	\bibitem[Ameijeiras-Alonso et~al., 2019b]{Ameijeiras-Alonso2019}
	Ameijeiras-Alonso, J., Lagona, F., Ranalli, M., and Crujeiras, R.~M. (2019b).
	\newblock A circular nonhomogeneous hidden {M}arkov field for the spatial
	segmentation of wildfire occurrences.
	\newblock {\em Environmetrics}, 30(2):e2501.
	\newblock \href {http://dx.doi.org/10.1002/env.2501}
	{\path{doi:10.1002/env.2501}}.
	
	\bibitem[Ameijeiras-Alonso and Ley, 2019]{Ameijeiras-Alonso2019a}
	Ameijeiras-Alonso, J. and Ley, C. (2019).
	\newblock Sine-skewed toroidal distributions and their application in protein
	bioinformatics.
	\newblock {\em arXiv:1910.13293}.
	
	\bibitem[Ameijeiras-Alonso et~al., 2020]{Ameijeiras-Alonso2020}
	Ameijeiras-Alonso, J., Ley, C., Pewsey, A., and Verdebout, T. (2020).
	\newblock On optimal tests for circular reflective symmetry about an unknown
	central direction.
	\newblock {\em Statistical Papers}, to appear.
	\newblock \href {http://dx.doi.org/10.1007/s00362-019-01150-7}
	{\path{doi:10.1007/s00362-019-01150-7}}.
	
	\bibitem[Amiri et~al., 2017]{Amiri2017}
	Amiri, A., Thiam, B., and Verdebout, T. (2017).
	\newblock On the estimation of the density of a directional data stream.
	\newblock {\em Scandinavian Journal of Statistics}, 44(1):249--267.
	\newblock \href {http://dx.doi.org/10.1111/sjos.12252}
	{\path{doi:10.1111/sjos.12252}}.
	
	\bibitem[Anderson, 1993]{Anderson1993}
	Anderson, C.~M. (1993).
	\newblock Graphical methods for circular and cylindrical data.
	\newblock Technical report, Univeristy of Waterloo.
	
	\bibitem[Anderson, 1963]{Anderson1963}
	Anderson, T.~W. (1963).
	\newblock Asymptotic theory for principal component analysis.
	\newblock {\em Annals of Mathematical Statistics}, 34(1):122--148.
	\newblock \href {http://dx.doi.org/10.1214/aoms/1177704248}
	{\path{doi:10.1214/aoms/1177704248}}.
	
	\bibitem[Arnold and SenGupta, 2006]{Arnold2006}
	Arnold, B.~C. and SenGupta, A. (2006).
	\newblock Recent advances in the analyses of directional data in ecological and
	environmental sciences.
	\newblock {\em Environmental and Ecological Statistics}, 13(3):253--256.
	\newblock \href {http://dx.doi.org/10.1007/s10651-006-0009-5}
	{\path{doi:10.1007/s10651-006-0009-5}}.
	
	\bibitem[Arnold and Jupp, 2018]{Arnold2018a}
	Arnold, R. and Jupp, P. (2018).
	\newblock Orientations of symmetrical objects.
	\newblock In Ley, C. and Verdebout, T. (Eds.), {\em Applied Directional
		Statistics}, Chapman \& Hall/CRC Interdisciplinary Statistics Series, pp.
	25--44. CRC Press, Boca Raton.
	
	\bibitem[Artes, 2008]{Artes2008}
	Artes, R. (2008).
	\newblock Hypothesis tests for covariance analysis models for circular data.
	\newblock {\em Communications in Statistics -- Theory and Methods},
	37(8-10):1632--1640.
	\newblock \href {http://dx.doi.org/10.1080/03610920801893962}
	{\path{doi:10.1080/03610920801893962}}.
	
	\bibitem[Artes et~al., 2000]{Artes2000}
	Artes, R., Paula, G.~A., and Ranvaud, R. (2000).
	\newblock Analysis of circular longitudinal data based on generalized
	estimating equations.
	\newblock {\em Australian \& New Zealand Journal of Statistics},
	42(3):347--358.
	\newblock \href {http://dx.doi.org/10.1111/1467-842x.00131}
	{\path{doi:10.1111/1467-842x.00131}}.
	
	\bibitem[Artes and Toloi, 2009]{Artes2009}
	Artes, R. and Toloi, C. M.~C. (2009).
	\newblock An autoregressive model for time series of circular data.
	\newblock {\em Communications in Statistics -- Theory and Methods},
	39(1):186--194.
	\newblock \href {http://dx.doi.org/10.1080/03610920802650338}
	{\path{doi:10.1080/03610920802650338}}.
	
	\bibitem[Azzalini, 1985]{Azzalini1985}
	Azzalini, A. (1985).
	\newblock A class of distributions which includes the normal ones.
	\newblock {\em Scandinavian Journal of Statistics}, 12(2):171--178.
	
	\bibitem[Baayen and Klugkist, 2014]{Baayen2014}
	Baayen, C. and Klugkist, I. (2014).
	\newblock Evaluating order-constrained hypotheses for circular data from a
	between-within subjects design.
	\newblock {\em Psychological Methods}, 19(3):398--408.
	\newblock \href {http://dx.doi.org/10.1037/a0037414}
	{\path{doi:10.1037/a0037414}}.
	
	\bibitem[Baayen et~al., 2012]{Baayen2012}
	Baayen, C., Klugkist, I., and Mechsner, F. (2012).
	\newblock A test of order-constrained hypotheses for circular data with
	applications to human movement science.
	\newblock {\em Journal of Motor Behavior}, 44(5):351--363.
	\newblock \href {http://dx.doi.org/10.1080/00222895.2012.709549}
	{\path{doi:10.1080/00222895.2012.709549}}.
	
	\bibitem[Baba, 1981]{Baba1981}
	Baba, Y. (1981).
	\newblock Statistics of angular data. {W}rapped normal distribution model.
	\newblock {\em Proceedings of the Institute of Statistical Mathematics},
	28(1):41--54.
	
	\bibitem[Bai et~al., 1988]{Bai1988}
	Bai, Z.~D., Rao, C.~R., and Zhao, L.~C. (1988).
	\newblock Kernel estimators of density function of directional data.
	\newblock {\em Journal of Multivariate Analysis}, 27(1):24--39.
	\newblock \href {http://dx.doi.org/10.1016/0047-259X(88)90113-3}
	{\path{doi:10.1016/0047-259X(88)90113-3}}.
	
	\bibitem[Bakshaev, 2010]{Bakshaev2010}
	Bakshaev, A. (2010).
	\newblock {$N$}-distance tests of uniformity on the hypersphere.
	\newblock {\em Nonlinear Analysis: Modelling and Control}, 15(1):15--8.
	\newblock \href {http://dx.doi.org/10.15388/na.2010.15.1.14361}
	{\path{doi:10.15388/na.2010.15.1.14361}}.
	
	\bibitem[Baldi et~al., 2009a]{Baldi2009}
	Baldi, P., Kerkyacharian, G., Marinucci, D., and Picard, D. (2009a).
	\newblock Adaptive density estimation for directional data using needlets.
	\newblock {\em The Annals of Statistics}, 37(6A):3362--3395.
	\newblock \href {http://dx.doi.org/10.1214/09-aos682}
	{\path{doi:10.1214/09-aos682}}.
	
	\bibitem[Baldi et~al., 2009b]{Baldi2009a}
	Baldi, P., Kerkyacharian, G., Marinucci, D., and Picard, D. (2009b).
	\newblock Asymptotics for spherical needlets.
	\newblock {\em The Annals of Statistics}, 37(3):1150--1171.
	\newblock \href {http://dx.doi.org/10.1214/08-AOS601}
	{\path{doi:10.1214/08-AOS601}}.
	
	\bibitem[Ball et~al., 2008]{Ball2008}
	Ball, F.~G., Dryden, I.~L., and Golalizadeh, M. (2008).
	\newblock Brownian motion and {O}rnstein--{U}hlenbeck processes in planar shape
	space.
	\newblock {\em Methodology and Computing in Applied Probability}, 10(1):1--22.
	\newblock \href {http://dx.doi.org/10.1007/s11009-007-9042-6}
	{\path{doi:10.1007/s11009-007-9042-6}}.
	
	\bibitem[Banerjee et~al., 2003]{Banerjee2003}
	Banerjee, A., Dhillon, I., Ghosh, J., and Sra, S. (2003).
	\newblock Generative model-based clustering of directional data.
	\newblock In {\em {KDD} '03}, pp. 19--28, New York. Association for Computing
	Machinery.
	\newblock \href {http://dx.doi.org/10.1145/956750.956757}
	{\path{doi:10.1145/956750.956757}}.
	
	\bibitem[Banerjee et~al., 2005]{Banerjee2005}
	Banerjee, A., Dhillon, I.~S., Ghosh, J., and Sra, S. (2005).
	\newblock Clustering on the unit hypersphere using von {M}ises-{F}isher
	distributions.
	\newblock {\em Journal of Machine Learning Research}, 6(Sep):1345--1382.
	
	\bibitem[Banerjee et~al., 2009]{Banerjee2009}
	Banerjee, A., Dhillon, I.~S., Ghosh, J., and Sra, S. (2009).
	\newblock Text clustering with mixture of von {M}ises-{F}isher distributions.
	\newblock In Srivastava, A.~N. and Sahami, M. (Eds.), {\em Text Mining},
	Chapman \& Hall/CRC Data Mining and Knowledge Discovery Series, pp. 151--184.
	CRC Press, New York.
	\newblock \href {http://dx.doi.org/10.1201/9781420059458}
	{\path{doi:10.1201/9781420059458}}.
	
	\bibitem[Bangert et~al., 2010]{Bangert2010}
	Bangert, M., Hennig, P., and Oelfke, U. (2010).
	\newblock Using an infinite von {M}ises-{F}isher mixture model to cluster
	treatment beam directions in external radiation therapy.
	\newblock In {\em ICMLA '10}, pp. 746--751, Washington D. C. IEEE Computer
	Society.
	\newblock \href {http://dx.doi.org/10.1109/icmla.2010.114}
	{\path{doi:10.1109/icmla.2010.114}}.
	
	\bibitem[Baragona, 2003]{Baragona2003}
	Baragona, R. (2003).
	\newblock Further results on {L}und's statistic for identifying cluster in a
	circular data set with application to time series.
	\newblock {\em Communications in Statistics -- Simulation and Computation},
	32(3):943--952.
	\newblock \href {http://dx.doi.org/10.1081/sac-120017869}
	{\path{doi:10.1081/sac-120017869}}.
	
	\bibitem[Barnett and Baker, 2020]{Barnett2020}
	Barnett, A. and Baker, P. (2020).
	\newblock {\em {season}: Analysing Seasonal Data {R} Functions}.
	\newblock {R} package version 0.3.12.
	\newblock URL: \url{https://CRAN.R-project.org/package=season}.
	
	\bibitem[Barrag\'an et~al., 2013]{Barragan2016}
	Barrag\'an, S., Fern\'andez, M.~A., Rueda, C., and Peddada, S. (2013).
	\newblock {isocir}: an {R} package for constrained inference using isotonic
	regression for circular data, with an application to cell biology.
	\newblock {\em Journal of Statistical Software}, 54(4):1--17.
	\newblock \href {http://dx.doi.org/10.18637/jss.v054.i04}
	{\path{doi:10.18637/jss.v054.i04}}.
	
	\bibitem[Barrag\'an et~al., 2017]{Barragan2017}
	Barrag\'an, S., Rueda, C., and Fern\'andez, M.~A. (2017).
	\newblock Circular order aggregation and its application to cell-cycle genes
	expressions.
	\newblock {\em IEEE/ACM Transactions on Computational Biology and
		Bioinformatics}, 14(4):819--829.
	
	\bibitem[Barrag\'an et~al., 2015]{Barragan2015}
	Barrag\'an, S., Rueda, C., Fern\'andez, M.~A., and Peddada, S.~D. (2015).
	\newblock Determination of temporal order among the components of an
	oscillatory system.
	\newblock {\em PLOS One}, 10:e0124842.
	\newblock \href {http://dx.doi.org/10.1371/journal.pone.0124842}
	{\path{doi:10.1371/journal.pone.0124842}}.
	
	\bibitem[Barros et~al., 2017]{Barros2017}
	Barros, C.~M., Amaral, G. J.~A., Nascimento, A. D.~C., and Cysneiros, A. H.
	M.~A. (2017).
	\newblock Detecting influential observations in {W}atson data.
	\newblock {\em Communications in Statistics -- Simulation and Computation},
	46(14):6882--6898.
	\newblock \href {http://dx.doi.org/10.1080/03610926.2016.1139130}
	{\path{doi:10.1080/03610926.2016.1139130}}.
	
	\bibitem[Batschelet, 1981]{Batschelet1981}
	Batschelet, E. (1981).
	\newblock {\em Circular Statistics in Biology}.
	\newblock Mathematics in Biology. Academic Press, London.
	
	\bibitem[Benjamin et~al., 2019]{Benjamin2019}
	Benjamin, J. B.~M., Hussain, I., and Yang, M.-S. (2019).
	\newblock Possibilistic c-means clustering on directional data.
	\newblock In {\em 2019 12th International Congress on Image and Signal
		Processing, BioMedical Engineering and Informatics ({CISP-BMEI})}, pp. 1--6,
	New York. IEEE.
	\newblock \href {http://dx.doi.org/10.1109/cisp-bmei48845.2019.8965703}
	{\path{doi:10.1109/cisp-bmei48845.2019.8965703}}.
	
	\bibitem[Beran and Ghosh, 2020]{Beran2020}
	Beran, J. and Ghosh, S. (2020).
	\newblock Estimating the mean direction of strongly dependent circular time
	series.
	\newblock {\em Journal of Time Series Analysis}, 41:210--228.
	\newblock \href {http://dx.doi.org/10.1111/jtsa.12500}
	{\path{doi:10.1111/jtsa.12500}}.
	
	\bibitem[Beran, 2016]{Beran2016}
	Beran, R. (2016).
	\newblock Nonparametric estimation of trend in directional data.
	\newblock {\em Stochastic Processes and their Applications},
	126(12):3808--3827.
	\newblock \href {http://dx.doi.org/10.1016/j.spa.2016.04.018}
	{\path{doi:10.1016/j.spa.2016.04.018}}.
	
	\bibitem[Beran, 1968]{Beran1968}
	Beran, R.~J. (1968).
	\newblock Testing for uniformity on a compact homogeneous space.
	\newblock {\em Journal of Applied Probability}, 5(1):177--195.
	\newblock \href {http://dx.doi.org/10.1017/s002190020003237x}
	{\path{doi:10.1017/s002190020003237x}}.
	
	\bibitem[Beran, 1969]{Beran1969}
	Beran, R.~J. (1969).
	\newblock Asymptotic theory of a class of tests for uniformity of a circular
	distribution.
	\newblock {\em Annals of Mathematical Statistics}, 40(4):1196--1206.
	\newblock \href {http://dx.doi.org/10.1214/aoms/1177697496}
	{\path{doi:10.1214/aoms/1177697496}}.
	
	\bibitem[Beran, 1979]{Beran1979}
	Beran, R.~J. (1979).
	\newblock Exponential models for directional data.
	\newblock {\em The Annals of Statistics}, 7(6):1162--1178.
	\newblock \href {http://dx.doi.org/10.1214/aos/1176344838}
	{\path{doi:10.1214/aos/1176344838}}.
	
	\bibitem[Berens, 2009]{Berens2009}
	Berens, P. (2009).
	\newblock {CircStat}: a {MATLAB} toolbox for circular statistics.
	\newblock {\em Journal of Statistical Software}, 31(10):1--21.
	\newblock \href {http://dx.doi.org/10.18637/jss.v031.i10}
	{\path{doi:10.18637/jss.v031.i10}}.
	
	\bibitem[Bhattacharya and Bhattacharya, 2012]{Bhattacharya2012}
	Bhattacharya, A. and Bhattacharya, R. (2012).
	\newblock {\em Nonparametric Inference on Manifolds}, volume~2 of {\em
		Institute of Mathematical Statistics Monographs}.
	\newblock Cambridge University Press, Cambridge.
	\newblock \href {http://dx.doi.org/10.1017/CBO9781139094764}
	{\path{doi:10.1017/CBO9781139094764}}.
	
	\bibitem[Bhattacharya and Dunson, 2012]{Bhattacharya2012a}
	Bhattacharya, A. and Dunson, D. (2012).
	\newblock Nonparametric {B}ayes classification and hypothesis testing on
	manifolds.
	\newblock {\em Journal of Multivariate Analysis}, 111:1--19.
	\newblock \href {http://dx.doi.org/10.1016/j.jmva.2012.02.020}
	{\path{doi:10.1016/j.jmva.2012.02.020}}.
	
	\bibitem[Bhattacharya and Patrangenaru, 2003]{Bhattacharya2003}
	Bhattacharya, R. and Patrangenaru, V. (2003).
	\newblock Large sample theory of intrinsic and extrinsic sample means on
	manifolds.
	\newblock {\em The Annals of Statistics}, 31(1):1--29.
	\newblock \href {http://dx.doi.org/10.1214/aos/1046294456}
	{\path{doi:10.1214/aos/1046294456}}.
	
	\bibitem[Bhattacharya and Patrangenaru, 2005]{Bhattacharya2005}
	Bhattacharya, R. and Patrangenaru, V. (2005).
	\newblock Large sample theory of intrinsic and extrinsic sample means on
	manifolds--{II}.
	\newblock {\em The Annals of Statistics}, 33(3):1225--1259.
	\newblock \href {http://dx.doi.org/10.1214/009053605000000093}
	{\path{doi:10.1214/009053605000000093}}.
	
	\bibitem[Bhattacharya and Patrangenaru, 2014]{Bhattacharya2014}
	Bhattacharya, R. and Patrangenaru, V. (2014).
	\newblock Statistics on manifolds and landmarks based image analysis: a
	nonparametric theory with applications.
	\newblock {\em Journal of Statistical Planning and Inference}, 145:1--22.
	\newblock \href {http://dx.doi.org/10.1016/j.jspi.2013.08.001}
	{\path{doi:10.1016/j.jspi.2013.08.001}}.
	
	\bibitem[Bhattacharya and SenGupta, 2009a]{Bhattacharya2009a}
	Bhattacharya, S. and SenGupta, A. (2009a).
	\newblock Bayesian analysis of semiparametric linear-circular models.
	\newblock {\em Journal of Agricultural, Biological, and Environmental
		Statistics}, 14(1):33--65.
	\newblock \href {http://dx.doi.org/10.1198/jabes.2009.0003}
	{\path{doi:10.1198/jabes.2009.0003}}.
	
	\bibitem[Bhattacharya and SenGupta, 2009b]{Bhattacharya2009}
	Bhattacharya, S. and SenGupta, A. (2009b).
	\newblock Bayesian inference for circular distributions with unknown
	normalising constants.
	\newblock {\em Journal of Statistical Planning and Inference},
	139(12):4179--4192.
	\newblock \href {http://dx.doi.org/10.1016/j.jspi.2009.06.008}
	{\path{doi:10.1016/j.jspi.2009.06.008}}.
	
	\bibitem[Bijral et~al., 2007]{Bijral2007}
	Bijral, A., Breitenbach, M., and Grudic, G.~Z. (2007).
	\newblock Mixture of {W}atson distributions: a generative model for
	hyperspherical embeddings.
	\newblock In Meila, M. and Shen, X. (Eds.), {\em Proceedings of the Eleventh
		International Conference on Artificial Intelligence and Statistics},
	Proceedings of Machine Learning Research, pp. 35--42, San Juan, Puerto Rico.
	PMLR.
	
	\bibitem[Bingham, 1974]{Bingham1974}
	Bingham, C. (1974).
	\newblock An antipodally symmetric distribution on the sphere.
	\newblock {\em The Annals of Statistics}, 2(6):1201--1225.
	\newblock \href {http://dx.doi.org/10.1214/aos/1176342874}
	{\path{doi:10.1214/aos/1176342874}}.
	
	\bibitem[Boente and Fraiman, 1991]{Boente1991}
	Boente, G. and Fraiman, R. (1991).
	\newblock Nonparametric regression for directional data.
	\newblock {\em Trabajos de Matem\'atica}, 176:1--13.
	
	\bibitem[Boente et~al., 2014]{Boente2014}
	Boente, G., Rodriguez, D., and Gonz\'alez-Manteiga, W. (2014).
	\newblock Goodness-of-fit test for directional data.
	\newblock {\em Scandinavian Journal of Statistics}, 41(1):259--275.
	\newblock \href {http://dx.doi.org/10.1111/sjos.12020}
	{\path{doi:10.1111/sjos.12020}}.
	
	\bibitem[Bogdan et~al., 2002]{Bogdan2002}
	Bogdan, M., Bogdan, K., and Futschik, A. (2002).
	\newblock A data driven smooth test for circular uniformity.
	\newblock {\em Annals of the Institute of Statistical Mathematics},
	54(1):29--44.
	\newblock \href {http://dx.doi.org/10.1023/A:1016109603897}
	{\path{doi:10.1023/A:1016109603897}}.
	
	\bibitem[Boomsma et~al., 2008]{Boomsma2008}
	Boomsma, W., Mardia, K.~V., Taylor, C.~C., Ferkinghoff-Borg, J., Krogh, A., and
	Hamelryck, T. (2008).
	\newblock A generative, probabilistic model of local protein structure.
	\newblock {\em Proceedings of the National Academy of Sciences of the United
		States of America}, 105(26):8932--8937.
	\newblock \href {http://dx.doi.org/10.1073/pnas.0801715105}
	{\path{doi:10.1073/pnas.0801715105}}.
	
	\bibitem[Boulerice and Ducharme, 1997]{Boulerice1997}
	Boulerice, B. and Ducharme, G.~R. (1997).
	\newblock Smooth tests of goodness-of-fit for directional and axial data.
	\newblock {\em Journal of Multivariate Analysis}, 60(1):154--174.
	\newblock \href {http://dx.doi.org/10.1006/jmva.1996.1650}
	{\path{doi:10.1006/jmva.1996.1650}}.
	
	\bibitem[Breckling, 1989]{Breckling1989}
	Breckling, J. (1989).
	\newblock {\em The Analysis of Directional Time Series: Applications to Wind
		Speed and Direction}, volume~61 of {\em Lecture Notes in Statistics}.
	\newblock Springer, London.
	
	\bibitem[Bulla et~al., 2012]{Bulla2012}
	Bulla, J., Lagona, F., Maruotti, A., and Picone, M. (2012).
	\newblock A multivariate hidden {M}arkov model for the identification of sea
	regimes from incomplete skewed and circular time series.
	\newblock {\em Journal of Agricultural, Biological, and Environmental
		Statistics}, 17(4):544--567.
	\newblock \href {http://dx.doi.org/10.1007/s13253-012-0110-1}
	{\path{doi:10.1007/s13253-012-0110-1}}.
	
	\bibitem[Buttarazzi, 2020]{Buttarazzi2020}
	Buttarazzi, D. (2020).
	\newblock {\em {bpDir}: Boxplots for Directional Data}.
	\newblock {R} package version 0.1.1.
	\newblock URL: \url{https://CRAN.R-project.org/package=bpDir}.
	
	\bibitem[Buttarazzi et~al., 2018]{Buttarazzi2018}
	Buttarazzi, D., Pandolfo, G., and Porzio, G.~C. (2018).
	\newblock A boxplot for circular data.
	\newblock {\em Biometrics}, 74(4):1492--1501.
	\newblock \href {http://dx.doi.org/10.1111/biom.12889}
	{\path{doi:10.1111/biom.12889}}.
	
	\bibitem[Byrne et~al., 2009]{Byrne2009}
	Byrne, R.~W., Noser, R., Bates, L.~A., and Jupp, P.~E. (2009).
	\newblock How did they get here from there? detecting changes of direction in
	terrestrial ranging.
	\newblock {\em Animal Behaviour}, 77(3):619--631.
	\newblock \href {http://dx.doi.org/10.1016/j.anbehav.2008.11.014}
	{\path{doi:10.1016/j.anbehav.2008.11.014}}.
	
	\bibitem[Byrne and Girolami, 2013]{Byrne2013}
	Byrne, S. and Girolami, M. (2013).
	\newblock Geodesic {M}onte {C}arlo on embedded manifolds.
	\newblock {\em Scandinavian Journal of Statistics}, 40(4):825--845.
	\newblock \href {http://dx.doi.org/10.1111/sjos.12036}
	{\path{doi:10.1111/sjos.12036}}.
	
	\bibitem[Cabella and Marinucci, 2009]{Cabella2009}
	Cabella, P. and Marinucci, D. (2009).
	\newblock Statistical challenges in the analysis of cosmic microwave background
	radiation.
	\newblock {\em The Annals of Applied Statistics}, 3(1):61--95.
	\newblock \href {http://dx.doi.org/10.1214/08-aoas190}
	{\path{doi:10.1214/08-aoas190}}.
	
	\bibitem[Cai et~al., 2013]{Cai2013}
	Cai, T., Fan, J., and Jiang, T. (2013).
	\newblock Distributions of angles in random packing on spheres.
	\newblock {\em Journal of Machine Learning Research}, 14(21):1837--1864.
	
	\bibitem[Cai and Jiang, 2012]{Cai2012}
	Cai, T. and Jiang, T. (2012).
	\newblock Phase transition in limiting distributions of coherence of
	high-dimensional random matrices.
	\newblock {\em Journal of Multivariate Analysis}, 107:24--39.
	\newblock \href {http://dx.doi.org/10.1016/j.jmva.2011.11.008}
	{\path{doi:10.1016/j.jmva.2011.11.008}}.
	
	\bibitem[Calderara et~al., 2011]{Calderara2011}
	Calderara, S., Prati, A., and Cucchiara, R. (2011).
	\newblock Mixtures of von {M}ises distributions for people trajectory shape
	analysis.
	\newblock {\em IEEE Transactions on Circuits and Systems for Video Technology},
	21(4):457--471.
	\newblock \href {http://dx.doi.org/10.1109/tcsvt.2011.2125550}
	{\path{doi:10.1109/tcsvt.2011.2125550}}.
	
	\bibitem[Carnicero et~al., 2018]{Carnicero2018}
	Carnicero, J.~A., Wiper, M.~P., and Aus\'in, M.~C. (2018).
	\newblock Density estimation of circular data with {B}ernstein polynomials.
	\newblock {\em Hacettepe Journal of Mathematics and Statistics},
	47(2):273--286.
	\newblock \href {http://dx.doi.org/10.15672/hjms.2014437525}
	{\path{doi:10.15672/hjms.2014437525}}.
	
	\bibitem[Carta et~al., 2008]{Carta2008}
	Carta, J.~A., Ramirez, P., and Bueno, C. (2008).
	\newblock A joint probability density function of wind speed and direction for
	wind energy analysis.
	\newblock {\em Energy Conversion and Management}, 49(6):1309--1320.
	\newblock \href {http://dx.doi.org/10.1016/j.enconman.2008.01.010}
	{\path{doi:10.1016/j.enconman.2008.01.010}}.
	
	\bibitem[Cetingul and Vidal, 2009]{Cetingul2009}
	Cetingul, H.~E. and Vidal, R. (2009).
	\newblock Intrinsic mean shift for clustering on {S}tiefel and {G}rassmann
	manifolds.
	\newblock In {\em 2009 {IEEE} Conference on Computer Vision and Pattern
		Recognition}, pp. 1896--1902, New York. IEEE.
	\newblock \href {http://dx.doi.org/10.1109/cvpr.2009.5206806}
	{\path{doi:10.1109/cvpr.2009.5206806}}.
	
	\bibitem[Chakraborty and Wong, 2019]{Chakraborty2019}
	Chakraborty, S. and Wong, S. W.~K. (2019).
	\newblock {\em {BAMBI}: Bivariate Angular Mixture Models}.
	\newblock {R} package version 2.3.0.
	\newblock URL: \url{https://CRAN.R-project.org/package=BAMBI}.
	
	\bibitem[Chang, 1986]{Chang1986}
	Chang, T. (1986).
	\newblock Spherical regression.
	\newblock {\em The Annals of Statistics}, 14(3):907--924.
	\newblock \href {http://dx.doi.org/10.1002/0471667196.ess0734}
	{\path{doi:10.1002/0471667196.ess0734}}.
	
	\bibitem[Chang-Chien et~al., 2010]{ChangChien2010}
	Chang-Chien, S.-J., Yang, M.-S., and Hung, W.-L. (2010).
	\newblock Mean shift-based clustering for directional data.
	\newblock In {\em Third International Workshop on Advanced Computational
		Intelligence}, pp. 367--372.
	\newblock \href {http://dx.doi.org/10.1109/IWACI.2010.5585203}
	{\path{doi:10.1109/IWACI.2010.5585203}}.
	
	\bibitem[Chaubey, 2018]{Chaubey2018}
	Chaubey, Y.~P. (2018).
	\newblock Smooth kernel estimation of a circular density function: a connection
	to orthogonal polynomials on the unit circle.
	\newblock {\em Journal of Probability and Statistics}, 2018:1--4.
	\newblock \href {http://dx.doi.org/10.1155/2018/5372803}
	{\path{doi:10.1155/2018/5372803}}.
	
	\bibitem[Chaudhuri and Marron, 1999]{Chaudhuri1999}
	Chaudhuri, P. and Marron, J.~S. (1999).
	\newblock Si{Z}er for exploration of structures in curves.
	\newblock {\em Journal of the American Statistical Association},
	94(447):807--823.
	\newblock \href {http://dx.doi.org/10.1080/01621459.1999.10474186}
	{\path{doi:10.1080/01621459.1999.10474186}}.
	
	\bibitem[Cheng et~al., 2020]{Cheng2020}
	Cheng, D., Cammarota, V., Fantaye, Y., Marinucci, D., and Schwartzman, A.
	(2020).
	\newblock Multiple testing of local maxima for detection of peaks on the
	(celestial) sphere.
	\newblock {\em Bernoulli}, 26(1):31--60.
	\newblock \href {http://dx.doi.org/10.3150/18-bej1068}
	{\path{doi:10.3150/18-bej1068}}.
	
	\bibitem[Cheng and Wu, 2013]{Cheng2013}
	Cheng, M.-Y. and Wu, H.-T. (2013).
	\newblock Local linear regression on manifolds and its geometric
	interpretation.
	\newblock {\em Journal of the American Statistical Association},
	108(504):1421--1434.
	\newblock \href {http://dx.doi.org/10.1080/01621459.2013.827984}
	{\path{doi:10.1080/01621459.2013.827984}}.
	
	\bibitem[Chikuse, 2012]{Chikuse2012}
	Chikuse, Y. (2012).
	\newblock {\em Statistics on Special Manifolds}, volume 174 of {\em Lecture
		Notes in Statistics}.
	\newblock Springer, Heidelberg.
	\newblock \href {http://dx.doi.org/10.1007/978-0-387-21540-2}
	{\path{doi:10.1007/978-0-387-21540-2}}.
	
	\bibitem[Chirikjian and Kyatkin, 2001]{Chirikjian2001}
	Chirikjian, G.~S. and Kyatkin, A. (2001).
	\newblock {\em Engineering Applications of Noncommutative Harmonic Analysis}.
	\newblock CRC Press, Boca Raton.
	\newblock \href {http://dx.doi.org/10.1115/1.1421108}
	{\path{doi:10.1115/1.1421108}}.
	
	\bibitem[Chiuso and Picci, 1998]{Chiuso1998}
	Chiuso, A. and Picci, G. (1998).
	\newblock Visual tracking of points as estimation on the unit sphere.
	\newblock In Kriegman, D.~J., Hager, G.~D., and Morse, A.~S. (Eds.), {\em The
		Confluence of Vision and Control}, volume 237 of {\em Lecture Notes in
		Control and Information Sciences}, pp. 90--105, London. Springer.
	\newblock \href {http://dx.doi.org/10.1007/BFb0109665}
	{\path{doi:10.1007/BFb0109665}}.
	
	\bibitem[Codling et~al., 2008]{Codling2008}
	Codling, E.~A., Plank, M.~J., and Benhamou, S. (2008).
	\newblock Random walk models in biology.
	\newblock {\em Journal of the Royal Society Interface}, 5(25):813--834.
	\newblock \href {http://dx.doi.org/10.1098/rsif.2008.0014}
	{\path{doi:10.1098/rsif.2008.0014}}.
	
	\bibitem[Comte and Taupin, 2003]{Comte2003}
	Comte, F. and Taupin, M.~L. (2003).
	\newblock Adaptive density deconvolution for circular data.
	\newblock Technical Report MAP5 2003-10, Universit\'e Paris Descartes.
	
	\bibitem[Cornea et~al., 2017]{Cornea2017}
	Cornea, E., Zhu, H., Kim, P., and Ibrahim, J.~G. (2017).
	\newblock Regression models on {R}iemannian symmetric spaces.
	\newblock {\em Journal of the Royal Statistical Society, Series B (Statistical
		Methodology)}, 79(2):463--482.
	\newblock \href {http://dx.doi.org/10.1111/rssb.12169}
	{\path{doi:10.1111/rssb.12169}}.
	
	\bibitem[Costa et~al., 2014]{Costa2014}
	Costa, M., Koivunen, V., and Poor, H.~V. (2014).
	\newblock Estimating directional statistics using wavefield modeling and
	mixtures of von-{M}ises distributions.
	\newblock {\em IEEE Signal Processing Letters}, 21(12):1496--1500.
	\newblock \href {http://dx.doi.org/10.1109/lsp.2014.2341651}
	{\path{doi:10.1109/lsp.2014.2341651}}.
	
	\bibitem[Cremers, 2020]{Cremers2020}
	Cremers, J. (2020).
	\newblock {\em {bpnreg}: {B}ayesian Projected Normal Regression Models for
		Circular Data}.
	\newblock {R} package version 1.0.3.
	\newblock URL: \url{https://CRAN.R-project.org/package=bpnreg}.
	
	\bibitem[Cremers et~al., 2018]{Cremers2018}
	Cremers, J., Mulder, K.~T., and Klugkist, I. (2018).
	\newblock Circular interpretation of regression coefficients.
	\newblock {\em British Journal of Mathematical and Statistical Psychology},
	71(1):75--95.
	\newblock \href {http://dx.doi.org/10.1111/bmsp.12108}
	{\path{doi:10.1111/bmsp.12108}}.
	
	\bibitem[Cremers et~al., 2020]{Cremers2020a}
	Cremers, J., Pennings, H. J.~M., and Ley, C. (2020).
	\newblock Regression models for cylindrical data in psychology.
	\newblock {\em Multivariate Behavioral Research}, to appear.
	\newblock \href {http://dx.doi.org/10.1080/00273171.2019.1693332}
	{\path{doi:10.1080/00273171.2019.1693332}}.
	
	\bibitem[Cuesta-Albertos et~al., 2009]{Cuesta-Albertos2009}
	Cuesta-Albertos, J.~A., Cuevas, A., and Fraiman, R. (2009).
	\newblock On projection-based tests for directional and compositional data.
	\newblock {\em Statistics and Computing}, 19(4):367--380.
	\newblock \href {http://dx.doi.org/10.1007/s11222-008-9098-3}
	{\path{doi:10.1007/s11222-008-9098-3}}.
	
	\bibitem[Curry et~al., 2019]{Curry2019}
	Curry, C., Marsland, S., and McLachlan, R.~I. (2019).
	\newblock Principal symmetric space analysis.
	\newblock {\em Journal of Computational Dynamics}, 6(2):251--276.
	\newblock \href {http://dx.doi.org/10.3934/jcd.2019013}
	{\path{doi:10.3934/jcd.2019013}}.
	
	\bibitem[Cutting et~al., 2017a]{Cutting2017}
	Cutting, C., Paindaveine, D., and Verdebout, T. (2017a).
	\newblock Testing uniformity on high-dimensional spheres against monotone
	rotationally symmetric alternatives.
	\newblock {\em The Annals of Statistics}, 45(3):1024--1058.
	\newblock \href {http://dx.doi.org/10.1214/16-aos1473}
	{\path{doi:10.1214/16-aos1473}}.
	
	\bibitem[Cutting et~al., 2017b]{Cutting2017a}
	Cutting, C., Paindaveine, D., and Verdebout, T. (2017b).
	\newblock Tests of concentration for low-dimensional and high-dimensional
	directional data.
	\newblock In Ahmed, S.~E. (Ed.), {\em Big and Complex Data Analysis},
	Contributions to Statistics. Springer, New York.
	\newblock \href {http://dx.doi.org/10.1007/978-3-319-41573-4_11}
	{\path{doi:10.1007/978-3-319-41573-4_11}}.
	
	\bibitem[Cutting et~al., 2020]{Cutting2020}
	Cutting, C., Paindaveine, D., and Verdebout, T. (2020).
	\newblock On the power of axial tests of uniformity on spheres.
	\newblock {\em Electronic Journal of Statistics}, 14(1):2123--2154.
	\newblock \href {http://dx.doi.org/10.1214/20-EJS1716}
	{\path{doi:10.1214/20-EJS1716}}.
	
	\bibitem[Dai and Xu, 2013]{Dai2013}
	Dai, F. and Xu, Y. (2013).
	\newblock {\em Approximation Theory and Harmonic Analysis on Spheres and
		Balls}.
	\newblock Springer Monographs in Mathematics. Springer, New York.
	\newblock \href {http://dx.doi.org/10.1007/978-1-4614-6660-4}
	{\path{doi:10.1007/978-1-4614-6660-4}}.
	
	\bibitem[Dai and M\"uller, 2018]{Dai2018}
	Dai, X. and M\"uller, H.-G. (2018).
	\newblock Principal component analysis for functional data on {R}iemannian
	manifolds and spheres.
	\newblock {\em The Annals of Statistics}, 46(6B):3334--3361.
	\newblock \href {http://dx.doi.org/10.1214/17-aos1660}
	{\path{doi:10.1214/17-aos1660}}.
	
	\bibitem[Damien and Walker, 1999]{Damien1999}
	Damien, P. and Walker, S. (1999).
	\newblock A full {B}ayesian analysis of circular data using the von {M}ises
	distribution.
	\newblock {\em The Canadian Journal of Statistics}, 27(2):291--298.
	\newblock \href {http://dx.doi.org/10.2307/3315639}
	{\path{doi:10.2307/3315639}}.
	
	\bibitem[Damon and Marron, 2014]{Damon2014}
	Damon, J. and Marron, J.~S. (2014).
	\newblock Backwards principal component analysis and principal nested
	relations.
	\newblock {\em Journal of Mathematical Imaging and Vision}, 50(1):107--114.
	\newblock \href {http://dx.doi.org/10.1007/s10851-013-0463-2}
	{\path{doi:10.1007/s10851-013-0463-2}}.
	
	\bibitem[D'Elia, 2001]{DElia2001}
	D'Elia, A. (2001).
	\newblock A statistical model for orientation mechanism.
	\newblock {\em Statistical Methods and Applications}, 10(1-3):157--174.
	\newblock \href {http://dx.doi.org/10.1007/BF02511646}
	{\path{doi:10.1007/BF02511646}}.
	
	\bibitem[Demni et~al., 2019]{Demni2019}
	Demni, H., Messaoud, A., and Porzio, G.~C. (2019).
	\newblock The cosine depth distribution classifier for directional data.
	\newblock In Bauer, N., Ickstadt, K., L\"ubke, K., Szepannek, G., Trautmann,
	H., and Vichi, M. (Eds.), {\em Applications in Statistical Computing},
	Studies in Classification, Data Analysis, and Knowledge Organization, pp.
	49--60. Springer, Cham.
	\newblock \href {http://dx.doi.org/10.1007/978-3-030-25147-5_4}
	{\path{doi:10.1007/978-3-030-25147-5_4}}.
	
	\bibitem[Deschepper et~al., 2008]{Deschepper2008}
	Deschepper, E., Thas, O., and Ottoy, J.-P. (2008).
	\newblock Tests and diagnostic plots for detecting lack-of-fit for
	circular-linear regression models.
	\newblock {\em Biometrics}, 64(3):912--920.
	\newblock \href {http://dx.doi.org/10.1111/j.1541-0420.2007.00950.x}
	{\path{doi:10.1111/j.1541-0420.2007.00950.x}}.
	
	\bibitem[Dette et~al., 2019]{Dette2019}
	Dette, H., Konstantinou, M., Schorning, K., and G\"osmann, J. (2019).
	\newblock Optimal designs for regression with spherical data.
	\newblock {\em Electronic Journal of Statistics}, 13(1):361--390.
	\newblock \href {http://dx.doi.org/10.1214/18-ejs1524}
	{\path{doi:10.1214/18-ejs1524}}.
	
	\bibitem[Dette and Melas, 2003]{Dette2003}
	Dette, H. and Melas, V.~B. (2003).
	\newblock Optimal designs for estimating individual coefficients in {F}ourier
	regression models.
	\newblock {\em The Annals of Statistics}, 31(5):1669--1692.
	\newblock \href {http://dx.doi.org/10.1214/aos/1065705122}
	{\path{doi:10.1214/aos/1065705122}}.
	
	\bibitem[Dette et~al., 2005]{Dette2005}
	Dette, H., Melas, V.~B., and Pepelyshev, A. (2005).
	\newblock Optimal designs for three-dimensional shape analysis with spherical
	harmonic descriptors.
	\newblock {\em The Annals of Statistics}, 33(6):2758--2788.
	\newblock \href {http://dx.doi.org/10.1214/009053605000000552}
	{\path{doi:10.1214/009053605000000552}}.
	
	\bibitem[Dette and Wiens, 2009]{Dette2009}
	Dette, H. and Wiens, D.~P. (2009).
	\newblock Robust designs for 3{D} shape analysis with spherical harmonic
	descriptors.
	\newblock {\em Statistica Sinica}, 19(1):83--102.
	
	\bibitem[Dhillon and Sra, 2003]{Dhillon2003a}
	Dhillon, I. and Sra, S. (2003).
	\newblock Modeling data using directional distributions.
	\newblock Technical Report TR-03-06, Department of Computer Sciences,
	University of Texas at Austin.
	
	\bibitem[Dhillon and Modha, 2001]{Dhillon2001}
	Dhillon, I.~S. and Modha, D.~S. (2001).
	\newblock Concept decompositions for large sparse text data using clustering.
	\newblock {\em Machine Learning}, 42(1):143--175.
	\newblock \href {http://dx.doi.org/10.1023/A:1007612920971}
	{\path{doi:10.1023/A:1007612920971}}.
	
	\bibitem[Di~Marzio et~al., 2016a]{DiMarzio2016}
	Di~Marzio, M., Fensore, S., Panzera, A., and Taylor, C.~C. (2016a).
	\newblock A note on nonparametric estimation of circular conditional densities.
	\newblock {\em Journal of Statistical Computation and Simulation},
	86(13):2573--2582.
	\newblock \href {http://dx.doi.org/10.1080/00949655.2016.1146279}
	{\path{doi:10.1080/00949655.2016.1146279}}.
	
	\bibitem[Di~Marzio et~al., 2016b]{DiMarzio2016b}
	Di~Marzio, M., Fensore, S., Panzera, A., and Taylor, C.~C. (2016b).
	\newblock Practical performance of local likelihood for circular density
	estimation.
	\newblock {\em Journal of Statistical Computation and Simulation},
	86(13):2560--2572.
	\newblock \href {http://dx.doi.org/10.1080/00949655.2016.1149588}
	{\path{doi:10.1080/00949655.2016.1149588}}.
	
	\bibitem[Di~Marzio et~al., 2017]{DiMarzio2017}
	Di~Marzio, M., Fensore, S., Panzera, A., and Taylor, C.~C. (2017).
	\newblock Nonparametric estimating equations for circular probability density
	functions and their derivatives.
	\newblock {\em Electronic Journal of Statistics}, 11(2):4323--4346.
	\newblock \href {http://dx.doi.org/10.1214/17-EJS1318}
	{\path{doi:10.1214/17-EJS1318}}.
	
	\bibitem[Di~Marzio et~al., 2018a]{DiMarzio2018a}
	Di~Marzio, M., Fensore, S., Panzera, A., and Taylor, C.~C. (2018a).
	\newblock Circular local likelihood.
	\newblock {\em TEST}, 27(4):921--945.
	\newblock \href {http://dx.doi.org/10.1007/s11749-017-0576-9}
	{\path{doi:10.1007/s11749-017-0576-9}}.
	
	\bibitem[Di~Marzio et~al., 2018b]{DiMarzio2018}
	Di~Marzio, M., Fensore, S., Panzera, A., and Taylor, C.~C. (2018b).
	\newblock Nonparametric classification for circular data.
	\newblock In Ley, C. and Verdebout, T. (Eds.), {\em Applied Directional
		Statistics}, Chapman \& Hall/CRC Interdisciplinary Statistics Series, pp.
	241--257. CRC Press, Boca Raton.
	
	\bibitem[Di~Marzio et~al., 2019a]{DiMarzio2019a}
	Di~Marzio, M., Fensore, S., Panzera, A., and Taylor, C.~C. (2019a).
	\newblock Kernel density classification for spherical data.
	\newblock {\em Statistics \& Probability Letters}, 144:23--29.
	\newblock \href {http://dx.doi.org/10.1016/j.spl.2018.07.018}
	{\path{doi:10.1016/j.spl.2018.07.018}}.
	
	\bibitem[Di~Marzio et~al., 2019b]{DiMarzio2019}
	Di~Marzio, M., Fensore, S., Panzera, A., and Taylor, C.~C. (2019b).
	\newblock Local binary regression with spherical predictors.
	\newblock {\em Statistics \& Probability Letters}, 144:30--36.
	\newblock \href {http://dx.doi.org/10.1016/j.spl.2018.07.019}
	{\path{doi:10.1016/j.spl.2018.07.019}}.
	
	\bibitem[Di~Marzio et~al., 2009]{DiMarzio2009}
	Di~Marzio, M., Panzera, A., and Taylor, C.~C. (2009).
	\newblock Local polynomial regression for circular predictors.
	\newblock {\em Statistics \& Probability Letters}, 79(19):2066--2075.
	\newblock \href {http://dx.doi.org/10.1016/j.spl.2009.06.014}
	{\path{doi:10.1016/j.spl.2009.06.014}}.
	
	\bibitem[Di~Marzio et~al., 2011]{DiMarzio2011}
	Di~Marzio, M., Panzera, A., and Taylor, C.~C. (2011).
	\newblock Kernel density estimation on the torus.
	\newblock {\em Journal of Statistical Planning and Inference},
	141(6):2156--2173.
	\newblock \href {http://dx.doi.org/10.1016/j.jspi.2011.01.002}
	{\path{doi:10.1016/j.jspi.2011.01.002}}.
	
	\bibitem[Di~Marzio et~al., 2012a]{DiMarzio2012a}
	Di~Marzio, M., Panzera, A., and Taylor, C.~C. (2012a).
	\newblock Non-parametric smoothing and prediction for nonlinear circular time
	series.
	\newblock {\em Journal of Time Series Analysis}, 33(4):620--630.
	\newblock \href {http://dx.doi.org/10.1111/j.1467-9892.2012.00794.x}
	{\path{doi:10.1111/j.1467-9892.2012.00794.x}}.
	
	\bibitem[Di~Marzio et~al., 2012b]{DiMarzio2012}
	Di~Marzio, M., Panzera, A., and Taylor, C.~C. (2012b).
	\newblock Smooth estimation of circular cumulative distribution functions and
	quantiles.
	\newblock {\em Journal of Nonparametric Statistics}, 24(4):935--949.
	\newblock \href {http://dx.doi.org/10.1080/10485252.2012.721517}
	{\path{doi:10.1080/10485252.2012.721517}}.
	
	\bibitem[Di~Marzio et~al., 2013]{DiMarzio2013}
	Di~Marzio, M., Panzera, A., and Taylor, C.~C. (2013).
	\newblock Non-parametric regression for circular responses.
	\newblock {\em Scandinavian Journal of Statistics}, 40(2):238--255.
	\newblock \href {http://dx.doi.org/10.1111/j.1467-9469.2012.00809.x}
	{\path{doi:10.1111/j.1467-9469.2012.00809.x}}.
	
	\bibitem[Di~Marzio et~al., 2014]{DiMarzio2014}
	Di~Marzio, M., Panzera, A., and Taylor, C.~C. (2014).
	\newblock Nonparametric regression for spherical data.
	\newblock {\em Journal of the American Statistical Association},
	109(506):748--763.
	\newblock \href {http://dx.doi.org/10.1080/01621459.2013.866567}
	{\path{doi:10.1080/01621459.2013.866567}}.
	
	\bibitem[Di~Marzio et~al., 2016c]{DiMarzio2016a}
	Di~Marzio, M., Panzera, A., and Taylor, C.~C. (2016c).
	\newblock Nonparametric circular quantile regression.
	\newblock {\em Journal of Statistical Planning and Inference}, 170:1--14.
	\newblock \href {http://dx.doi.org/10.1016/j.jspi.2015.08.004}
	{\path{doi:10.1016/j.jspi.2015.08.004}}.
	
	\bibitem[Di~Marzio et~al., 2019c]{DiMarzio2019b}
	Di~Marzio, M., Panzera, A., and Taylor, C.~C. (2019c).
	\newblock Nonparametric rotations for sphere-sphere regression.
	\newblock {\em Journal of the American Statistical Association},
	114(525):466--476.
	\newblock \href {http://dx.doi.org/10.1080/01621459.2017.1421542}
	{\path{doi:10.1080/01621459.2017.1421542}}.
	
	\bibitem[Dokmanic and Petrinovic, 2010]{Dokmanic2010}
	Dokmanic, I. and Petrinovic, D. (2010).
	\newblock Convolution on the $n$-sphere with application to pdf modeling.
	\newblock {\em IEEE Transactions on Signal Processing}, 58(3):1157--1170.
	\newblock \href {http://dx.doi.org/10.1109/TSP.2009.2033329}
	{\path{doi:10.1109/TSP.2009.2033329}}.
	
	\bibitem[Dortet-Bernadet and Wicker, 2008]{Dortet-Bernadet2008}
	Dortet-Bernadet, J.-L. and Wicker, N. (2008).
	\newblock Model-based clustering on the unit sphere with an illustration using
	gene expression profiles.
	\newblock {\em Biostatistics}, 9(1):66--80.
	\newblock \href {http://dx.doi.org/10.1093/biostatistics/kxm012}
	{\path{doi:10.1093/biostatistics/kxm012}}.
	
	\bibitem[Downs, 2003]{Downs2003}
	Downs, T.~D. (2003).
	\newblock Spherical regression.
	\newblock {\em Biometrika}, 90(3):655--668.
	\newblock \href {http://dx.doi.org/10.1093/biomet/90.3.655}
	{\path{doi:10.1093/biomet/90.3.655}}.
	
	\bibitem[Downs and Mardia, 2002]{Downs2002}
	Downs, T.~D. and Mardia, K.~V. (2002).
	\newblock Circular regression.
	\newblock {\em Biometrika}, 89(3):683--697.
	\newblock \href {http://dx.doi.org/10.1093/biomet/89.3.683}
	{\path{doi:10.1093/biomet/89.3.683}}.
	
	\bibitem[Dryden, 2019]{Dryden2019}
	Dryden, I.~L. (2019).
	\newblock {\em {shapes}: Statistical Shape Analysis}.
	\newblock {R} package version 1.2.5.
	\newblock URL: \url{https://CRAN.R-project.org/package=shapes}.
	
	\bibitem[Dryden and Mardia, 2016]{Dryden2016}
	Dryden, I.~L. and Mardia, K.~V. (2016).
	\newblock {\em Statistical Shape Analysis with Applications in {R}}.
	\newblock Wiley Series in Probability and Statistics. Wiley, Chichester, second
	edition.
	\newblock \href {http://dx.doi.org/10.1002/9781119072492}
	{\path{doi:10.1002/9781119072492}}.
	
	\bibitem[Ducharme et~al., 2012]{Ducharme2012}
	Ducharme, G.~R., Vincent, C., and Aliaume, C. (2012).
	\newblock A statistical test to detect vortices in the current fields of bodies
	of water.
	\newblock {\em Environmental and Ecological Statistics}, 19(3):345--367.
	\newblock \href {http://dx.doi.org/10.1007/s10651-012-0190-7}
	{\path{doi:10.1007/s10651-012-0190-7}}.
	
	\bibitem[Ebner et~al., 2018]{Ebner2018}
	Ebner, B., Henze, N., and Yukich, J.~E. (2018).
	\newblock Multivariate goodness-of-fit on flat and curved spaces via nearest
	neighbor distances.
	\newblock {\em Journal of Multivariate Analysis}, 165:231--242.
	\newblock \href {http://dx.doi.org/10.1016/j.jmva.2017.12.009}
	{\path{doi:10.1016/j.jmva.2017.12.009}}.
	
	\bibitem[Efromovich, 1997]{Efromovich1997}
	Efromovich, S. (1997).
	\newblock Density estimation for the case of supersmooth measurement error.
	\newblock {\em Journal of the American Statistical Association},
	92(438):526--535.
	\newblock \href {http://dx.doi.org/10.2307/2965701}
	{\path{doi:10.2307/2965701}}.
	
	\bibitem[Ehler and Galanis, 2011]{Ehler2011}
	Ehler, M. and Galanis, J. (2011).
	\newblock Frame theory in directional statistics.
	\newblock {\em Statistics \& Probability Letters}, 81(8):1046--1051.
	\newblock \href {http://dx.doi.org/10.1016/j.spl.2011.02.027}
	{\path{doi:10.1016/j.spl.2011.02.027}}.
	
	\bibitem[Eisen et~al., 1998]{Eisen1998}
	Eisen, M.~B., Spellman, P.~T., Brown, P.~O., and Botstein, D. (1998).
	\newblock Cluster analysis and display of genome-wide expression patterns.
	\newblock {\em Proceedings of the National Academy of Sciences of the United
		States of America}, 95(25):14863--14868.
	\newblock \href {http://dx.doi.org/10.1073/pnas.95.25.14863}
	{\path{doi:10.1073/pnas.95.25.14863}}.
	
	\bibitem[Elad et~al., 2005]{Elad2005}
	Elad, A., Keller, Y., and Kimmel, R. (2005).
	\newblock Texture mapping via spherical multi-dimensional scaling.
	\newblock In Kimmel, R., Sochen, N.~A., and Weickert, J. (Eds.), {\em Scale
		Space and {PDE} Methods in Computer Vision}, volume 3459 of {\em Lecture
		Notes in Computer Science}, pp. 443--455, Berlin. Springer.
	\newblock \href {http://dx.doi.org/10.1007/11408031_38}
	{\path{doi:10.1007/11408031_38}}.
	
	\bibitem[Eltzner et~al., 2018]{Eltzner2018}
	Eltzner, B., Huckemann, S., and Mardia, K.~V. (2018).
	\newblock Torus principal component analysis with applications to {RNA}
	structure.
	\newblock {\em The Annals of Applied Statistics}, 12(2):1332--1359.
	\newblock \href {http://dx.doi.org/10.1214/17-AOAS1115}
	{\path{doi:10.1214/17-AOAS1115}}.
	
	\bibitem[Eltzner and Huckemann, 2019]{Eltzner2019}
	Eltzner, B. and Huckemann, S.~F. (2019).
	\newblock A smeary central limit theorem for manifolds with application to
	high-dimensional spheres.
	\newblock {\em The Annals of Statistics}, 47(6):3360--3381.
	\newblock \href {http://dx.doi.org/10.1214/18-AOS1781}
	{\path{doi:10.1214/18-AOS1781}}.
	
	\bibitem[Eltzner et~al., 2015]{Eltzner2015}
	Eltzner, B., Jung, S., and Huckemann, S. (2015).
	\newblock Dimension reduction on polyspheres with application to skeletal
	representations.
	\newblock In Nielsen, F. and Barbaresco, F. (Eds.), {\em Geometric Science of
		Information}, volume 9389 of {\em Lecture Notes in Computer Science}, pp.
	22--29, Cham. Springer.
	\newblock \href {http://dx.doi.org/10.1007/978-3-319-25040-3_3}
	{\path{doi:10.1007/978-3-319-25040-3_3}}.
	
	\bibitem[Erdem and Shi, 2011]{Erdem2011}
	Erdem, E. and Shi, J. (2011).
	\newblock Comparison of bivariate distribution construction approaches for
	analysing wind speed and direction data.
	\newblock {\em Wind Energy}, 14(1):27--41.
	\newblock \href {http://dx.doi.org/10.1002/we.400} {\path{doi:10.1002/we.400}}.
	
	\bibitem[Esteves et~al., 2020]{Esteves2020}
	Esteves, C., Allen-Blanchette, C., Makadia, A., and Daniilidis, K. (2020).
	\newblock Learning {SO}(3) equivariant representations with spherical {CNN}s.
	\newblock {\em International Journal of Computer Vision}, 128:588--600.
	\newblock \href {http://dx.doi.org/10.1007/s11263-019-01220-1}
	{\path{doi:10.1007/s11263-019-01220-1}}.
	
	\bibitem[E\u{g}ecio\u{g}lu and Srinivasan, 2000]{Egecioglu2000}
	E\u{g}ecio\u{g}lu, O. and Srinivasan, A. (2000).
	\newblock Efficient nonparametric density estimation on the sphere with
	applications in fluid mechanics.
	\newblock {\em SIAM Journal on Scientific Computing}, 22(1):152--176.
	\newblock \href {http://dx.doi.org/10.1137/S1064827595290462}
	{\path{doi:10.1137/S1064827595290462}}.
	
	\bibitem[Fallaize and Kypraios, 2016]{Fallaize2016}
	Fallaize, C.~J. and Kypraios, T. (2016).
	\newblock Exact {B}ayesian inference for the {B}ingham distribution.
	\newblock {\em Statistics and Computing}, 26(1-2):349--360.
	\newblock \href {http://dx.doi.org/10.1007/s11222-014-9508-7}
	{\path{doi:10.1007/s11222-014-9508-7}}.
	
	\bibitem[Fa\"y et~al., 2013]{Fay2013}
	Fa\"y, G., Delabrouille, J., Kerkyacharian, G., and Picard, D. (2013).
	\newblock Testing the isotropy of high energy cosmic rays using spherical
	needlets.
	\newblock {\em The Annals of Applied Statistics}, 7(2):1040--1073.
	\newblock \href {http://dx.doi.org/10.1214/12-aoas619}
	{\path{doi:10.1214/12-aoas619}}.
	
	\bibitem[Fej\'er, 1916]{Fejer1916}
	Fej\'er, L. (1916).
	\newblock \"{U}ber trigonometrische {P}olynome.
	\newblock {\em Journal f\"ur die reine und angewandte Mathematik}, 146:53--82.
	\newblock \href {http://dx.doi.org/10.1515/crll.1916.146.53}
	{\path{doi:10.1515/crll.1916.146.53}}.
	
	\bibitem[Feltz and Goldin, 2001]{Feltz2001}
	Feltz, C.~J. and Goldin, G.~A. (2001).
	\newblock Partition-based goodness-of-fit tests on the line and the circle.
	\newblock {\em Australian \& New Zealand Journal of Statistics},
	43(2):207--220.
	\newblock \href {http://dx.doi.org/10.1111/1467-842X.00166}
	{\path{doi:10.1111/1467-842X.00166}}.
	
	\bibitem[Fernandes and Cardoso, 2016]{Fernandes2016}
	Fernandes, K. and Cardoso, J.~S. (2016).
	\newblock Discriminative directional classifiers.
	\newblock {\em Neurocomputing}, 207:141--149.
	\newblock \href {http://dx.doi.org/10.1016/j.neucom.2016.03.076}
	{\path{doi:10.1016/j.neucom.2016.03.076}}.
	
	\bibitem[Fern\'andez et~al., 2012]{Fernandez2012}
	Fern\'andez, M.~A., Rueda, C., and Peddada, S.~D. (2012).
	\newblock Identification of a core set of signature cell cycle genes whose
	relative order of time to peak expression is conserved across species.
	\newblock {\em Nucleic Acids Research}, 40(7):2823--2832.
	\newblock \href {http://dx.doi.org/10.1093/nar/gkr1077}
	{\path{doi:10.1093/nar/gkr1077}}.
	
	\bibitem[Fern\'andez-Dur\'an, 2004]{Fernandez-Duran2004}
	Fern\'andez-Dur\'an, J.~J. (2004).
	\newblock Circular distributions based on nonnegative trigonometric sums.
	\newblock {\em Biometrics}, 60(2):499--503.
	\newblock \href {http://dx.doi.org/10.1111/j.0006-341x.2004.00195.x}
	{\path{doi:10.1111/j.0006-341x.2004.00195.x}}.
	
	\bibitem[Fern\'andez-Dur\'an and Gregorio-Dom\'inguez,
	2014a]{Fernandez-Duran2014a}
	Fern\'andez-Dur\'an, J.~J. and Gregorio-Dom\'inguez, M.~M. (2014a).
	\newblock Distributions for spherical data based on nonnegative trigonometric
	sums.
	\newblock {\em Statistical Papers}, 55(4):983--1000.
	\newblock \href {http://dx.doi.org/10.1007/s00362-013-0547-5}
	{\path{doi:10.1007/s00362-013-0547-5}}.
	
	\bibitem[Fern\'andez-Dur\'an and Gregorio-Dom\'inguez,
	2014b]{Fernandez-Duran2014}
	Fern\'andez-Dur\'an, J.~J. and Gregorio-Dom\'inguez, M.~M. (2014b).
	\newblock Modeling angles in proteins and circular genomes using multivariate
	angular distributions based on multiple nonnegative trigonometric sums.
	\newblock {\em Statistical Applications in Genetics and Molecular Biology},
	13(1):1--18.
	\newblock \href {http://dx.doi.org/10.1515/sagmb-2012-0012}
	{\path{doi:10.1515/sagmb-2012-0012}}.
	
	\bibitem[Fern\'andez-Dur\'an and Gregorio-Dom\'inguez,
	2016]{Fernandez-Duran2016}
	Fern\'andez-Dur\'an, J.~J. and Gregorio-Dom\'inguez, M.~M. (2016).
	\newblock {CircNNTSR}: an {R} package for the statistical analysis of circular,
	multivariate circular, and spherical data using nonnegative trigonometric
	sums.
	\newblock {\em Journal of Statistical Software}, 70(6):1--19.
	\newblock \href {http://dx.doi.org/10.18637/jss.v070.i06}
	{\path{doi:10.18637/jss.v070.i06}}.
	
	\bibitem[Ferreira et~al., 2008]{Ferreira2008}
	Ferreira, J. T. A.~S., Ju\'arez, M.~A., and Steel, M. F.~J. (2008).
	\newblock Directional log-spline distributions.
	\newblock {\em Bayesian Analysis}, 3(2):297--316.
	\newblock \href {http://dx.doi.org/10.1214/08-ba311}
	{\path{doi:10.1214/08-ba311}}.
	
	\bibitem[Figueiredo, 2007]{Figueiredo2007}
	Figueiredo, A. (2007).
	\newblock Comparison of tests of uniformity defined on the hypersphere.
	\newblock {\em Statistics \& Probability Letters}, 77(3):329--334.
	\newblock \href {http://dx.doi.org/10.1016/j.spl.2006.07.012}
	{\path{doi:10.1016/j.spl.2006.07.012}}.
	
	\bibitem[Figueiredo, 2009]{Figueiredo2009}
	Figueiredo, A. (2009).
	\newblock Discriminant analysis for the von {M}ises-{F}isher distribution.
	\newblock {\em Communications in Statistics -- Simulation and Computation},
	38(9):1991--2003.
	\newblock \href {http://dx.doi.org/10.1080/03610910903200281}
	{\path{doi:10.1080/03610910903200281}}.
	
	\bibitem[Figueiredo, 2017]{Figueiredo2017}
	Figueiredo, A. (2017).
	\newblock Bootstrap and permutation tests in {ANOVA} for directional data.
	\newblock {\em Computational Statistics}, 32(4):1213--1240.
	\newblock \href {http://dx.doi.org/10.1007/s00180-017-0739-x}
	{\path{doi:10.1007/s00180-017-0739-x}}.
	
	\bibitem[Figueiredo and Gomes, 2003]{Figueiredo2003}
	Figueiredo, A. and Gomes, P. (2003).
	\newblock Power of tests of uniformity defined on the hypersphere.
	\newblock {\em Communications in Statistics -- Simulation and Computation},
	32(1):87--94.
	\newblock \href {http://dx.doi.org/10.1081/sac-120013113}
	{\path{doi:10.1081/sac-120013113}}.
	
	\bibitem[Figueiredo and Gomes, 2005]{Figueiredo2005}
	Figueiredo, A. and Gomes, P. (2005).
	\newblock Discordancy test for the bipolar {W}atson distribution defined on the
	hypersphere.
	\newblock {\em Communications in Statistics -- Simulation and Computation},
	34(1):145--153.
	\newblock \href {http://dx.doi.org/10.1081/sac-200047092}
	{\path{doi:10.1081/sac-200047092}}.
	
	\bibitem[Figueiredo and Gomes, 2006]{Figueiredo2006}
	Figueiredo, A. and Gomes, P. (2006).
	\newblock Discriminant analysis based on the {W}atson distribution defined on
	the hypersphere.
	\newblock {\em Statistics}, 40(5):435--445.
	\newblock \href {http://dx.doi.org/10.1080/02331880600766662}
	{\path{doi:10.1080/02331880600766662}}.
	
	\bibitem[Fisher, 1993]{Fisher1993}
	Fisher, N.~I. (1993).
	\newblock {\em Statistical Analysis of Circular Data}.
	\newblock Cambridge University Press, Cambridge.
	\newblock \href {http://dx.doi.org/10.1017/cbo9780511564345}
	{\path{doi:10.1017/cbo9780511564345}}.
	
	\bibitem[Fisher and Lee, 1992]{Fisher1992}
	Fisher, N.~I. and Lee, A.~J. (1992).
	\newblock Regression models for an angular response.
	\newblock {\em Biometrics}, 48(3):665--677.
	\newblock \href {http://dx.doi.org/10.2307/2532334}
	{\path{doi:10.2307/2532334}}.
	
	\bibitem[Fisher and Lee, 1994]{Fisher1994}
	Fisher, N.~I. and Lee, A.~J. (1994).
	\newblock Time series analysis of circular data.
	\newblock {\em Journal of the Royal Statistical Society, Series B
		(Methodological)}, 56(2):327--339.
	\newblock \href {http://dx.doi.org/10.1111/j.2517-6161.1994.tb01981.x}
	{\path{doi:10.1111/j.2517-6161.1994.tb01981.x}}.
	
	\bibitem[Fisher et~al., 1987]{Fisher1987}
	Fisher, N.~I., Lewis, T., and Embleton, B.~J. (1987).
	\newblock {\em Statistical Analysis of Spherical Data}.
	\newblock Cambridge University Press, Cambridge.
	\newblock \href {http://dx.doi.org/10.1017/cbo9780511623059}
	{\path{doi:10.1017/cbo9780511623059}}.
	
	\bibitem[Fisher and Marron, 2001]{Fisher2001}
	Fisher, N.~I. and Marron, J.~S. (2001).
	\newblock Mode testing via the excess mass estimate.
	\newblock {\em Biometrika}, 88(2):499--517.
	\newblock \href {http://dx.doi.org/10.1093/biomet/88.2.499}
	{\path{doi:10.1093/biomet/88.2.499}}.
	
	\bibitem[Fitak and Johnsen, 2017]{Fitak2017}
	Fitak, R.~R. and Johnsen, S. (2017).
	\newblock Bringing the analysis of animal orientation data full circle:
	model-based approaches with maximum likelihood.
	\newblock {\em Journal of Experimental Biology}, 220(21):3878--3882.
	\newblock \href {http://dx.doi.org/10.1242/jeb.167056}
	{\path{doi:10.1242/jeb.167056}}.
	
	\bibitem[Fletcher et~al., 2004]{Fletcher2004}
	Fletcher, P.~T., Lu, C., Pizer, S.~M., and Joshi, S. (2004).
	\newblock Principal geodesic analysis for the study of nonlinear statistics of
	shape.
	\newblock {\em IEEE Transactions on Medical Imaging}, 23(8):995--1005.
	\newblock \href {http://dx.doi.org/10.1109/tmi.2004.831793}
	{\path{doi:10.1109/tmi.2004.831793}}.
	
	\bibitem[Franke et~al., 2016]{Franke2016}
	Franke, J., Redenbach, C., and Zhang, N. (2016).
	\newblock On a mixture model for directional data on the sphere.
	\newblock {\em Scandinavian Journal of Statistics}, 43(1):139--155.
	\newblock \href {http://dx.doi.org/10.1111/sjos.12169}
	{\path{doi:10.1111/sjos.12169}}.
	
	\bibitem[Fryer et~al., 2020]{Fryer2020}
	Fryer, D., Olenko, A., Li, M., and Wang, Y. (2020).
	\newblock {\em {rcosmo}: Cosmic Microwave Background Data Analysis}.
	\newblock {R} package version 1.1.2.
	\newblock URL: \url{https://CRAN.R-project.org/package=rcosmo}.
	
	\bibitem[Fu et~al., 2008]{Fu2008}
	Fu, Y., Chen, J., and Li, P. (2008).
	\newblock Modified likelihood ratio test for homogeneity in a mixture of von
	{M}ises distributions.
	\newblock {\em Journal of Statistical Planning and Inference}, 138(3):667--681.
	\newblock \href {http://dx.doi.org/10.1016/j.jspi.2007.01.003}
	{\path{doi:10.1016/j.jspi.2007.01.003}}.
	
	\bibitem[Gao and Li, 2010]{Gao2010}
	Gao, F.~Q. and Li, L.~N. (2010).
	\newblock Large deviations and moderate deviations for kernel density
	estimators of directional data.
	\newblock {\em Acta Mathematica Sinica, English Series}, 26(5):937--950.
	\newblock \href {http://dx.doi.org/10.1007/s10114-010-7205-9}
	{\path{doi:10.1007/s10114-010-7205-9}}.
	
	\bibitem[Garc\'ia-Portugu\'es, 2013]{Garcia-Portugues2013a}
	Garc\'ia-Portugu\'es, E. (2013).
	\newblock Exact risk improvement of bandwidth selectors for kernel density
	estimation with directional data.
	\newblock {\em Electronic Journal of Statistics}, 7:1655--1685.
	\newblock \href {http://dx.doi.org/10.1214/13-ejs821}
	{\path{doi:10.1214/13-ejs821}}.
	
	\bibitem[Garc\'ia-Portugu\'es, 2020a]{Garcia-Portugues2020f}
	Garc\'ia-Portugu\'es, E. (2020a).
	\newblock {\em {DirStats}: Nonparametric Methods for Directional Data}.
	\newblock {R} package version 0.1.6.
	\newblock URL: \url{https://CRAN.R-project.org/package=DirStats}.
	
	\bibitem[Garc\'ia-Portugu\'es, 2020b]{Garcia-Portugues2020d}
	Garc\'ia-Portugu\'es, E. (2020b).
	\newblock {\em {sdetorus}: Statistical Tools for Toroidal Diffusions}.
	\newblock {R} package version 0.1.7.
	\newblock URL: \url{https://CRAN.R-project.org/package=sdetorus}.
	
	\bibitem[Garc\'ia-Portugu\'es et~al., 2014]{Garcia-Portugues2014}
	Garc\'ia-Portugu\'es, E., Barros, A. M.~G., Crujeiras, R.~M.,
	Gonz\'alez-Manteiga, W., and Pereira, J. (2014).
	\newblock A test for directional-linear independence, with applications to
	wildfire orientation and size.
	\newblock {\em Stochastic Environmental Research and Risk Assessment},
	28(5):1261--1275.
	\newblock \href {http://dx.doi.org/10.1007/s00477-013-0819-6}
	{\path{doi:10.1007/s00477-013-0819-6}}.
	
	\bibitem[Garc\'ia-Portugu\'es et~al., 2013a]{Garcia-Portugues2013}
	Garc\'ia-Portugu\'es, E., Crujeiras, R.~M., and Gonz\'alez-Manteiga, W.
	(2013a).
	\newblock Exploring wind direction and {SO$_2$} concentration by
	circular-linear density estimation.
	\newblock {\em Stochastic Environmental Research and Risk Assessment},
	27(5):1055--1067.
	\newblock \href {http://dx.doi.org/10.1007/s00477-012-0642-5}
	{\path{doi:10.1007/s00477-012-0642-5}}.
	
	\bibitem[Garc\'ia-Portugu\'es et~al., 2013b]{Garcia-Portugues2013b}
	Garc\'ia-Portugu\'es, E., Crujeiras, R.~M., and Gonz\'alez-Manteiga, W.
	(2013b).
	\newblock Kernel density estimation for directional-linear data.
	\newblock {\em Journal of Multivariate Analysis}, 121:152--175.
	\newblock \href {http://dx.doi.org/10.1016/j.jmva.2013.06.009}
	{\path{doi:10.1016/j.jmva.2013.06.009}}.
	
	\bibitem[Garc\'ia-Portugu\'es et~al., 2015]{Garcia-Portugues2015}
	Garc\'ia-Portugu\'es, E., Crujeiras, R.~M., and Gonz\'alez-Manteiga, W. (2015).
	\newblock Central limit theorems for directional and linear random variables
	with applications.
	\newblock {\em Statistica Sinica}, 25(3):1207--1229.
	\newblock \href {http://dx.doi.org/10.5705/ss.2014.153}
	{\path{doi:10.5705/ss.2014.153}}.
	
	\bibitem[Garc\'ia-Portugu\'es et~al., 2020a]{Garcia-Portugues2020b}
	Garc\'ia-Portugu\'es, E., Navarro-Esteban, P., and Cuesta-Albertos, J.~A.
	(2020a).
	\newblock On a projection-based class of uniformity tests on the hypersphere.
	\newblock {\em arXiv:2008.09897}.
	
	\bibitem[Garc\'ia-Portugu\'es et~al., 2020b]{Garcia-Portugues2020}
	Garc\'ia-Portugu\'es, E., Paindaveine, D., and Verdebout, T. (2020b).
	\newblock On optimal tests for rotational symmetry against new classes of
	hyperspherical distributions.
	\newblock {\em Journal of the American Statistical Association}, to appear.
	\newblock \href {http://dx.doi.org/10.1080/01621459.2019.1665527}
	{\path{doi:10.1080/01621459.2019.1665527}}.
	
	\bibitem[Garc\'ia-Portugu\'es et~al., 2020c]{Garcia-Portugues2020e}
	Garc\'ia-Portugu\'es, E., Paindaveine, D., and Verdebout, T. (2020c).
	\newblock {\em {rotasym}: Tests for Rotational Symmetry on the Hypersphere}.
	\newblock {R} package version 1.0.9.
	\newblock URL: \url{https://CRAN.R-project.org/package=rotasym}.
	
	\bibitem[Garc\'ia-Portugu\'es et~al., 2019]{Garcia-Portugues2019}
	Garc\'ia-Portugu\'es, E., S{\o}rensen, M., Mardia, K.~V., and Hamelryck, T.
	(2019).
	\newblock Langevin diffusions on the torus: estimation and applications.
	\newblock {\em Statistics and Computing}, 29(1):1--22.
	\newblock \href {http://dx.doi.org/10.1007/s11222-017-9790-2}
	{\path{doi:10.1007/s11222-017-9790-2}}.
	
	\bibitem[Garc\'ia-Portugu\'es et~al., 2016]{Garcia-Portugues2016}
	Garc\'ia-Portugu\'es, E., Van~Keilegom, I., Crujeiras, R.~M., and
	Gonz\'alez-Manteiga, W. (2016).
	\newblock Testing parametric models in linear-directional regression.
	\newblock {\em Scandinavian Journal of Statistics}, 43(4):1178--1191.
	\newblock \href {http://dx.doi.org/10.1111/sjos.12236}
	{\path{doi:10.1111/sjos.12236}}.
	
	\bibitem[Garc\'ia-Portugu\'es and Verdebout, 2018]{Garcia-Portugues2020a}
	Garc\'ia-Portugu\'es, E. and Verdebout, T. (2018).
	\newblock A review of uniformity tests on the hypersphere.
	\newblock {\em arXiv:1804.00286}.
	
	\bibitem[Garc\'ia-Portugu\'es and Verdebout, 2020]{Garcia-Portugues2020c}
	Garc\'ia-Portugu\'es, E. and Verdebout, T. (2020).
	\newblock {\em {sphunif}: Uniformity Tests on the Circle, Sphere, and
		Hypersphere}.
	\newblock URL: \url{https://github.com/egarpor/sphunif}.
	
	\bibitem[Gatto, 2000]{Gatto2000}
	Gatto, R. (2000).
	\newblock Multivariate saddlepoint test for the wrapped normal model.
	\newblock {\em Journal of Statistical Computation and Simulation},
	65(1-4):271--285.
	\newblock \href {http://dx.doi.org/10.1080/00949650008812002}
	{\path{doi:10.1080/00949650008812002}}.
	
	\bibitem[Gatto, 2006]{Gatto2006}
	Gatto, R. (2006).
	\newblock A bootstrap test for circular data.
	\newblock {\em Communications in Statistics -- Theory and Methods},
	35(1-3):281--292.
	\newblock \href {http://dx.doi.org/10.1080/03610920500440057}
	{\path{doi:10.1080/03610920500440057}}.
	
	\bibitem[Gatto, 2008]{Gatto2008}
	Gatto, R. (2008).
	\newblock Some computational aspects of the generalized von {M}ises
	distribution.
	\newblock {\em Statistics and Computing}, 18(3):321--331.
	\newblock \href {http://dx.doi.org/10.1007/s11222-008-9060-4}
	{\path{doi:10.1007/s11222-008-9060-4}}.
	
	\bibitem[Gatto, 2009]{Gatto2009}
	Gatto, R. (2009).
	\newblock Information theoretic results for circular distributions.
	\newblock {\em Statistics}, 43(4):409--421.
	\newblock \href {http://dx.doi.org/10.1080/09603100802395947}
	{\path{doi:10.1080/09603100802395947}}.
	
	\bibitem[Gatto, 2017]{Gatto2017}
	Gatto, R. (2017).
	\newblock Multivariate saddlepoint tests on the mean direction of the von
	{M}ises-{F}isher distribution.
	\newblock {\em Metrika}, 80(6-8):733--747.
	\newblock \href {http://dx.doi.org/10.1007/s00184-017-0625-0}
	{\path{doi:10.1007/s00184-017-0625-0}}.
	
	\bibitem[Genest et~al., 2019]{Genest2019}
	Genest, M., Masse, J.-C., and Plante, J.-F. (2019).
	\newblock {\em {depth}: Nonparametric Depth Functions for Multivariate
		Analysis}.
	\newblock {R} package version 2.1-1.1.
	\newblock URL: \url{https://CRAN.R-project.org/package=depth}.
	
	\bibitem[Genton and Hall, 2007]{Genton2007}
	Genton, M.~G. and Hall, P. (2007).
	\newblock Statistical inference for evolving periodic functions.
	\newblock {\em Journal of the Royal Statistical Society, Series B (Statistical
		Methodology)}, 69(4):643--657.
	\newblock \href {http://dx.doi.org/10.1111/j.1467-9868.2007.00604.x}
	{\path{doi:10.1111/j.1467-9868.2007.00604.x}}.
	
	\bibitem[George and Ghosh, 2006]{George2006}
	George, B.~J. and Ghosh, K. (2006).
	\newblock A semiparametric {B}ayesian model for circular-linear regression.
	\newblock {\em Communications in Statistics -- Simulation and Computation},
	35(4):911--923.
	\newblock \href {http://dx.doi.org/10.1080/03610910600880302}
	{\path{doi:10.1080/03610910600880302}}.
	
	\bibitem[Ghazanfarihesari and Sarmad, 2016]{Ghazanfarihesari2016}
	Ghazanfarihesari, A. and Sarmad, M. (2016).
	\newblock {\em {CircOutlier}: Detection of Outliers in Circular-Circular
		Regression}.
	\newblock {R} package version 3.2.3.
	\newblock URL: \url{https://CRAN.R-project.org/package=CircOutlier}.
	
	\bibitem[Ghosh et~al., 1999]{Ghosh1999}
	Ghosh, K., Jammalamadaka, S.~R., and Vasudaven, M. (1999).
	\newblock Change-point problems for the von {M}ises distribution.
	\newblock {\em Journal of Applied Statistics}, 26(4):423--434.
	\newblock \href {http://dx.doi.org/10.1080/02664769922313}
	{\path{doi:10.1080/02664769922313}}.
	
	\bibitem[Ghosh et~al., 2019]{Ghosh2019}
	Ghosh, M., Zhong, X., SenGupta, A., and Zhang, R. (2019).
	\newblock Non-subjective priors for wrapped {C}auchy distributions.
	\newblock {\em Statistics \& Probability Letters}, 153:90--97.
	\newblock \href {http://dx.doi.org/10.1016/j.spl.2019.05.016}
	{\path{doi:10.1016/j.spl.2019.05.016}}.
	
	\bibitem[Gill and Hangartner, 2010]{Gill2010}
	Gill, J. and Hangartner, D. (2010).
	\newblock Circular data in political science and how to handle it.
	\newblock {\em Political Analysis}, 18(3):316--336.
	\newblock \href {http://dx.doi.org/10.1093/pan/mpq009}
	{\path{doi:10.1093/pan/mpq009}}.
	
	\bibitem[Gin\'e, 1975]{Gine1975}
	Gin\'e, E. (1975).
	\newblock Invariant tests for uniformity on compact {R}iemannian manifolds
	based on {S}obolev norms.
	\newblock {\em The Annals of Statistics}, 3(6):1243--1266.
	\newblock \href {http://dx.doi.org/10.1214/aos/1176343283}
	{\path{doi:10.1214/aos/1176343283}}.
	
	\bibitem[Giummol\`e et~al., 2019]{Giummole2019}
	Giummol\`e, F., Mameli, V., Ruli, E., and Ventura, L. (2019).
	\newblock Objective {B}ayesian inference with proper scoring rules.
	\newblock {\em TEST}, 28(3):728--755.
	\newblock \href {http://dx.doi.org/10.1007/s11749-018-0597-z}
	{\path{doi:10.1007/s11749-018-0597-z}}.
	
	\bibitem[Gneiting, 2013]{Gneiting2013}
	Gneiting, T. (2013).
	\newblock Strictly and non-strictly positive definite functions on spheres.
	\newblock {\em Bernoulli}, 19(4):1327--1349.
	\newblock \href {http://dx.doi.org/10.3150/12-bejsp06}
	{\path{doi:10.3150/12-bejsp06}}.
	
	\bibitem[Godtliebsen et~al., 2002]{Godtliebsen2002}
	Godtliebsen, F., Marron, J.~S., and Chaudhuri, P. (2002).
	\newblock Significance in scale space for bivariate density estimation.
	\newblock {\em Journal of Computational and Graphical Statistics}, 11(1):1--21.
	\newblock \href {http://dx.doi.org/10.1198/106186002317375596}
	{\path{doi:10.1198/106186002317375596}}.
	
	\bibitem[Golden et~al., 2017]{Golden2017}
	Golden, M., Garc\'ia-Portugu\'es, E., S{\o}rensen, M., Mardia, K.~V.,
	Hamelryck, T., and Hein, J. (2017).
	\newblock A generative angular model of protein structure evolution.
	\newblock {\em Molecular Biology and Evolution}, 34(8):2085--2100.
	\newblock \href {http://dx.doi.org/10.1093/molbev/msx137}
	{\path{doi:10.1093/molbev/msx137}}.
	
	\bibitem[Gopal and Yang, 2014]{Gopal2014}
	Gopal, S. and Yang, Y. (2014).
	\newblock Von {M}ises-{F}isher clustering models.
	\newblock In Xing, E.~P. and Jebara, T. (Eds.), {\em Proceedings of the 31st
		International Conference Machine Learning}, volume~32 of {\em Proceedings of
		Machine Learning Research}, pp. 154--162, Bejing. PMLR.
	
	\bibitem[Graul and Poppinga, 2018]{Graul2018}
	Graul, C. and Poppinga, C. (2018).
	\newblock {\em {bReeze}: Functions for Wind Resource Assessment}.
	\newblock {R} package version 0.4-3.
	\newblock URL: \url{https://CRAN.R-project.org/package=bReeze}.
	
	\bibitem[Gu et~al., 2004]{Gu2004}
	Gu, X., Wang, Y., Chan, T.~F., Thompson, P.~M., and Yau, S.-T. (2004).
	\newblock Genus zero surface conformal mapping and its application to brain
	surface mapping.
	\newblock {\em IEEE Transactions on Medical Imaging}, 23(8):949--958.
	
	\bibitem[Guella et~al., 2018]{Guella2018}
	Guella, J.~C., Menegatto, V.~A., and Porcu, E. (2018).
	\newblock Strictly positive definite multivariate covariance functions on
	spheres.
	\newblock {\em Journal of Multivariate Analysis}, 166:150--159.
	\newblock \href {http://dx.doi.org/10.1016/j.jmva.2018.03.001}
	{\path{doi:10.1016/j.jmva.2018.03.001}}.
	
	\bibitem[Hall et~al., 2000]{Hall2000}
	Hall, P., Reimann, J., and Rice, J. (2000).
	\newblock Nonparametric estimation of a periodic function.
	\newblock {\em Biometrika}, 87(3):545--557.
	\newblock \href {http://dx.doi.org/10.1093/biomet/87.3.545}
	{\path{doi:10.1093/biomet/87.3.545}}.
	
	\bibitem[Hall et~al., 1987]{Hall1987}
	Hall, P., Watson, G.~S., and Cabrera, J. (1987).
	\newblock Kernel density estimation with spherical data.
	\newblock {\em Biometrika}, 74(4):751--762.
	\newblock \href {http://dx.doi.org/10.1093/biomet/74.4.751}
	{\path{doi:10.1093/biomet/74.4.751}}.
	
	\bibitem[Hall and Yin, 2003]{Hall2003}
	Hall, P. and Yin, J. (2003).
	\newblock Nonparametric methods for deconvolving multiperiodic functions.
	\newblock {\em Journal of the Royal Statistical Society, Series B (Statistical
		Methodology)}, 65(4):869--886.
	\newblock \href {http://dx.doi.org/10.1046/j.1369-7412.2003.00420.x}
	{\path{doi:10.1046/j.1369-7412.2003.00420.x}}.
	
	\bibitem[Hamsici and Martinez, 2007]{Hamsici2007}
	Hamsici, O.~C. and Martinez, A.~M. (2007).
	\newblock Spherical-homoscedastic distributions: the equivalency of spherical
	and normal distributions in classification.
	\newblock {\em Journal of Machine Learning Research}, 8(Jul):1583--1623.
	
	\bibitem[Hara et~al., 2008]{Hara2008}
	Hara, K., Nishino, K., and Ikeuchi, K. (2008).
	\newblock Mixture of spherical distributions for single-view relighting.
	\newblock {\em IEEE Transactions on Pattern Analysis and Machine Intelligence},
	30(1):25--35.
	\newblock \href {http://dx.doi.org/10.1109/tpami.2007.1164}
	{\path{doi:10.1109/tpami.2007.1164}}.
	
	\bibitem[Hartman and Watson, 1974]{Hartman1974}
	Hartman, P. and Watson, G.~S. (1974).
	\newblock ``normal'' distribution functions on spheres and the modified
	{B}essel functions.
	\newblock {\em The Annals of Probability}, 2(4):593--607.
	\newblock \href {http://dx.doi.org/10.1214/aop/1176996606}
	{\path{doi:10.1214/aop/1176996606}}.
	
	\bibitem[Hasnat et~al., 2014]{Hasnat2014}
	Hasnat, M.~A., Alata, O., and Tr\'emeau, A. (2014).
	\newblock Unsupervised clustering of depth images using {W}atson mixture model.
	\newblock In {\em 2014 22nd International Conference on Pattern Recognition},
	pp. 214--219, New York. IEEE.
	\newblock \href {http://dx.doi.org/10.1109/icpr.2014.46}
	{\path{doi:10.1109/icpr.2014.46}}.
	
	\bibitem[Hassanzadeh and Kalaylioglu, 2018]{Hassanzadeh2018}
	Hassanzadeh, F. and Kalaylioglu, Z. (2018).
	\newblock A new multimodal and asymmetric bivariate circular distribution.
	\newblock {\em Environmental and Ecological Statistics}, 25(3):363--385.
	\newblock \href {http://dx.doi.org/10.1007/s10651-018-0409-3}
	{\path{doi:10.1007/s10651-018-0409-3}}.
	
	\bibitem[Hawkins and Lombard, 2015]{Hawkins2015}
	Hawkins, D.~M. and Lombard, F. (2015).
	\newblock Segmentation of circular data.
	\newblock {\em Journal of Applied Statistics}, 42(1):88--97.
	\newblock \href {http://dx.doi.org/10.1080/02664763.2014.934665}
	{\path{doi:10.1080/02664763.2014.934665}}.
	
	\bibitem[Hawkins and Lombard, 2017]{Hawkins2017}
	Hawkins, D.~M. and Lombard, F. (2017).
	\newblock Cusum control for data following the von {M}ises distribution.
	\newblock {\em Journal of Applied Statistics}, 44(8):1319--1332.
	
	\bibitem[Healy et~al., 1998]{Healy1998}
	Healy, D. M.~J., Hendriks, H., and Kim, P.~T. (1998).
	\newblock Spherical deconvolution.
	\newblock {\em Journal of Multivariate Analysis}, 67(1):1--22.
	\newblock \href {http://dx.doi.org/10.1006/jmva.1998.1757}
	{\path{doi:10.1006/jmva.1998.1757}}.
	
	\bibitem[Henry and Rodriguez, 2009]{Henry2009}
	Henry, G. and Rodriguez, D. (2009).
	\newblock Kernel density estimation on {R}iemannian manifolds: asymptotic
	results.
	\newblock {\em Journal of Mathematical Imaging and Vision}, 34(3):235--239.
	\newblock \href {http://dx.doi.org/10.1007/s10851-009-0145-2}
	{\path{doi:10.1007/s10851-009-0145-2}}.
	
	\bibitem[Hernandez-Stumpfhauser et~al., 2016]{Hernandez-Stumpfhauser2016}
	Hernandez-Stumpfhauser, D., Breidt, F.~J., and Opsomer, J.~D. (2016).
	\newblock Hierarchical {B}ayesian small area estimation for circular data.
	\newblock {\em The Canadian Journal of Statistics}, 44(4):416--430.
	\newblock \href {http://dx.doi.org/10.1002/cjs.11303}
	{\path{doi:10.1002/cjs.11303}}.
	
	\bibitem[Hernandez-Stumpfhauser et~al., 2017]{Hernandez-Stumpfhauser2017}
	Hernandez-Stumpfhauser, D., Breidt, F.~J., and van~der Woerd, M.~J. (2017).
	\newblock The general projected normal distribution of arbitrary dimension:
	modeling and {B}ayesian inference.
	\newblock {\em Bayesian Analysis}, 12(1):113--133.
	\newblock \href {http://dx.doi.org/10.1214/15-ba989}
	{\path{doi:10.1214/15-ba989}}.
	
	\bibitem[Hill and H\"ader, 1997]{Hill1997}
	Hill, N.~A. and H\"ader, D.-P. (1997).
	\newblock A biased random walk model for the trajectories of swimming
	micro-organisms.
	\newblock {\em Journal of Theoretical Biology}, 186(4):503--526.
	\newblock \href {http://dx.doi.org/10.1006/jtbi.1997.0421}
	{\path{doi:10.1006/jtbi.1997.0421}}.
	
	\bibitem[Hinkle et~al., 2014]{Hinkle2014}
	Hinkle, J., Fletcher, P.~T., and Joshi, S. (2014).
	\newblock Intrinsic polynomials for regression on {R}iemannian manifolds.
	\newblock {\em Journal of Mathematical Imaging and Vision}, 50(1):32--52.
	\newblock \href {http://dx.doi.org/10.1007/s10851-013-0489-5}
	{\path{doi:10.1007/s10851-013-0489-5}}.
	
	\bibitem[Hokimoto and Shimizu, 2008]{Hokimoto2008}
	Hokimoto, T. and Shimizu, K. (2008).
	\newblock An angular-linear time series model for waveheight prediction.
	\newblock {\em Annals of the Institute of Statistical Mathematics},
	60(4):781--800.
	\newblock \href {http://dx.doi.org/10.1007/s10463-008-0207-z}
	{\path{doi:10.1007/s10463-008-0207-z}}.
	
	\bibitem[Hokimoto and Shimizu, 2014]{Hokimoto2014}
	Hokimoto, T. and Shimizu, K. (2014).
	\newblock A non-homogeneous hidden {M}arkov model for predicting the
	distribution of sea surface elevation.
	\newblock {\em Journal of Applied Statistics}, 41(2):294--319.
	
	\bibitem[Holmquist and Gustafsson, 2017]{Holmquist2017}
	Holmquist, B. and Gustafsson, P. (2017).
	\newblock A two-level directional model for dependence in circular data.
	\newblock {\em The Canadian Journal of Statistics}, 45(4):461--478.
	\newblock \href {http://dx.doi.org/10.1002/cjs.11345}
	{\path{doi:10.1002/cjs.11345}}.
	
	\bibitem[Holzmann et~al., 2004]{Holzmann2004}
	Holzmann, H., Munk, A., and Stratmann, B. (2004).
	\newblock Identifiability of finite mixtures - with applications to circular
	distributions.
	\newblock {\em Sankhy\=a}, 66(3):440--449.
	
	\bibitem[Holzmann et~al., 2006]{Holzmann2006}
	Holzmann, H., Munk, A., Suster, M., and Zucchini, W. (2006).
	\newblock Hidden {M}arkov models for circular and linear-circular time series.
	\newblock {\em Environmental and Ecological Statistics}, 13(3):325--347.
	\newblock \href {http://dx.doi.org/10.1007/s10651-006-0015-7}
	{\path{doi:10.1007/s10651-006-0015-7}}.
	
	\bibitem[Hornik et~al., 2012]{Hornik2012}
	Hornik, K., Feinerer, I., Kober, M., and Buchta, C. (2012).
	\newblock Spherical {$k$}-means clustering.
	\newblock {\em Journal of Statistical Software}, 50(10):1--22.
	\newblock \href {http://dx.doi.org/10.18637/jss.v050.i10}
	{\path{doi:10.18637/jss.v050.i10}}.
	
	\bibitem[Hornik and Gr\"un, 2013]{Hornik2013}
	Hornik, K. and Gr\"un, B. (2013).
	\newblock On conjugate families and {J}effreys priors for von {M}ises-{F}isher
	distributions.
	\newblock {\em Journal of Statistical Planning and Inference}, 143(5):992--999.
	\newblock \href {http://dx.doi.org/10.1016/j.jspi.2012.11.003}
	{\path{doi:10.1016/j.jspi.2012.11.003}}.
	
	\bibitem[Hornik and Gr\"un, 2014]{Hornik2014}
	Hornik, K. and Gr\"un, B. (2014).
	\newblock {movMF}: an {R} package for fitting mixtures of von {M}ises-{F}isher
	distributions.
	\newblock {\em Journal of Statistical Software}, 58(10):1--31.
	\newblock \href {http://dx.doi.org/10.18637/jss.v058.i10}
	{\path{doi:10.18637/jss.v058.i10}}.
	
	\bibitem[Horwood and Poore, 2014]{Horwood2014}
	Horwood, J.~T. and Poore, A.~B. (2014).
	\newblock Gauss von {M}ises distribution for improved uncertainty realism in
	space situational awareness.
	\newblock {\em SIAM/ASA Journal on Uncertainty Quantification}, 2(1):276--304.
	\newblock \href {http://dx.doi.org/10.1137/130917296}
	{\path{doi:10.1137/130917296}}.
	
	\bibitem[Hotz, 2013]{Hotz2013}
	Hotz, T. (2013).
	\newblock Extrinsic vs intrinsic means on the circle.
	\newblock In Nielsen, F. and Barbaresco, F. (Eds.), {\em Geometric Science of
		Information}, volume 8085 of {\em Lecture Notes in Computer Science}, pp.
	433--440, Berlin. Springer.
	\newblock \href {http://dx.doi.org/10.1007/978-3-642-40020-9_7}
	{\path{doi:10.1007/978-3-642-40020-9_7}}.
	
	\bibitem[Hotz and Huckemann, 2015]{Hotz2015}
	Hotz, T. and Huckemann, S. (2015).
	\newblock Intrinsic means on the circle: uniqueness, locus and asymptotics.
	\newblock {\em Annals of the Institute of Statistical Mathematics},
	67(1):177--193.
	
	\bibitem[Huckemann et~al., 2010]{Huckemann2010a}
	Huckemann, S., Hotz, T., and Munk, A. (2010).
	\newblock Intrinsic shape analysis: geodesic {PCA} for {R}iemannian manifolds
	modulo isometric {L}ie group actions.
	\newblock {\em Statistica Sinica}, 20(1):1--58.
	
	\bibitem[Huckemann et~al., 2016]{Huckemann2016}
	Huckemann, S., Kim, K.-R., Munk, A., Rehfeldt, F., Sommerfeld, M., Weickert,
	J., and Wollnik, C. (2016).
	\newblock The circular {S}i{Z}er, inferred persistence of shape parameters and
	application to early stem cell differentiation.
	\newblock {\em Bernoulli}, 22(4):2113--2142.
	\newblock \href {http://dx.doi.org/10.3150/15-BEJ722}
	{\path{doi:10.3150/15-BEJ722}}.
	
	\bibitem[Huckemann and Ziezold, 2006]{Huckemann2006}
	Huckemann, S. and Ziezold, H. (2006).
	\newblock Principal component analysis for {R}iemannian manifolds, with an
	application to triangular shape spaces.
	\newblock {\em Advances in Applied Probability}, 38(2):299--319.
	\newblock \href {http://dx.doi.org/10.1239/aap/1151337073}
	{\path{doi:10.1239/aap/1151337073}}.
	
	\bibitem[Huckemann and Eltzner, 2018]{Huckemann2018}
	Huckemann, S.~F. and Eltzner, B. (2018).
	\newblock Backward nested descriptors asymptotics with inference on stem cell
	differentiation.
	\newblock {\em The Annals of Statistics}, 46(5):1994--2019.
	\newblock \href {http://dx.doi.org/10.1214/17-AOS1609}
	{\path{doi:10.1214/17-AOS1609}}.
	
	\bibitem[Hughes, 2007]{Hughes2007}
	Hughes, G. (2007).
	\newblock {\em Multivariate and Time Series Models for Circular Data with
		Applications to Protein Conformational Angles}.
	\newblock PhD thesis, University of Leeds.
	
	\bibitem[Humphreys and Ruxton, 2017]{Humphreys2017}
	Humphreys, R.~K. and Ruxton, G.~D. (2017).
	\newblock Consequences of grouped data for testing for departure from circular
	uniformity.
	\newblock {\em Behavioral Ecology and Sociobiology}, 71(11):167.
	\newblock \href {http://dx.doi.org/10.1007/s00265-017-2393-2}
	{\path{doi:10.1007/s00265-017-2393-2}}.
	
	\bibitem[Hundrieser et~al., 2020]{Hundrieser2020}
	Hundrieser, S., Eltzner, B., and Huckemann, S.~F. (2020).
	\newblock Finite sample smeariness of {F}r\'echet means and application to
	climate.
	\newblock {\em arXiv:2005.02321}.
	
	\bibitem[Hung et~al., 2015]{Hung2015}
	Hung, W.-L., Chang-Chien, S.-J., and Yang, M.-S. (2015).
	\newblock An intuitive clustering algorithm for spherical data with application
	to extrasolar planets.
	\newblock {\em Journal of Applied Statistics}, 42(10):2220--2232.
	\newblock \href {http://dx.doi.org/10.1080/02664763.2015.1023271}
	{\path{doi:10.1080/02664763.2015.1023271}}.
	
	\bibitem[Hyv\"arinen, 2005]{Hyvarinen2005}
	Hyv\"arinen, A. (2005).
	\newblock Estimation of non-normalized statistical models by score matching.
	\newblock {\em Journal of Machine Learning Research}, 6(Apr):695--709.
	
	\bibitem[Imoto et~al., 2019]{Imoto2019}
	Imoto, T., Shimizu, K., and Abe, T. (2019).
	\newblock A cylindrical distribution with heavy-tailed linear part.
	\newblock {\em Japanese Journal of Statistics and Data Science}, 2(1):129--154.
	\newblock \href {http://dx.doi.org/10.1007/s42081-019-00031-5}
	{\path{doi:10.1007/s42081-019-00031-5}}.
	
	\bibitem[Irwin et~al., 2002]{Irwin2002}
	Irwin, M.~E., Cressie, N., and Johannesson, G. (2002).
	\newblock Spatial-temporal nonlinear filtering based on hierarchical
	statistical models.
	\newblock {\em TEST}, 11(2):249--302.
	\newblock \href {http://dx.doi.org/10.1007/BF02595708}
	{\path{doi:10.1007/BF02595708}}.
	
	\bibitem[Jacimovic and Crnki\'c, 2017]{Jacimovic2017}
	Jacimovic, V. and Crnki\'c, A. (2017).
	\newblock Collective motions of globally coupled oscillators and some
	probability distributions on circle.
	\newblock {\em Physics Letters A}, 381(24):1989--1994.
	\newblock \href {http://dx.doi.org/10.1016/j.physleta.2017.04.024}
	{\path{doi:10.1016/j.physleta.2017.04.024}}.
	
	\bibitem[Jammalamadaka and Kozubowski, 2004]{Jammalamadaka2004a}
	Jammalamadaka, S.~R. and Kozubowski, T.~J. (2004).
	\newblock New families of wrapped distributions for modeling skew circular
	data.
	\newblock {\em Communications in Statistics -- Theory and Methods},
	33(9):2059--2074.
	\newblock \href {http://dx.doi.org/10.1081/sta-200026570}
	{\path{doi:10.1081/sta-200026570}}.
	
	\bibitem[Jammalamadaka et~al., 2020]{Jammalamadaka2020}
	Jammalamadaka, S.~R., Meintanis, S., and Verdebout, T. (2020).
	\newblock On new {S}obolev tests of uniformity on the circle with extension to
	the sphere.
	\newblock {\em Bernoulli}, 26(3):2226--2252.
	\newblock \href {http://dx.doi.org/10.3150/19-BEJ1191}
	{\path{doi:10.3150/19-BEJ1191}}.
	
	\bibitem[Jammalamadaka and SenGupta, 2001]{Jammalamadaka2001}
	Jammalamadaka, S.~R. and SenGupta, A. (2001).
	\newblock {\em Topics in Circular Statistics}, volume~5 of {\em Series on
		Multivariate Analysis}.
	\newblock World Scientific, Singapore.
	\newblock \href {http://dx.doi.org/10.1142/4031} {\path{doi:10.1142/4031}}.
	
	\bibitem[Jammalamadaka and Terdik, 2019]{Jammalamadaka2019a}
	Jammalamadaka, S.~R. and Terdik, G.~H. (2019).
	\newblock Harmonic analysis and distribution-free inference for spherical
	distributions.
	\newblock {\em Journal of Multivariate Analysis}, 171:436--451.
	\newblock \href {http://dx.doi.org/10.1016/j.jmva.2019.01.012}
	{\path{doi:10.1016/j.jmva.2019.01.012}}.
	
	\bibitem[Jensen et~al., 2019]{Jensen2019}
	Jensen, M.~H., Mallasto, A., and Sommer, S. (2019).
	\newblock Simulation of conditioned diffusions on the flat torus.
	\newblock In Nielsen, F. and Barbaresco, F. (Eds.), {\em Geometric Science of
		Information}, volume 11712 of {\em Lecture Notes in Computer Science}, pp.
	685--694, Cham. Springer.
	\newblock \href {http://dx.doi.org/10.1007/978-3-030-26980-7_71}
	{\path{doi:10.1007/978-3-030-26980-7_71}}.
	
	\bibitem[Jeong et~al., 2017]{Jeong2017}
	Jeong, J., Jun, M., and Genton, M.~G. (2017).
	\newblock Spherical process models for global spatial statistics.
	\newblock {\em Statistical Science}, 32(4):501--513.
	\newblock \href {http://dx.doi.org/10.1214/17-STS620}
	{\path{doi:10.1214/17-STS620}}.
	
	\bibitem[Johannes and Schwarz, 2013]{Johannes2013}
	Johannes, J. and Schwarz, M. (2013).
	\newblock Adaptive circular deconvolution by model selection under unknown
	error distribution.
	\newblock {\em Bernoulli}, 19(5A):1576--1611.
	\newblock \href {http://dx.doi.org/10.3150/12-BEJ422}
	{\path{doi:10.3150/12-BEJ422}}.
	
	\bibitem[Johnson and Wehrly, 1977]{Johnson1977}
	Johnson, R.~A. and Wehrly, T.~E. (1977).
	\newblock Measures and models for angular correlation and angular-linear
	correlation.
	\newblock {\em Journal of the Royal Statistical Society, Series B
		(Methodological)}, 39(2):222--229.
	\newblock \href {http://dx.doi.org/10.1111/j.2517-6161.1977.tb01619.x}
	{\path{doi:10.1111/j.2517-6161.1977.tb01619.x}}.
	
	\bibitem[Johnson and Wehrly, 1978]{Johnson1978}
	Johnson, R.~A. and Wehrly, T.~E. (1978).
	\newblock Some angular-linear distributions and related regression models.
	\newblock {\em Journal of the American Statistical Association},
	73(363):602--606.
	\newblock \href {http://dx.doi.org/10.1080/01621459.1978.10480062}
	{\path{doi:10.1080/01621459.1978.10480062}}.
	
	\bibitem[Jona-Lasinio et~al., 2012]{Jona-Lasinio2012}
	Jona-Lasinio, G., Gelfand, A., and Jona-Lasinio, M. (2012).
	\newblock Spatial analysis of wave direction data using wrapped {G}aussian
	processes.
	\newblock {\em The Annals of Applied Statistics}, 6(4):1478--1498.
	\newblock \href {http://dx.doi.org/10.1214/12-aoas576}
	{\path{doi:10.1214/12-aoas576}}.
	
	\bibitem[Jona-Lasinio et~al., 2018]{Jona-Lasinio2018}
	Jona-Lasinio, G., Gelfand, A.~E., and Mastrantonio, G. (2018).
	\newblock Spatial and spatio-temporal circular processes with application to
	wave directions.
	\newblock In Ley, C. and Verdebout, T. (Eds.), {\em Applied Directional
		Statistics}, Chapman \& Hall/CRC Interdisciplinary Statistics Series, pp.
	129--162. CRC Press, Boca Raton.
	
	\bibitem[Jona~Lasinio et~al., 2020]{Jona-Lasinio2020}
	Jona~Lasinio, G., Santoro, M., and Mastrantonio, G. (2020).
	\newblock Circ{S}pace{T}ime: an {R} package for spatial and spatio-temporal
	modelling of circular data.
	\newblock {\em Journal of Statistical Computation and Simulation},
	90(7):1315--1345.
	\newblock \href {http://dx.doi.org/10.1080/00949655.2020.1725008}
	{\path{doi:10.1080/00949655.2020.1725008}}.
	
	\bibitem[Jones and Pewsey, 2005]{Jones2005}
	Jones, M.~C. and Pewsey, A. (2005).
	\newblock A family of symmetric distributions on the circle.
	\newblock {\em Journal of the American Statistical Association},
	100(472):1422--1428.
	\newblock \href {http://dx.doi.org/10.1198/016214505000000286}
	{\path{doi:10.1198/016214505000000286}}.
	
	\bibitem[Jones and Pewsey, 2012]{Jones2012}
	Jones, M.~C. and Pewsey, A. (2012).
	\newblock Inverse {B}atschelet distributions for circular data.
	\newblock {\em Biometrics}, 68(1):183--193.
	\newblock \href {http://dx.doi.org/10.1111/j.1541-0420.2011.01651.x}
	{\path{doi:10.1111/j.1541-0420.2011.01651.x}}.
	
	\bibitem[Jones et~al., 2015]{Jones2015}
	Jones, M.~C., Pewsey, A., and Kato, S. (2015).
	\newblock On a class of circulas: copulas for circular distributions.
	\newblock {\em Annals of the Institute of Statistical Mathematics},
	67(5):843--862.
	
	\bibitem[Jung et~al., 2012]{Jung2012}
	Jung, S., Dryden, I.~L., and Marron, J.~S. (2012).
	\newblock Analysis of principal nested spheres.
	\newblock {\em Biometrika}, 99(3):551--568.
	\newblock \href {http://dx.doi.org/10.1093/biomet/ass022}
	{\path{doi:10.1093/biomet/ass022}}.
	
	\bibitem[Jung et~al., 2011]{Jung2011}
	Jung, S., Foskey, M., and Marron, J.~S. (2011).
	\newblock Principal arc analysis on direct product manifolds.
	\newblock {\em The Annals of Applied Statistics}, 5(1):578--603.
	\newblock \href {http://dx.doi.org/10.1214/10-aoas370}
	{\path{doi:10.1214/10-aoas370}}.
	
	\bibitem[Jupp, 2001]{Jupp2001}
	Jupp, P.~E. (2001).
	\newblock Modifications of the {R}ayleigh and {B}ingham tests for uniformity of
	directions.
	\newblock {\em Journal of Multivariate Analysis}, 77(1):1--20.
	\newblock \href {http://dx.doi.org/10.1006/jmva.2000.1922}
	{\path{doi:10.1006/jmva.2000.1922}}.
	
	\bibitem[Jupp, 2005]{Jupp2005}
	Jupp, P.~E. (2005).
	\newblock Sobolev tests of goodness of fit of distributions on compact
	{R}iemannian manifolds.
	\newblock {\em The Annals of Statistics}, 33(6):2957--2966.
	\newblock \href {http://dx.doi.org/10.1214/009053605000000697}
	{\path{doi:10.1214/009053605000000697}}.
	
	\bibitem[Jupp, 2008]{Jupp2008}
	Jupp, P.~E. (2008).
	\newblock Data-driven {S}obolev tests of uniformity on compact {R}iemannian
	manifolds.
	\newblock {\em The Annals of Statistics}, 36(3):1246--1260.
	\newblock \href {http://dx.doi.org/10.1214/009053607000000541}
	{\path{doi:10.1214/009053607000000541}}.
	
	\bibitem[Jupp, 2009]{Jupp2009}
	Jupp, P.~E. (2009).
	\newblock Data-driven tests of uniformity on product manifolds.
	\newblock {\em Journal of Statistical Planning and Inference},
	139(11):3820--3829.
	\newblock \href {http://dx.doi.org/10.1016/j.jspi.2009.05.019}
	{\path{doi:10.1016/j.jspi.2009.05.019}}.
	
	\bibitem[Jupp, 2015]{Jupp2015}
	Jupp, P.~E. (2015).
	\newblock Copulae on products of compact {R}iemannian manifolds.
	\newblock {\em Journal of Multivariate Analysis}, 140:92--98.
	\newblock \href {http://dx.doi.org/10.1016/j.jmva.2015.04.008}
	{\path{doi:10.1016/j.jmva.2015.04.008}}.
	
	\bibitem[Jupp and Kume, 2020]{Jupp2020}
	Jupp, P.~E. and Kume, A. (2020).
	\newblock Measures of goodness of fit obtained by almost-canonical
	transformations on {R}iemannian manifolds.
	\newblock {\em Journal of Multivariate Analysis}, 176:104579.
	\newblock \href {http://dx.doi.org/10.1016/j.jmva.2019.104579}
	{\path{doi:10.1016/j.jmva.2019.104579}}.
	
	\bibitem[Jupp and Mardia, 1989]{Jupp1989}
	Jupp, P.~E. and Mardia, K.~V. (1989).
	\newblock A unified view of the theory of directional statistics.
	\newblock {\em International Statistical Review}, 57(3):261--294.
	
	\bibitem[Jupp et~al., 2016]{Jupp2016}
	Jupp, P.~E., Regoli, G., and Azzalini, A. (2016).
	\newblock A general setting for symmetric distributions and their relationship
	to general distributions.
	\newblock {\em Journal of Multivariate Analysis}, 148:107--119.
	\newblock \href {http://dx.doi.org/10.1016/j.jmva.2016.02.011}
	{\path{doi:10.1016/j.jmva.2016.02.011}}.
	
	\bibitem[Kasarapu and Allison, 2015]{Kasarapu2015}
	Kasarapu, P. and Allison, L. (2015).
	\newblock Minimum message length estimation of mixtures of multivariate
	{G}aussian and von {M}ises-{F}isher distributions.
	\newblock {\em Machine Learning}, 100(2-3):333--378.
	\newblock \href {http://dx.doi.org/10.1007/s10994-015-5493-0}
	{\path{doi:10.1007/s10994-015-5493-0}}.
	
	\bibitem[Kato, 2009]{Kato2009}
	Kato, S. (2009).
	\newblock A distribution for a pair of unit vectors generated by {B}rownian
	motion.
	\newblock {\em Bernoulli}, 15(3):898--921.
	\newblock \href {http://dx.doi.org/10.3150/08-bej178}
	{\path{doi:10.3150/08-bej178}}.
	
	\bibitem[Kato, 2010]{Kato2010}
	Kato, S. (2010).
	\newblock A {M}arkov process for circular data.
	\newblock {\em Journal of the Royal Statistical Society, Series B (Statistical
		Methodology)}, 72(5):655--672.
	\newblock \href {http://dx.doi.org/10.1111/j.1467-9868.2010.00748.x}
	{\path{doi:10.1111/j.1467-9868.2010.00748.x}}.
	
	\bibitem[Kato and Eguchi, 2016]{Kato2016}
	Kato, S. and Eguchi, S. (2016).
	\newblock Robust estimation of location and concentration parameters for the
	von {M}ises-{F}isher distribution.
	\newblock {\em Statistical Papers}, 57(1):205--234.
	
	\bibitem[Kato and Jones, 2010]{Kato2010a}
	Kato, S. and Jones, M.~C. (2010).
	\newblock A family of distributions on the circle with links to, and
	applications arising from, {M}\"obius transformation.
	\newblock {\em Journal of the American Statistical Association},
	105(489):249--262.
	\newblock \href {http://dx.doi.org/10.1198/jasa.2009.tm08313}
	{\path{doi:10.1198/jasa.2009.tm08313}}.
	
	\bibitem[Kato and Jones, 2013]{Kato2013}
	Kato, S. and Jones, M.~C. (2013).
	\newblock An extended family of circular distributions related to wrapped
	{C}auchy distributions via {B}rownian motion.
	\newblock {\em Bernoulli}, 19(1):154--171.
	\newblock \href {http://dx.doi.org/10.3150/11-bej397}
	{\path{doi:10.3150/11-bej397}}.
	
	\bibitem[Kato and Jones, 2015]{Kato2015}
	Kato, S. and Jones, M.~C. (2015).
	\newblock A tractable and interpretable four-parameter family of unimodal
	distributions on the circle.
	\newblock {\em Biometrika}, 102(1):181--190.
	\newblock \href {http://dx.doi.org/10.1093/biomet/asu059}
	{\path{doi:10.1093/biomet/asu059}}.
	
	\bibitem[Kato and McCullagh, 2020]{Kato2020}
	Kato, S. and McCullagh, P. (2020).
	\newblock Some properties of a {C}auchy family on the sphere derived from the
	{M}\"obius transformations.
	\newblock {\em Bernoulli}, 266(4):3224--3248.
	\newblock \href {http://dx.doi.org/10.3150/20-BEJ1222}
	{\path{doi:10.3150/20-BEJ1222}}.
	
	\bibitem[Kato and Pewsey, 2015]{Kato2015a}
	Kato, S. and Pewsey, A. (2015).
	\newblock A {M}\"obius transformation-induced distribution on the torus.
	\newblock {\em Biometrika}, 102(2):359--370.
	\newblock \href {http://dx.doi.org/10.1093/biomet/asv003}
	{\path{doi:10.1093/biomet/asv003}}.
	
	\bibitem[Kato et~al., 2018]{Kato2018}
	Kato, S., Pewsey, A., and Jones, M.~C. (2018).
	\newblock Circulas from {F}ourier series.
	\newblock Technical Report~7, School of Mathematics and Statistics, Open
	University.
	
	\bibitem[Kato and Shimizu, 2008]{Kato2008}
	Kato, S. and Shimizu, K. (2008).
	\newblock Dependent models for observations which include angular ones.
	\newblock {\em Journal of Statistical Planning and Inference},
	138(11):3538--3549.
	\newblock \href {http://dx.doi.org/10.1016/j.jspi.2006.12.009}
	{\path{doi:10.1016/j.jspi.2006.12.009}}.
	
	\bibitem[Kato et~al., 2008]{Kato2008a}
	Kato, S., Shimizu, K., and Shieh, G.~S. (2008).
	\newblock A circular-circular regression model.
	\newblock {\em Statistica Sinica}, 18(2):633--645.
	
	\bibitem[Kaufman et~al., 2005]{Kaufman2005}
	Kaufman, C.~G., Ventura, V., and Kass, R.~E. (2005).
	\newblock Spline-based non-parametric regression for periodic functions and its
	application to directional tuning of neurons.
	\newblock {\em Statistics in Medicine}, 24(14):2255--2265.
	\newblock \href {http://dx.doi.org/10.1002/sim.2104}
	{\path{doi:10.1002/sim.2104}}.
	
	\bibitem[Kendall et~al., 1999]{Kendall1999}
	Kendall, D.~G., Barden, D., Carne, T.~K., and Le, H. (1999).
	\newblock {\em Shape and Shape Theory}.
	\newblock Wiley Series in Probability and Statistics. Wiley, Chichester.
	\newblock \href {http://dx.doi.org/10.1002/9780470317006}
	{\path{doi:10.1002/9780470317006}}.
	
	\bibitem[Kent, 1975]{Kent1975}
	Kent, J.~T. (1975).
	\newblock Discussion of ``{S}tatistics of directional data''.
	\newblock {\em Journal of the Royal Statistical Society, Series B
		(Methodological)}, 37(3):377--378.
	\newblock \href {http://dx.doi.org/10.1111/j.2517-6161.1975.tb01550.x}
	{\path{doi:10.1111/j.2517-6161.1975.tb01550.x}}.
	
	\bibitem[Kent et~al., 2018]{Kent2018}
	Kent, J.~T., Ganeiber, A.~M., and Mardia, K.~V. (2018).
	\newblock A new unified approach for the simulation of a wide class of
	directional distributions.
	\newblock {\em Journal of Computational and Graphical Statistics},
	27(2):291--301.
	\newblock \href {http://dx.doi.org/10.1080/10618600.2017.1390468}
	{\path{doi:10.1080/10618600.2017.1390468}}.
	
	\bibitem[Kent et~al., 2016]{Kent2016}
	Kent, J.~T., I.~Hussein, I., and Jah, M.~K. (2016).
	\newblock Directional distributions in tracking of space debris.
	\newblock In {\em 2016 19th International Conference on Information Fusion
		({FUSION})}, pp. 2081--2086.
	
	\bibitem[Kent and Mardia, 2009]{Kent2009}
	Kent, J.~T. and Mardia, K.~V. (2009).
	\newblock Principal component analysis for the wrapped normal torus model.
	\newblock In Gusnanto, A., Mardia, K.~V., and Fallaize, C.~J. (Eds.), {\em
		{LASR} 2009 -- Statistical Tools for Challenges in Bioinformatics}, pp.
	39--41, Leeds. Department of Statistics, University of Leeds.
	
	\bibitem[Kent and Mardia, 2015]{Kent2015}
	Kent, J.~T. and Mardia, K.~V. (2015).
	\newblock The winding number for circular data.
	\newblock In Mardia, K.~V., Gusnanto, A., Nooney, C., and Voss, J. (Eds.), {\em
		{LASR} 2015 -- Geometry-Driven Statistics and its Cutting Edge Applications:
		Celebrating Four Decades of {L}eeds Statistics Workshops}, pp. 47--50, Leeds.
	Department of Statistics, University of Leeds.
	
	\bibitem[Kent et~al., 2008]{Kent2008}
	Kent, J.~T., Mardia, K.~V., and Taylor, C.~C. (2008).
	\newblock Modelling strategies for bivariate circular data.
	\newblock In Barber, S., Baxter, P.~D., Gusnanto, A., and Mardia, K.~V. (Eds.),
	{\em {LASR} 2008 -- The Art \& Science of Statistical Bioinformatics}, pp.
	70--73, Leeds. Department of Statistics, University of Leeds.
	
	\bibitem[Kerkyacharian et~al., 2011]{Kerkyacharian2011}
	Kerkyacharian, G., Pham~Ngoc, T.~M., and Picard, D. (2011).
	\newblock Localized spherical deconvolution.
	\newblock {\em The Annals of Statistics}, 39(2):1042--1068.
	\newblock \href {http://dx.doi.org/10.1214/10-aos858}
	{\path{doi:10.1214/10-aos858}}.
	
	\bibitem[Kesemen et~al., 2016]{Kesemen2016}
	Kesemen, O., Tezel, {\"O}., and \"Ozkul, E. (2016).
	\newblock Fuzzy {$c$}-means clustering algorithm for directional data
	({FCM4DD}).
	\newblock {\em Expert systems with applications}, 58:76--82.
	\newblock \href {http://dx.doi.org/10.1016/j.eswa.2016.03.034}
	{\path{doi:10.1016/j.eswa.2016.03.034}}.
	
	\bibitem[Kim et~al., 2019]{Kim2019}
	Kim, B., Huckemann, S., Schulz, J., and Jung, S. (2019).
	\newblock Small-sphere distributions for directional data with application to
	medical imaging.
	\newblock {\em Scandinavian Journal of Statistics}, 46(4):1047--1071.
	\newblock \href {http://dx.doi.org/10.1111/sjos.12381}
	{\path{doi:10.1111/sjos.12381}}.
	
	\bibitem[Kim and So, 2018]{Kim2018}
	Kim, N.~C. and So, H.~J. (2018).
	\newblock Directional statistical {G}abor features for texture classification.
	\newblock {\em Pattern Recognition Letters}, 112:18--26.
	\newblock \href {http://dx.doi.org/10.1016/j.patrec.2018.05.010}
	{\path{doi:10.1016/j.patrec.2018.05.010}}.
	
	\bibitem[Kim and Koo, 2002]{Kim2002}
	Kim, P.~T. and Koo, J.-Y. (2002).
	\newblock Optimal spherical deconvolution.
	\newblock {\em Journal of Multivariate Analysis}, 80(1):21--42.
	\newblock \href {http://dx.doi.org/10.1006/jmva.2000.1968}
	{\path{doi:10.1006/jmva.2000.1968}}.
	
	\bibitem[Kim et~al., 2004]{Kim2004}
	Kim, P.~T., Koo, J.~Y., and Park, H.~J. (2004).
	\newblock Sharp minimaxity and spherical deconvolution for super-smooth error
	distributions.
	\newblock {\em Journal of Multivariate Analysis}, 90(2):384--392.
	\newblock \href {http://dx.doi.org/10.1016/j.jmva.2003.08.004}
	{\path{doi:10.1016/j.jmva.2003.08.004}}.
	
	\bibitem[Kim et~al., 2016]{Kim2016}
	Kim, P.~T., Koo, J.-Y., and Pham~Ngoc, T.~M. (2016).
	\newblock Supersmooth testing on the sphere over analytic classes.
	\newblock {\em Journal of Nonparametric Statistics}, 28(1):84--115.
	\newblock \href {http://dx.doi.org/10.1080/10485252.2015.1113284}
	{\path{doi:10.1080/10485252.2015.1113284}}.
	
	\bibitem[Klemel\"a, 1999]{Klemela1999}
	Klemel\"a, J. (1999).
	\newblock Asymptotic minimax risk for the white noise model on the sphere.
	\newblock {\em Scandinavian Journal of Statistics}, 26(3):465--473.
	\newblock \href {http://dx.doi.org/10.1111/1467-9469.00160}
	{\path{doi:10.1111/1467-9469.00160}}.
	
	\bibitem[Klemel\"a, 2000]{Klemela2000}
	Klemel\"a, J. (2000).
	\newblock Estimation of densities and derivatives of densities with directional
	data.
	\newblock {\em Journal of Multivariate Analysis}, 73(1):18--40.
	\newblock \href {http://dx.doi.org/10.1006/jmva.1999.1861}
	{\path{doi:10.1006/jmva.1999.1861}}.
	
	\bibitem[Klemel\"a, 2003]{Klemela2003}
	Klemel\"a, J. (2003).
	\newblock Lower bounds for the asymptotic minimax risk with spherical data.
	\newblock {\em Journal of Statistical Planning and Inference}, 113(1):113--136.
	\newblock \href {http://dx.doi.org/10.1016/S0378-3758(01)00303-2}
	{\path{doi:10.1016/S0378-3758(01)00303-2}}.
	
	\bibitem[Klugkist et~al., 2012]{Klugkist2012}
	Klugkist, I., Bullens, J., and Postma, A. (2012).
	\newblock Evaluating order-constrained hypotheses for circular data using
	permutation tests.
	\newblock {\em British Journal of Mathematical and Statistical Psychology},
	65(2):222--236.
	\newblock \href {http://dx.doi.org/10.1111/j.2044-8317.2011.02018.x}
	{\path{doi:10.1111/j.2044-8317.2011.02018.x}}.
	
	\bibitem[Kranstauber et~al., 2020]{Kranstauber2020}
	Kranstauber, B., Smolla, M., and Scharf, A.~K. (2020).
	\newblock {\em {move}: Visualizing and Analyzing Animal Track Data}.
	\newblock {R} package version 4.0.4.
	\newblock URL: \url{https://CRAN.R-project.org/package=move}.
	
	\bibitem[Kueh, 2012]{Kueh2012}
	Kueh, A. (2012).
	\newblock Locally adaptive density estimation on the unit sphere using
	needlets.
	\newblock {\em Constructive Approximation}, 36(3):433--458.
	\newblock \href {http://dx.doi.org/10.1007/s00365-012-9170-2}
	{\path{doi:10.1007/s00365-012-9170-2}}.
	
	\bibitem[Kume and Sei, 2018]{Kume2018}
	Kume, A. and Sei, T. (2018).
	\newblock On the exact maximum likelihood inference of {F}isher-{B}ingham
	distributions using an adjusted holonomic gradient method.
	\newblock {\em Statistics and Computing}, 28(4):835--847.
	
	\bibitem[Kurz et~al., 2014]{Kurz2014}
	Kurz, G., Gilitschenski, I., Julier, S.~J., and Hanebeck, U.~D. (2014).
	\newblock Recursive {B}ingham filter for directional estimation involving 180
	degree symmetry.
	\newblock {\em Journal of Advances in Information Fusion}, 9(2):90--105.
	
	\bibitem[Kurz et~al., 2019]{Kurz2019}
	Kurz, G., Gilitschenski, I., Pfaff, F., Drude, L., Hanebeck, U.~D.,
	Haeb-Umbach, R., and Siegwart, R.~Y. (2019).
	\newblock Directional statistics and filtering using {libDirectional}.
	\newblock {\em Journal of Statistical Software}, 89(4):1--31.
	\newblock \href {http://dx.doi.org/10.18637/jss.v089.i04}
	{\path{doi:10.18637/jss.v089.i04}}.
	
	\bibitem[Lacour and Pham~Ngoc, 2014]{Lacour2014}
	Lacour, C. and Pham~Ngoc, T.~M. (2014).
	\newblock Goodness-of-fit test for noisy directional data.
	\newblock {\em Bernoulli}, 20(4):2131--2168.
	\newblock \href {http://dx.doi.org/10.3150/13-bej553}
	{\path{doi:10.3150/13-bej553}}.
	
	\bibitem[Lagona, 2016]{Lagona2016a}
	Lagona, F. (2016).
	\newblock Regression analysis of correlated circular data based on the
	multivariate von {M}ises distribution.
	\newblock {\em Environmental and Ecological Statistics}, 23(1):89--113.
	\newblock \href {http://dx.doi.org/10.1007/s10651-015-0330-y}
	{\path{doi:10.1007/s10651-015-0330-y}}.
	
	\bibitem[Lagona, 2018]{Lagona2018}
	Lagona, F. (2018).
	\newblock Correlated cylindrical data.
	\newblock In Ley, C. and Verdebout, T. (Eds.), {\em Applied Directional
		Statistics}, Chapman \& Hall/CRC Interdisciplinary Statistics Series, pp.
	45--59. CRC Press, Boca Raton.
	
	\bibitem[Lagona and Picone, 2011]{Lagona2011}
	Lagona, F. and Picone, M. (2011).
	\newblock A latent-class model for clustering incomplete linear and circular
	data in marine studies.
	\newblock {\em Journal of Data Science}, 9(4):585--605.
	
	\bibitem[Lagona and Picone, 2012]{Lagona2012}
	Lagona, F. and Picone, M. (2012).
	\newblock Model-based clustering of multivariate skew data with circular
	components and missing values.
	\newblock {\em Journal of Applied Statistics}, 39(5):927--945.
	\newblock \href {http://dx.doi.org/10.1080/02664763.2011.626850}
	{\path{doi:10.1080/02664763.2011.626850}}.
	
	\bibitem[Lagona and Picone, 2013]{Lagona2013}
	Lagona, F. and Picone, M. (2013).
	\newblock Maximum likelihood estimation of bivariate circular hidden {M}arkov
	models from incomplete data.
	\newblock {\em Journal of Statistical Computation and Simulation},
	83(7):1223--1237.
	\newblock \href {http://dx.doi.org/10.1080/00949655.2012.656642}
	{\path{doi:10.1080/00949655.2012.656642}}.
	
	\bibitem[Lagona and Picone, 2016]{Lagona2016}
	Lagona, F. and Picone, M. (2016).
	\newblock Model-based segmentation of spatial cylindrical data.
	\newblock {\em Journal of Statistical Computation and Simulation},
	86(13):2598--2610.
	
	\bibitem[Lagona et~al., 2015a]{Lagona2015a}
	Lagona, F., Picone, M., and Maruotti, A. (2015a).
	\newblock A hidden {M}arkov model for the analysis of cylindrical time series.
	\newblock {\em Environmetrics}, 26(8):534--544.
	\newblock \href {http://dx.doi.org/10.1002/env.2355}
	{\path{doi:10.1002/env.2355}}.
	
	\bibitem[Lagona et~al., 2015b]{Lagona2015}
	Lagona, F., Picone, M., Maruotti, A., and Cosoli, S. (2015b).
	\newblock A hidden {M}arkov approach to the analysis of space-time
	environmental data with linear and circular components.
	\newblock {\em Stochastic Environmental Research and Risk Assessment},
	29(2):397--409.
	
	\bibitem[Laha and Mahesh, 2015]{Laha2015}
	Laha, A.~K. and Mahesh, K.~C. (2015).
	\newblock Robustness of tests for directional mean.
	\newblock {\em Statistics}, 49(3):522--536.
	
	\bibitem[Laha et~al., 2019]{Laha2019}
	Laha, A.~K., Raja, A. C.~P., and Mahesh, K.~C. (2019).
	\newblock {SB}-robust estimation of mean direction for some new circular
	distributions.
	\newblock {\em Statistical Papers}, 60(3):527--552.
	\newblock \href {http://dx.doi.org/10.1007/s00362-016-0853-9}
	{\path{doi:10.1007/s00362-016-0853-9}}.
	
	\bibitem[Landler et~al., 2018]{Landler2018}
	Landler, L., Ruxton, G.~D., and Malkemper, E.~P. (2018).
	\newblock Circular data in biology: advice for effectively implementing
	statistical procedures.
	\newblock {\em Behavioral Ecology and Sociobiology}, 72(8):128.
	\newblock \href {http://dx.doi.org/10.1007/s00265-018-2538-y}
	{\path{doi:10.1007/s00265-018-2538-y}}.
	
	\bibitem[Landler et~al., 2019]{Landler2019}
	Landler, L., Ruxton, G.~D., and Malkemper, E.~P. (2019).
	\newblock Circular statistics meets practical limitations: a simulation-based
	{R}ao's spacing test for non-continuous data.
	\newblock {\em Movement Ecology}, 7(1):15.
	\newblock \href {http://dx.doi.org/10.1186/s40462-019-0160-x}
	{\path{doi:10.1186/s40462-019-0160-x}}.
	
	\bibitem[Larriba et~al., 2020]{Larriba2020}
	Larriba, Y., Rueda, C., Fern\'andez, M.~A., and Peddada, S.~D. (2020).
	\newblock Order restricted inference in chronobiology.
	\newblock {\em Statistics in Medicine}, 39(3):265--278.
	\newblock \href {http://dx.doi.org/10.1002/sim.8397}
	{\path{doi:10.1002/sim.8397}}.
	
	\bibitem[Larsen et~al., 2002]{Larsen2002}
	Larsen, P.~V., Blaesild, P., and S{\o}rensen, M.~K. (2002).
	\newblock Improved likelihood ratio tests on the von {M}ises-{F}isher
	distribution.
	\newblock {\em Biometrika}, 89(4):947--951.
	\newblock \href {http://dx.doi.org/10.1093/biomet/89.4.947}
	{\path{doi:10.1093/biomet/89.4.947}}.
	
	\bibitem[Le~Bihan et~al., 2016]{LeBihan2016}
	Le~Bihan, N., Chatelain, F., and Manton, J.~H. (2016).
	\newblock Isotropic multiple scattering processes on hyperspheres.
	\newblock {\em IEEE Transactions on Information Theory}, 62(10):5740--5752.
	\newblock \href {http://dx.doi.org/10.1109/TIT.2015.2508932}
	{\path{doi:10.1109/TIT.2015.2508932}}.
	
	\bibitem[Leguey et~al., 2019a]{Leguey2019a}
	Leguey, I., Bielza, C., and Larra\~{n}aga, P. (2019a).
	\newblock Circular {B}ayesian classifiers using wrapped {C}auchy distributions.
	\newblock {\em Data \& Knowledge Engineering}, 122:101--115.
	\newblock \href {http://dx.doi.org/10.1016/j.datak.2019.05.005}
	{\path{doi:10.1016/j.datak.2019.05.005}}.
	
	\bibitem[Leguey et~al., 2019b]{Leguey2019}
	Leguey, I., Larra\~{n}aga, P., Bielza, C., and Kato, S. (2019b).
	\newblock A circular-linear dependence measure under {J}ohnson-{W}ehrly
	distributions and its application in {B}ayesian networks.
	\newblock {\em Information Sciences}, 486:240--253.
	\newblock \href {http://dx.doi.org/10.1016/j.ins.2019.01.080}
	{\path{doi:10.1016/j.ins.2019.01.080}}.
	
	\bibitem[Lennox et~al., 2010]{Lennox2010}
	Lennox, K.~P., Dahl, D.~B., Vannucci, M., Day, R., and Tsai, J.~W. (2010).
	\newblock A {D}irichlet process mixture of hidden {M}arkov models for protein
	structure prediction.
	\newblock {\em The Annals of Applied Statistics}, 4(2):916--942.
	\newblock \href {http://dx.doi.org/10.1214/09-AOAS296}
	{\path{doi:10.1214/09-AOAS296}}.
	
	\bibitem[Lennox et~al., 2009]{Lennox2009}
	Lennox, K.~P., Dahl, D.~B., Vannucci, M., and Tsai, J.~W. (2009).
	\newblock Density estimation for protein conformation angles using a bivariate
	von {M}ises distribution and {B}ayesian nonparametrics.
	\newblock {\em Journal of the American Statistical Association},
	104(486):586--596.
	\newblock \href {http://dx.doi.org/10.1198/jasa.2009.0024}
	{\path{doi:10.1198/jasa.2009.0024}}.
	
	\bibitem[Leonenko and Ruiz-Medina, 2018]{Leonenko2017}
	Leonenko, N.~N. and Ruiz-Medina, M.~D. (2018).
	\newblock Increasing domain asymptotics for the first {M}inkowski functional of
	spherical random fields.
	\newblock {\em Theory of Probability and Mathematical Statistics}, 97:127--149.
	\newblock \href {http://dx.doi.org/10.1090/tpms/1053}
	{\path{doi:10.1090/tpms/1053}}.
	
	\bibitem[Ley et~al., 2015]{Ley2015}
	Ley, C., Paindaveine, D., and Verdebout, T. (2015).
	\newblock High-dimensional tests for spherical location and spiked covariance.
	\newblock {\em Journal of Multivariate Analysis}, 139:79--91.
	\newblock \href {http://dx.doi.org/10.1016/j.jmva.2015.02.019}
	{\path{doi:10.1016/j.jmva.2015.02.019}}.
	
	\bibitem[Ley et~al., 2014]{Ley2014}
	Ley, C., Sabbah, C., and Verdebout, T. (2014).
	\newblock A new concept of quantiles for directional data and the angular
	{M}ahalanobis depth.
	\newblock {\em Electronic Journal of Statistics}, 8(1):795--816.
	\newblock \href {http://dx.doi.org/10.1214/14-ejs904}
	{\path{doi:10.1214/14-ejs904}}.
	
	\bibitem[Ley et~al., 2013]{Ley2013}
	Ley, C., Swan, Y., Thiam, B., and Verdebout, T. (2013).
	\newblock Optimal {R}-estimation of a spherical location.
	\newblock {\em Statistica Sinica}, 23(1):305--332.
	\newblock \href {http://dx.doi.org/10.5705/ss.2011.206}
	{\path{doi:10.5705/ss.2011.206}}.
	
	\bibitem[Ley et~al., 2017]{Ley2017}
	Ley, C., Swan, Y., and Verdebout, T. (2017).
	\newblock Efficient {ANOVA} for directional data.
	\newblock {\em Annals of the Institute of Statistical Mathematics},
	69(1):39--62.
	\newblock \href {http://dx.doi.org/10.1007/s10463-015-0533-x}
	{\path{doi:10.1007/s10463-015-0533-x}}.
	
	\bibitem[Ley and Verdebout, 2014a]{Ley2014a}
	Ley, C. and Verdebout, T. (2014a).
	\newblock Local powers of one- and multi-sample tests for the concentration of
	{F}isher-von {M}ises-{L}angevin distributions.
	\newblock {\em International Statistical Review}, 82(3):440--456.
	
	\bibitem[Ley and Verdebout, 2014b]{Ley2014b}
	Ley, C. and Verdebout, T. (2014b).
	\newblock Simple optimal tests for circular reflective symmetry about a
	specified median direction.
	\newblock {\em Statistica Sinica}, 24(3):1319--1339.
	\newblock \href {http://dx.doi.org/10.5705/ss.2013.083}
	{\path{doi:10.5705/ss.2013.083}}.
	
	\bibitem[Ley and Verdebout, 2017a]{Ley2017a}
	Ley, C. and Verdebout, T. (2017a).
	\newblock {\em Modern Directional Statistics}.
	\newblock Chapman \& Hall/CRC Interdisciplinary Statistics Series. CRC Press,
	Boca Raton.
	\newblock \href {http://dx.doi.org/10.1201/9781315119472}
	{\path{doi:10.1201/9781315119472}}.
	
	\bibitem[Ley and Verdebout, 2017b]{Ley2017b}
	Ley, C. and Verdebout, T. (2017b).
	\newblock Skew-rotationally-symmetric distributions and related efficient
	inferential procedures.
	\newblock {\em Journal of Multivariate Analysis}, 159:67--81.
	\newblock \href {http://dx.doi.org/10.1016/j.jmva.2017.02.010}
	{\path{doi:10.1016/j.jmva.2017.02.010}}.
	
	\bibitem[Ley and Verdebout, 2018]{Ley2018}
	Ley, C. and Verdebout, T. (Eds.) (2018).
	\newblock {\em Applied Directional Statistics}.
	\newblock Chapman \& Hall/CRC Interdisciplinary Statistics Series. CRC Press,
	Boca Raton.
	\newblock \href {http://dx.doi.org/10.1201/9781315228570}
	{\path{doi:10.1201/9781315228570}}.
	
	\bibitem[Li, 2014]{Li2014}
	Li, L. (2014).
	\newblock Moderate deviations results for a symmetry testing statistic based on
	the kernel density estimator for directional data.
	\newblock {\em Communications in Statistics -- Theory and Methods},
	43(14):3007--3018.
	\newblock \href {http://dx.doi.org/10.1080/03610926.2012.694545}
	{\path{doi:10.1080/03610926.2012.694545}}.
	
	\bibitem[Lin et~al., 2017]{Lin2017}
	Lin, L., St.~Thomas, B., Zhu, H., and Dunson, D.~B. (2017).
	\newblock Extrinsic local regression on manifold-valued data.
	\newblock {\em Journal of the American Statistical Association},
	112(519):1261--1273.
	\newblock \href {http://dx.doi.org/10.1080/01621459.2016.1208615}
	{\path{doi:10.1080/01621459.2016.1208615}}.
	
	\bibitem[Lin, 2019]{Lin2019}
	Lin, S.-B. (2019).
	\newblock Nonparametric regression using needlet kernels for spherical data.
	\newblock {\em Journal of Complexity}, 50:66--83.
	\newblock \href {http://dx.doi.org/10.1016/j.jco.2018.09.003}
	{\path{doi:10.1016/j.jco.2018.09.003}}.
	
	\bibitem[Liu et~al., 2004]{Liu2004}
	Liu, D., Umbach, D.~M., Peddada, S.~D., Li, L., Crockett, P.~W., and Weinberg,
	C.~R. (2004).
	\newblock A random-periods model for expression of cell-cycle genes.
	\newblock {\em Proceedings of the National Academy of Sciences of the United
		States of America}, 101(19):7240--7245.
	\newblock \href {http://dx.doi.org/10.1073/pnas.0402285101}
	{\path{doi:10.1073/pnas.0402285101}}.
	
	\bibitem[Liu and Singh, 1992]{Liu1992}
	Liu, R.~Y. and Singh, K. (1992).
	\newblock Ordering directional data: concepts of data depth on circles and
	spheres.
	\newblock {\em The Annals of Statistics}, 20(3):1468--1484.
	\newblock \href {http://dx.doi.org/10.1214/aos/1176348779}
	{\path{doi:10.1214/aos/1176348779}}.
	
	\bibitem[Loader, 1996]{Loader1996}
	Loader, C.~R. (1996).
	\newblock Local likelihood density estimation.
	\newblock {\em The Annals of Statistics}, 24(4):1602--1618.
	\newblock \href {http://dx.doi.org/10.1214/aos/1032298287}
	{\path{doi:10.1214/aos/1032298287}}.
	
	\bibitem[Lombard et~al., 2017]{Lombard2017}
	Lombard, F., Hawkins, D.~M., and Potgieter, C.~J. (2017).
	\newblock Sequential rank {CUSUM} charts for angular data.
	\newblock {\em Computational Statistics \& Data Analysis}, 105:268--279.
	\newblock \href {http://dx.doi.org/10.1016/j.csda.2016.08.001}
	{\path{doi:10.1016/j.csda.2016.08.001}}.
	
	\bibitem[Lombard and Maxwell, 2012]{Lombard2012}
	Lombard, F. and Maxwell, R.~K. (2012).
	\newblock A cusum procedure to detect deviations from uniformity in angular
	data.
	\newblock {\em Journal of Applied Statistics}, 39(9):1871--1880.
	\newblock \href {http://dx.doi.org/10.1080/02664763.2012.683857}
	{\path{doi:10.1080/02664763.2012.683857}}.
	
	\bibitem[L\'opez-Cruz et~al., 2015]{Lopez-Cruz2015}
	L\'opez-Cruz, P.~L., Bielza, C., and Larra\~{n}aga, P. (2015).
	\newblock Directional naive {B}ayes classifiers.
	\newblock {\em Pattern Analysis and Applications}, 18(2):225--246.
	\newblock \href {http://dx.doi.org/10.1007/s10044-013-0340-z}
	{\path{doi:10.1007/s10044-013-0340-z}}.
	
	\bibitem[Lu et~al., 2019]{Lu2019}
	Lu, Y., Corander, J., and Yang, Z. (2019).
	\newblock Doubly stochastic neighbor embedding on spheres.
	\newblock {\em Pattern Recognition Letters}, 128:100--106.
	\newblock \href {http://dx.doi.org/10.1016/j.patrec.2019.08.026}
	{\path{doi:10.1016/j.patrec.2019.08.026}}.
	
	\bibitem[Lund, 1999]{Lund1999a}
	Lund, U. (1999).
	\newblock Cluster analysis for directional data.
	\newblock {\em Communications in Statistics -- Simulation and Computation},
	28(4):1001--1009.
	\newblock \href {http://dx.doi.org/10.1080/03610919908813589}
	{\path{doi:10.1080/03610919908813589}}.
	
	\bibitem[Lund, 2002]{Lund2002}
	Lund, U. (2002).
	\newblock Tree-based regression for a circular response.
	\newblock {\em Communications in Statistics -- Theory and Methods},
	31(9):1549--1560.
	\newblock \href {http://dx.doi.org/10.1081/sta-120013011}
	{\path{doi:10.1081/sta-120013011}}.
	
	\bibitem[Lunga and Ersoy, 2013]{Lunga2013}
	Lunga, D. and Ersoy, O. (2013).
	\newblock Spherical stochastic neighbor embedding of hyperspectral data.
	\newblock {\em IEEE Transactions on Geoscience and Remote Sensing},
	51(2):857--871.
	\newblock \href {http://dx.doi.org/10.1109/tgrs.2012.2205004}
	{\path{doi:10.1109/tgrs.2012.2205004}}.
	
	\bibitem[van~der Maaten and Hinton, 2008]{Maaten2008}
	van~der Maaten, L. J.~P. and Hinton, G.~E. (2008).
	\newblock Visualizing high-dimensional data using t-{SNE}.
	\newblock {\em Journal of Machine Learning Research}, 9(Nov):2579--2605.
	
	\bibitem[Mahmood et~al., 2017]{Mahmood2017a}
	Mahmood, E.~A., Rana, S., Midi, H., and Hussin, A.~G. (2017).
	\newblock Detection of outliers in univariate circular data using robust
	circular distance.
	\newblock {\em Journal of Modern Applied Statistical Methods}, 16(2):22.
	\newblock \href {http://dx.doi.org/10.22237/jmasm/1509495720}
	{\path{doi:10.22237/jmasm/1509495720}}.
	
	\bibitem[Maitra and Ramler, 2010]{Maitra2010}
	Maitra, R. and Ramler, I.~P. (2010).
	\newblock A {$k$}-mean-directions algorithm for fast clustering of data on the
	sphere.
	\newblock {\em Journal of Computational and Graphical Statistics},
	19(2):377--396.
	\newblock \href {http://dx.doi.org/10.1198/jcgs.2009.08155}
	{\path{doi:10.1198/jcgs.2009.08155}}.
	
	\bibitem[Maksimov, 1967]{Maksimov1967}
	Maksimov, V.~M. (1967).
	\newblock Necessary and sufficient statistics for the family of shifts of
	probability distributions on continuous bicompact groups (in {R}ussian).
	\newblock {\em Theoria Verojatna}, 12(2):307--321.
	
	\bibitem[Mardia, 2018]{Mardia2018a}
	Mardia, K. (2018).
	\newblock A new estimation methodology for standard directional distributions.
	\newblock In {\em 2018 21st International Conference on Information Fusion
		({FUSION})}, pp. 724--729, New York. IEEE.
	
	\bibitem[Mardia, 1972]{Mardia1972}
	Mardia, K.~V. (1972).
	\newblock {\em Statistics of Directional Data}.
	\newblock Probability and Mathematical Statistics. Academic Press, London.
	
	\bibitem[Mardia, 1975]{Mardia1975b}
	Mardia, K.~V. (1975).
	\newblock Statistics of directional data.
	\newblock {\em Journal of the Royal Statistical Society, Series B
		(Methodological)}, 37(3):349--393.
	\newblock \href {http://dx.doi.org/10.1111/j.2517-6161.1975.tb01550.x}
	{\path{doi:10.1111/j.2517-6161.1975.tb01550.x}}.
	
	\bibitem[Mardia, 2010]{Mardia2010}
	Mardia, K.~V. (2010).
	\newblock Bayesian analysis for bivariate von {M}ises distributions.
	\newblock {\em Journal of Applied Statistics}, 37(3):515--528.
	\newblock \href {http://dx.doi.org/10.1080/02664760903551267}
	{\path{doi:10.1080/02664760903551267}}.
	
	\bibitem[Mardia et~al., 2018]{Mardia2018}
	Mardia, K.~V., Foldager, J.~I., and Frellsen, J. (2018).
	\newblock Directional statistics in protein bioinformatics.
	\newblock In Ley, C. and Verdebout, T. (Eds.), {\em Applied Directional
		Statistics}, Chapman \& Hall/CRC Interdisciplinary Statistics Series, pp.
	1--23. CRC Press, Boca Raton.
	
	\bibitem[Mardia and Frellsen, 2012]{Mardia2012}
	Mardia, K.~V. and Frellsen, J. (2012).
	\newblock Statistics of bivariate von {M}ises distributions.
	\newblock In Hamelryck, T., Mardia, K., and Ferkinghoff-Borg, J. (Eds.), {\em
		Bayesian Methods in Structural Bioinformatics}, Statistics for Biology and
	Health, pp. 159--178. Springer, Berlin.
	\newblock \href {http://dx.doi.org/10.1007/978-3-642-27225-7_6}
	{\path{doi:10.1007/978-3-642-27225-7_6}}.
	
	\bibitem[Mardia et~al., 2008]{Mardia2008}
	Mardia, K.~V., Hughes, G., Taylor, C.~C., and Singh, H. (2008).
	\newblock A multivariate von {M}ises distribution with applications to
	bioinformatics.
	\newblock {\em The Canadian Journal of Statistics}, 36(1):99--109.
	\newblock \href {http://dx.doi.org/10.1002/cjs.5550360110}
	{\path{doi:10.1002/cjs.5550360110}}.
	
	\bibitem[Mardia and Jupp, 1999]{Mardia1999a}
	Mardia, K.~V. and Jupp, P.~E. (1999).
	\newblock {\em Directional Statistics}.
	\newblock Wiley Series in Probability and Statistics. Wiley, Chichester.
	\newblock \href {http://dx.doi.org/10.1002/0471667196.ess7086}
	{\path{doi:10.1002/0471667196.ess7086}}.
	
	\bibitem[Mardia et~al., 2016]{Mardia2016}
	Mardia, K.~V., Kent, J.~T., and Laha, A.~K. (2016).
	\newblock Score matching estimators for directional distributions.
	\newblock {\em arXiv:1604.08470}.
	
	\bibitem[Mardia et~al., 2012]{Mardia2012a}
	Mardia, K.~V., Kent, J.~T., Zhang, Z., Taylor, C.~C., and Hamelryck, T. (2012).
	\newblock Mixtures of concentrated multivariate sine distributions with
	applications to bioinformatics.
	\newblock {\em Journal of Applied Statistics}, 39(11):2475--2492.
	\newblock \href {http://dx.doi.org/10.1080/02664763.2012.719221}
	{\path{doi:10.1080/02664763.2012.719221}}.
	
	\bibitem[Mardia and Patrangenaru, 2005]{Mardia2005}
	Mardia, K.~V. and Patrangenaru, V. (2005).
	\newblock Directions and projective shapes.
	\newblock {\em The Annals of Statistics}, 33(4):1666--1699.
	\newblock \href {http://dx.doi.org/10.1214/009053605000000273}
	{\path{doi:10.1214/009053605000000273}}.
	
	\bibitem[Mardia and Sutton, 1978]{Mardia1978a}
	Mardia, K.~V. and Sutton, T.~W. (1978).
	\newblock A model for cylindrical variables with applications.
	\newblock {\em Journal of the Royal Statistical Society, Series B
		(Methodological)}, 40(2):229--233.
	\newblock \href {http://dx.doi.org/10.1111/j.2517-6161.1978.tb01668.x}
	{\path{doi:10.1111/j.2517-6161.1978.tb01668.x}}.
	
	\bibitem[Mardia et~al., 2007]{Mardia2007}
	Mardia, K.~V., Taylor, C.~C., and Subramaniam, G.~K. (2007).
	\newblock Protein bioinformatics and mixtures of bivariate von {M}ises
	distributions for angular data.
	\newblock {\em Biometrics}, 63(2):505--512.
	\newblock \href {http://dx.doi.org/10.1111/j.1541-0420.2006.00682.x}
	{\path{doi:10.1111/j.1541-0420.2006.00682.x}}.
	
	\bibitem[Mardia and Voss, 2014]{Mardia2014}
	Mardia, K.~V. and Voss, J. (2014).
	\newblock Some fundamental properties of a multivariate von {M}ises
	distribution.
	\newblock {\em Communications in Statistics -- Theory and Methods},
	43(6):1132--1144.
	\newblock \href {http://dx.doi.org/10.1080/03610926.2012.670353}
	{\path{doi:10.1080/03610926.2012.670353}}.
	
	\bibitem[Marinucci and Peccati, 2011]{Marinucci2011}
	Marinucci, D. and Peccati, G. (2011).
	\newblock {\em Random Fields on the Sphere}.
	\newblock London Mathematical Society Lecture Note Series. Cambridge University
	Press, Cambridge.
	
	\bibitem[Marinucci et~al., 2008]{Marinucci2008}
	Marinucci, D., Pietrobon, D., Balbi, A., Baldi, P., Cabella, P., Kerkyacharian,
	G., Natoli, P., Picard, D., and Vittorio, N. (2008).
	\newblock Spherical needlets for cosmic microwave background data analysis.
	\newblock {\em Monthly Notices of the Royal Astronomical Society},
	383(2):539--545.
	\newblock \href {http://dx.doi.org/10.1111/j.1365-2966.2007.12550.x}
	{\path{doi:10.1111/j.1365-2966.2007.12550.x}}.
	
	\bibitem[Marron and Alonso, 2014]{Marron2014}
	Marron, J.~S. and Alonso, A.~M. (2014).
	\newblock Overview of object oriented data analysis.
	\newblock {\em Biometrical Journal}, 56(5):732--753.
	\newblock \href {http://dx.doi.org/10.1002/bimj.201300072}
	{\path{doi:10.1002/bimj.201300072}}.
	
	\bibitem[Maruotti, 2016]{Maruotti2016a}
	Maruotti, A. (2016).
	\newblock Analyzing longitudinal circular data by projected normal models: a
	semi-parametric approach based on finite mixture models.
	\newblock {\em Environmental and Ecological Statistics}, 23(2):257--277.
	\newblock \href {http://dx.doi.org/10.1007/s10651-015-0338-3}
	{\path{doi:10.1007/s10651-015-0338-3}}.
	
	\bibitem[Maruotti et~al., 2016]{Maruotti2016}
	Maruotti, A., Punzo, A., Mastrantonio, G., and Lagona, F. (2016).
	\newblock A time-dependent extension of the projected normal regression model
	for longitudinal circular data based on a hidden {M}arkov heterogeneity
	structure.
	\newblock {\em Stochastic Environmental Research and Risk Assessment},
	30(6):1725--1740.
	\newblock \href {http://dx.doi.org/10.1007/s00477-015-1183-5}
	{\path{doi:10.1007/s00477-015-1183-5}}.
	
	\bibitem[Mash'al and Hosseini, 2015]{Mashal2015}
	Mash'al, M. and Hosseini, R. (2015).
	\newblock {$K$}-means++ for mixtures of von {M}ises-{F}isher distributions.
	\newblock In {\em 2015 7th Conference on Information and Knowledge Technology
		({IKT})}, pp. 1--6, New York. IEEE.
	\newblock \href {http://dx.doi.org/10.1109/ikt.2015.7288786}
	{\path{doi:10.1109/ikt.2015.7288786}}.
	
	\bibitem[Mastrantonio, 2018]{Mastrantonio2018}
	Mastrantonio, G. (2018).
	\newblock The joint projected normal and skew-normal: a distribution for
	poly-cylindrical data.
	\newblock {\em Journal of Multivariate Analysis}, 165:14--26.
	\newblock \href {http://dx.doi.org/10.1016/j.jmva.2017.11.006}
	{\path{doi:10.1016/j.jmva.2017.11.006}}.
	
	\bibitem[Mastrantonio and Calise, 2016]{Mastrantonio2016b}
	Mastrantonio, G. and Calise, G. (2016).
	\newblock Hidden {M}arkov model for discrete circular-linear wind data time
	series.
	\newblock {\em Journal of Statistical Computation and Simulation},
	86(13):2611--2624.
	\newblock \href {http://dx.doi.org/10.1080/00949655.2016.1142544}
	{\path{doi:10.1080/00949655.2016.1142544}}.
	
	\bibitem[Mastrantonio et~al., 2016a]{Mastrantonio2016}
	Mastrantonio, G., Gelfand, A.~E., and Jona~Lasinio, G. (2016a).
	\newblock The wrapped skew {G}aussian process for analyzing spatio-temporal
	data.
	\newblock {\em Stochastic Environmental Research and Risk Assessment},
	30(8):2231--2242.
	\newblock \href {http://dx.doi.org/10.1007/s00477-015-1163-9}
	{\path{doi:10.1007/s00477-015-1163-9}}.
	
	\bibitem[Mastrantonio et~al., 2016b]{Mastrantonio2016a}
	Mastrantonio, G., Jona~Lasinio, G., and Gelfand, A.~E. (2016b).
	\newblock Spatio-temporal circular models with non-separable covariance
	structure.
	\newblock {\em TEST}, 25(2):331--350.
	\newblock \href {http://dx.doi.org/10.1007/s11749-015-0458-y}
	{\path{doi:10.1007/s11749-015-0458-y}}.
	
	\bibitem[Mastrantonio et~al., 2019]{Mastrantonio2019}
	Mastrantonio, G., Jona~Lasinio, G., Maruotti, A., and Calise, G. (2019).
	\newblock Invariance properties and statistical inference for circular data.
	\newblock {\em Statistica Sinica}, 29(1):67--80.
	\newblock \href {http://dx.doi.org/10.5705/ss.202016.0067}
	{\path{doi:10.5705/ss.202016.0067}}.
	
	\bibitem[Mastrantonio et~al., 2015]{Mastrantonio2015a}
	Mastrantonio, G., Maruotti, A., and Jona~Lasinio, G. (2015).
	\newblock Bayesian hidden {M}arkov modelling using circular-linear general
	projected normal distribution.
	\newblock {\em Environmetrics}, 26(2):145--158.
	\newblock \href {http://dx.doi.org/10.1002/env.2326}
	{\path{doi:10.1002/env.2326}}.
	
	\bibitem[Mazumder and Bhattacharya, 2016]{Mazumder2016}
	Mazumder, S. and Bhattacharya, S. (2016).
	\newblock Bayesian nonparametric dynamic state space modeling with circular
	latent states.
	\newblock {\em Journal of Statistical Theory and Practice}, 10(1):154--178.
	\newblock \href {http://dx.doi.org/10.1080/15598608.2015.1100562}
	{\path{doi:10.1080/15598608.2015.1100562}}.
	
	\bibitem[Mazumder and Bhattacharya, 2017]{Mazumder2017}
	Mazumder, S. and Bhattacharya, S. (2017).
	\newblock Nonparametric dynamic state space modeling of observed circular time
	series with circular latent states: a {B}ayesian perspective.
	\newblock {\em Journal of Statistical Theory and Practice}, 11(4):693--718.
	\newblock \href {http://dx.doi.org/10.1080/15598608.2017.1305922}
	{\path{doi:10.1080/15598608.2017.1305922}}.
	
	\bibitem[McClintock et~al., 2012]{McClintock2012}
	McClintock, B.~T., King, R., Thomas, L., Matthiopoulos, J., McConnell, B.~J.,
	and Morales, J.~M. (2012).
	\newblock A general discrete-time modeling framework for animal movement using
	multistate random walks.
	\newblock {\em Ecological Monographs}, 82(3):335--349.
	\newblock \href {http://dx.doi.org/10.1890/11-0326.1}
	{\path{doi:10.1890/11-0326.1}}.
	
	\bibitem[McCullagh, 1996]{McCullagh1996}
	McCullagh, P. (1996).
	\newblock M\"obius transformation and {C}auchy parameter estimation.
	\newblock {\em The Annals of Statistics}, 24(2):787--808.
	\newblock \href {http://dx.doi.org/10.1214/aos/1032894465}
	{\path{doi:10.1214/aos/1032894465}}.
	
	\bibitem[McMillan et~al., 2013]{McMillan2013}
	McMillan, G.~P., Hanson, T.~E., Saunders, G., and Gallun, F.~J. (2013).
	\newblock A two-component circular regression model for repeated measures
	auditory localization data.
	\newblock {\em Journal of the Royal Statistical Society, Series C (Applied
		Statistics)}, 62(4):515--534.
	\newblock \href {http://dx.doi.org/10.1111/rssc.12004}
	{\path{doi:10.1111/rssc.12004}}.
	
	\bibitem[McVinish and Mengersen, 2008]{McVinish2008}
	McVinish, R. and Mengersen, K. (2008).
	\newblock Semiparametric {B}ayesian circular statistics.
	\newblock {\em Computational Statistics \& Data Analysis}, 52(10):4722--4730.
	\newblock \href {http://dx.doi.org/10.1016/j.csda.2008.03.016}
	{\path{doi:10.1016/j.csda.2008.03.016}}.
	
	\bibitem[Meil\'an-Vila et~al., 2020]{Meilan-Vila2020}
	Meil\'an-Vila, A., Francisco-Fern\'andez, M., Crujeiras, R.~M., and Panzera, A.
	(2020).
	\newblock Nonparametric multivariate regression estimation for circular
	responses.
	\newblock {\em arXiv:2001.10317}.
	
	\bibitem[Meintanis and Verdebout, 2019]{Meintanis2019}
	Meintanis, S. and Verdebout, T. (2019).
	\newblock Le {C}am maximin tests for symmetry of circular data based on the
	characteristic function.
	\newblock {\em Statistica Sinica}, 29(3):1301--1320.
	\newblock \href {http://dx.doi.org/10.5705/ss.202016.0016}
	{\path{doi:10.5705/ss.202016.0016}}.
	
	\bibitem[Michelot et~al., 2016]{Michelot2019}
	Michelot, T., Langrock, R., Patterson, T., and McClintock, B. (2016).
	\newblock {moveHMM}: an {R} package for the statistical modelling of animal
	movement data using hidden {M}arkov models.
	\newblock {\em Methods in Ecology and Evolution}, 7(11):1308--1315.
	\newblock \href {http://dx.doi.org/10.1111/2041-210X.12578}
	{\path{doi:10.1111/2041-210X.12578}}.
	
	\bibitem[Miolane et~al., 2020]{Miolane2020}
	Miolane, N., Le~Brigant, A., Mathe, J., Hou, B., Guigui, N., Thanwerdas, Y.,
	Heyder, S., Peltre, O., Koep, N., Zaatiti, H., Hajri, H., Cabanes, Y.,
	Gerald, T., Chauchat, P., Shewmake, C., Kainz, B., Donnat, C., Holmes, S.,
	and Pennec, X. (2020).
	\newblock {geomstats}: A {P}ython package for {R}iemannian geometry in machine
	learning.
	\newblock {\em arXiv:2004.04667}.
	
	\bibitem[Miyata et~al., 2019]{Miyata2019}
	Miyata, Y., Shiohama, T., and Abe, T. (2019).
	\newblock Estimation of finite mixture models of skew-symmetric circular
	distributions.
	\newblock {\em Metrika}.
	\newblock \href {http://dx.doi.org/10.1007/s00184-019-00756-z}
	{\path{doi:10.1007/s00184-019-00756-z}}.
	
	\bibitem[Modlin et~al., 2012]{Modlin2012}
	Modlin, D., Fuentes, M., and Reich, B. (2012).
	\newblock Circular conditional autoregressive modeling of vector fields.
	\newblock {\em Environmetrics}, 23(1):46--53.
	\newblock \href {http://dx.doi.org/10.1002/env.1133}
	{\path{doi:10.1002/env.1133}}.
	
	\bibitem[Moghimbeygi and Golalizadeh, 2020]{Moghimbeygi2020}
	Moghimbeygi, M. and Golalizadeh, M. (2020).
	\newblock Spherical logistic distribution.
	\newblock {\em Communications in Mathematics and Statistics}, 8(2):151--166.
	\newblock \href {http://dx.doi.org/10.1007/s40304-018-00171-2}
	{\path{doi:10.1007/s40304-018-00171-2}}.
	
	\bibitem[Monbet, 2020]{Monbet2020}
	Monbet, V. (2020).
	\newblock {\em {NHMSAR}: Non-Homogeneous {M}arkov Switching Autoregressive
		Models}.
	\newblock {R} package version 1.17.
	\newblock URL: \url{https://CRAN.R-project.org/package=NHMSAR}.
	
	\bibitem[Monnier, 2011]{Monnier2011}
	Monnier, J.-B. (2011).
	\newblock Non-parametric regression on the hypersphere with uniform design.
	\newblock {\em TEST}, 20(2):412--446.
	\newblock \href {http://dx.doi.org/10.1007/s11749-011-0233-7}
	{\path{doi:10.1007/s11749-011-0233-7}}.
	
	\bibitem[Montanari and Cal\`o, 2013]{Montanari2013}
	Montanari, A. and Cal\`o, D.~G. (2013).
	\newblock Model-based clustering of probability density functions.
	\newblock {\em Advances in Data Analysis and Classification}, 7(3):301--319.
	\newblock \href {http://dx.doi.org/10.1007/s11634-013-0140-8}
	{\path{doi:10.1007/s11634-013-0140-8}}.
	
	\bibitem[Mooney et~al., 2003]{Mooney2003}
	Mooney, J.~A., Helms, P.~J., and Jolliffe, I.~T. (2003).
	\newblock Fitting mixtures of von {M}ises distributions: a case study involving
	sudden infant death syndrome.
	\newblock {\em Computational Statistics \& Data Analysis}, 41(3-4):505--513.
	\newblock \href {http://dx.doi.org/10.1016/s0167-9473(02)00181-0}
	{\path{doi:10.1016/s0167-9473(02)00181-0}}.
	
	\bibitem[Morales et~al., 2004]{Morales2004}
	Morales, J.~M., Haydon, D.~T., Frair, J., Holsinger, K.~E., and Fryxell, J.~M.
	(2004).
	\newblock Extracting more out of relocation data: building movement models as
	mixtures of random walks.
	\newblock {\em Ecology}, 85(9):2436--2445.
	\newblock \href {http://dx.doi.org/10.1890/03-0269}
	{\path{doi:10.1890/03-0269}}.
	
	\bibitem[Morphet and Symanzik, 2010]{Morphet2010}
	Morphet, W.~J. and Symanzik, J. (2010).
	\newblock The circular dataimage, a graph for high-resolution circular-spatial
	data.
	\newblock {\em International Journal of Digital Earth}, 3(1):47--71.
	\newblock \href {http://dx.doi.org/10.1080/17538940903277657}
	{\path{doi:10.1080/17538940903277657}}.
	
	\bibitem[Mu et~al., 2005]{Mu2005}
	Mu, Y., Nguyen, P.~H., and Stock, G. (2005).
	\newblock Energy landscape of a small peptide revealed by dihedral angle
	principal component analysis.
	\newblock {\em Proteins: Structure, Function, and Bioinformatics},
	58(1):45--52.
	\newblock \href {http://dx.doi.org/10.1002/prot.20310}
	{\path{doi:10.1002/prot.20310}}.
	
	\bibitem[Mulder et~al., 2020a]{Mulder2020a}
	Mulder, K., Jongsma, P., and Klugkist, I. (2020a).
	\newblock Bayesian inference for mixtures of von {M}ises distributions using
	reversible jump {MCMC} sampler.
	\newblock {\em Journal of Statistical Computation and Simulation},
	90(9):1539--1556.
	\newblock \href {http://dx.doi.org/10.1080/00949655.2020.1740997}
	{\path{doi:10.1080/00949655.2020.1740997}}.
	
	\bibitem[Mulder and Klugkist, 2017]{Mulder2017}
	Mulder, K. and Klugkist, I. (2017).
	\newblock Bayesian estimation and hypothesis tests for a circular generalized
	linear model.
	\newblock {\em Journal of Mathematical Psychology}, 80:4--14.
	\newblock \href {http://dx.doi.org/10.1016/j.jmp.2017.07.001}
	{\path{doi:10.1016/j.jmp.2017.07.001}}.
	
	\bibitem[Mulder et~al., 2020b]{Mulder2020}
	Mulder, K., Klugkist, I., van Renswoude, D., and Visser, I. (2020b).
	\newblock Mixtures of peaked power {B}atschelet distributions for circular data
	with application to saccade directions.
	\newblock {\em Journal of Mathematical Psychology}, 95:102309.
	\newblock \href {http://dx.doi.org/10.1016/j.jmp.2019.102309}
	{\path{doi:10.1016/j.jmp.2019.102309}}.
	
	\bibitem[Mulder and Klugkist, 2021]{Mulder2021}
	Mulder, K.~T. and Klugkist, I. (2021).
	\newblock Bayesian tests for circular uniformity.
	\newblock {\em Journal of Statistical Planning and Inference}, 211:315--325.
	\newblock \href {http://dx.doi.org/10.1016/j.jspi.2020.06.002}
	{\path{doi:10.1016/j.jspi.2020.06.002}}.
	
	\bibitem[Munro and Blenkinsop, 2012]{Munro2012}
	Munro, M.~A. and Blenkinsop, T.~G. (2012).
	\newblock {MARD}---{A} moving average rose diagram application for the
	geosciences.
	\newblock {\em Computers \& Geosciences}, 49:112--120.
	\newblock \href {http://dx.doi.org/10.1016/j.cageo.2012.07.012}
	{\path{doi:10.1016/j.cageo.2012.07.012}}.
	
	\bibitem[Mushkudiani, 2002]{Mushkudiani2002}
	Mushkudiani, N.~A. (2002).
	\newblock Small nonparametric tolerance regions for directional data.
	\newblock {\em Journal of Statistical Planning and Inference}, 100(1):67--80.
	\newblock \href {http://dx.doi.org/10.1016/S0378-3758(01)00093-3}
	{\path{doi:10.1016/S0378-3758(01)00093-3}}.
	
	\bibitem[Narcowich et~al., 2006]{Narcowich2006}
	Narcowich, F.~J., Petrushev, P., and Ward, J.~D. (2006).
	\newblock Localized tight frames on spheres.
	\newblock {\em SIAM Journal on Mathematical Analysis}, 38(2):574--594.
	\newblock \href {http://dx.doi.org/10.1137/040614359}
	{\path{doi:10.1137/040614359}}.
	
	\bibitem[Navarro et~al., 2017]{Navarro2017}
	Navarro, A. K.~W., Frellsen, J., and Turner, R.~E. (2017).
	\newblock The multivariate generalised von {M}ises distribution: inference and
	applications.
	\newblock In {\em Proceedings of the Thirty-First {AAAI} Conference on
		Artificial Intelligence ({AAAI}-17)}, pp. 2394--2400, San Francisco.
	Association for the Advancement of Artificial Intelligence.
	
	\bibitem[Nicosia et~al., 2017]{Nicosia2017}
	Nicosia, A., Duchesne, T., Rivest, L.-P., and Fortin, D. (2017).
	\newblock A general hidden state random walk model for animal movement.
	\newblock {\em Computational Statistics \& Data Analysis}, 105:76--95.
	\newblock \href {http://dx.doi.org/10.1016/j.csda.2016.07.009}
	{\path{doi:10.1016/j.csda.2016.07.009}}.
	
	\bibitem[Nodehi et~al., 2015]{Nodehi2015}
	Nodehi, A., Golalizadeh, M., and Heydari, A. (2015).
	\newblock Dihedral angles principal geodesic analysis using nonlinear
	statistics.
	\newblock {\em Journal of Applied Statistics}, 42(9):1962--1972.
	\newblock \href {http://dx.doi.org/10.1080/02664763.2015.1014892}
	{\path{doi:10.1080/02664763.2015.1014892}}.
	
	\bibitem[{N\'u\~{n}ez-Antonio} and Geneyro, 2020]{Nunez-Antonio2020}
	{N\'u\~{n}ez-Antonio}, G. and Geneyro, E. (2020).
	\newblock A multivariate projected gamma model for directional data.
	\newblock {\em Communications in Statistics: Case Studies, Data Analysis and
		Applications}, to appear.
	\newblock \href {http://dx.doi.org/10.1080/03610918.2019.1612910}
	{\path{doi:10.1080/03610918.2019.1612910}}.
	
	\bibitem[{N\'u\~{n}ez-Antonio} and Guti\'errez-Pe\~{n}a,
	2005a]{NunezAntonio2005}
	{N\'u\~{n}ez-Antonio}, G. and Guti\'errez-Pe\~{n}a, E. (2005a).
	\newblock A {B}ayesian analysis of directional data using the projected normal
	distribution.
	\newblock {\em Journal of Applied Statistics}, 32(10):995--1001.
	\newblock \href {http://dx.doi.org/10.1080/02664760500164886}
	{\path{doi:10.1080/02664760500164886}}.
	
	\bibitem[{N\'u\~{n}ez-Antonio} and Guti\'errez-Pe\~{n}a,
	2005b]{NunezAntonio2005a}
	{N\'u\~{n}ez-Antonio}, G. and Guti\'errez-Pe\~{n}a, E. (2005b).
	\newblock A {B}ayesian analysis of directional data using the von
	{M}ises-{F}isher distribution.
	\newblock {\em Communications in Statistics -- Simulation and Computation},
	34(4):989--999.
	\newblock \href {http://dx.doi.org/10.1080/03610910500308495}
	{\path{doi:10.1080/03610910500308495}}.
	
	\bibitem[{N\'u\~{n}ez-Antonio} and Guti\'errez-Pe\~{n}a,
	2014]{Nunez-Antonio2014}
	{N\'u\~{n}ez-Antonio}, G. and Guti\'errez-Pe\~{n}a, E. (2014).
	\newblock A {B}ayesian model for longitudinal circular data based on the
	projected normal distribution.
	\newblock {\em Computational Statistics \& Data Analysis}, 71:506--519.
	\newblock \href {http://dx.doi.org/10.1016/j.csda.2012.07.025}
	{\path{doi:10.1016/j.csda.2012.07.025}}.
	
	\bibitem[{N\'u\~{n}ez-Antonio} et~al., 2011]{Nunez-Antonio2011}
	{N\'u\~{n}ez-Antonio}, G., Guti\'errez-Pe\~{n}a, E., and Escarela, G. (2011).
	\newblock A {B}ayesian regression model for circular data based on the
	projected normal distribution.
	\newblock {\em Statistical Modelling}, 11(3):185--201.
	\newblock \href {http://dx.doi.org/10.1177/1471082x1001100301}
	{\path{doi:10.1177/1471082x1001100301}}.
	
	\bibitem[{N\'u\~{n}ez-Antonio} et~al., 2018]{NunezAntonio2018}
	{N\'u\~{n}ez-Antonio}, G., Mendoza, M., Contreras-Crist\'an, A.,
	Guti\'errez-Pe\~{n}a, E., and Mendoza, E. (2018).
	\newblock Bayesian nonparametric inference for the overlap of daily animal
	activity patterns.
	\newblock {\em Environmental and Ecological Statistics}, 25(4):471--494.
	\newblock \href {http://dx.doi.org/10.1007/s10651-018-0414-6}
	{\path{doi:10.1007/s10651-018-0414-6}}.
	
	\bibitem[Oba et~al., 2005]{Oba2005}
	Oba, S., Kato, K., and Ishii, S. (2005).
	\newblock Multi-scale clustering for gene expression profiling data.
	\newblock In {\em Fifth {IEEE} Symposium on Bioinformatics and Bioengineering
		({BIBE} '05)}, pp. 210--217.
	\newblock \href {http://dx.doi.org/10.1109/BIBE.2005.41}
	{\path{doi:10.1109/BIBE.2005.41}}.
	
	\bibitem[Oliveira et~al., 2012]{Oliveira2012}
	Oliveira, M., Crujeiras, R.~M., and Rodr\'iguez-Casal, A. (2012).
	\newblock A plug-in rule for bandwidth selection in circular density
	estimation.
	\newblock {\em Computational Statistics \& Data Analysis}, 56(12):3898--3908.
	\newblock \href {http://dx.doi.org/10.1016/j.csda.2012.05.021}
	{\path{doi:10.1016/j.csda.2012.05.021}}.
	
	\bibitem[Oliveira et~al., 2014]{Oliveira2014}
	Oliveira, M., Crujeiras, R.~M., and Rodr\'iguez-Casal, A. (2014).
	\newblock Circ{S}i{Z}er: an exploratory tool for circular data.
	\newblock {\em Environmental and Ecological Statistics}, 21(1):143--159.
	\newblock \href {http://dx.doi.org/10.1007/s10651-013-0249-0}
	{\path{doi:10.1007/s10651-013-0249-0}}.
	
	\bibitem[Otieno and Anderson-Cook, 2012]{Otieno2012}
	Otieno, S.~B. and Anderson-Cook, C.~M. (2012).
	\newblock Design and analysis of experiments for directional data.
	\newblock In Hinkelmann, K. (Ed.), {\em Design and Analysis of Experiments},
	Wiley Series in Probability and Statistic, pp. 501--532. Wiley, Hoboken.
	\newblock \href {http://dx.doi.org/10.1002/9781118147634.ch15}
	{\path{doi:10.1002/9781118147634.ch15}}.
	
	\bibitem[Oualkacha and Rivest, 2009]{Oualkacha2009}
	Oualkacha, K. and Rivest, L.-P. (2009).
	\newblock A new statistical model for random unit vectors.
	\newblock {\em Journal of Multivariate Analysis}, 100(1):70--80.
	\newblock \href {http://dx.doi.org/10.1016/j.jmva.2008.03.004}
	{\path{doi:10.1016/j.jmva.2008.03.004}}.
	
	\bibitem[Paindaveine and Verdebout, 2015]{Paindaveine2015}
	Paindaveine, D. and Verdebout, T. (2015).
	\newblock Optimal rank-based tests for the location parameter of a rotationally
	symmetric distribution on the hypersphere.
	\newblock In Hallin, M., Mason, D., Pfeifer, D., and Steinebach, J. (Eds.),
	{\em Mathematical Statistics and Limit Theorems}, pp. 249--269. Springer,
	Cham.
	\newblock \href {http://dx.doi.org/10.1007/978-3-319-12442-1_14}
	{\path{doi:10.1007/978-3-319-12442-1_14}}.
	
	\bibitem[Paindaveine and Verdebout, 2016]{Paindaveine2016}
	Paindaveine, D. and Verdebout, T. (2016).
	\newblock On high-dimensional sign tests.
	\newblock {\em Bernoulli}, 22(3):1745--1769.
	\newblock \href {http://dx.doi.org/10.3150/15-bej710}
	{\path{doi:10.3150/15-bej710}}.
	
	\bibitem[Paindaveine and Verdebout, 2017]{Paindaveine2017}
	Paindaveine, D. and Verdebout, T. (2017).
	\newblock Inference on the mode of weak directional signals: a {L}e {C}am
	perspective on hypothesis testing near singularities.
	\newblock {\em The Annals of Statistics}, 45(2):800--832.
	\newblock \href {http://dx.doi.org/10.1214/16-aos1468}
	{\path{doi:10.1214/16-aos1468}}.
	
	\bibitem[Paindaveine and Verdebout, 2020]{Paindaveine2020}
	Paindaveine, D. and Verdebout, T. (2020).
	\newblock Inference for spherical location under high concentration.
	\newblock {\em The Annals of Statistics}, to appear.
	
	\bibitem[Paine et~al., 2018]{Paine2018}
	Paine, P.~J., Preston, S.~P., Tsagris, M., and Wood, A. T.~A. (2018).
	\newblock An elliptically symmetric angular {G}aussian distribution.
	\newblock {\em Statistics and Computing}, 28(3):689--697.
	\newblock \href {http://dx.doi.org/10.1007/s11222-017-9756-4}
	{\path{doi:10.1007/s11222-017-9756-4}}.
	
	\bibitem[Paine et~al., 2020]{Paine2020}
	Paine, P.~J., Preston, S.~P., Tsagris, M., and Wood, A. T.~A. (2020).
	\newblock Spherical regression models with general covariates and anisotropic
	errors.
	\newblock {\em Statistics and Computing}, 30(1):153--165.
	\newblock \href {http://dx.doi.org/10.1007/s11222-019-09872-2}
	{\path{doi:10.1007/s11222-019-09872-2}}.
	
	\bibitem[Paluszewski and Hamelryck, 2010]{Paluszewski2010}
	Paluszewski, M. and Hamelryck, T. (2010).
	\newblock Mocapy++ - a toolkit for inference and learning in dynamic {B}ayesian
	networks.
	\newblock {\em BMC Bioinformatics}, 11(126):1--6.
	\newblock \href {http://dx.doi.org/10.1186/1471-2105-11-126}
	{\path{doi:10.1186/1471-2105-11-126}}.
	
	\bibitem[Panaretos et~al., 2014]{Panaretos2014}
	Panaretos, V.~M., Pham, T., and Yao, Z. (2014).
	\newblock Principal flows.
	\newblock {\em Journal of the American Statistical Association},
	109(505):424--436.
	\newblock \href {http://dx.doi.org/10.1080/01621459.2013.849199}
	{\path{doi:10.1080/01621459.2013.849199}}.
	
	\bibitem[Pandolfo et~al., 2018a]{Pandolfo2018a}
	Pandolfo, G., D'Ambrosio, A., and Porzio, G.~C. (2018a).
	\newblock A note on depth-based classification of circular data.
	\newblock {\em Electronic Journal of Applied Statistical Analysis},
	11(2):447--462.
	\newblock \href {http://dx.doi.org/10.1285/i20705948v11n2p447}
	{\path{doi:10.1285/i20705948v11n2p447}}.
	
	\bibitem[Pandolfo et~al., 2018b]{Pandolfo2018}
	Pandolfo, G., Paindaveine, D., and Porzio, G.~C. (2018b).
	\newblock Distance-based depths for directional data.
	\newblock {\em The Canadian Journal of Statistics}, 46(4):593--609.
	\newblock \href {http://dx.doi.org/10.1002/cjs.11479}
	{\path{doi:10.1002/cjs.11479}}.
	
	\bibitem[Pardo et~al., 2017]{Pardo2017}
	Pardo, A., Real, E., Krishnaswamy, V., L\'opez-Higuera, J.~M., Pogue, B.~W.,
	and Conde, O.~M. (2017).
	\newblock Directional kernel density estimation for classification of breast
	tissue spectra.
	\newblock {\em IEEE Transactions on Medical Imaging}, 36(1):64--73.
	\newblock \href {http://dx.doi.org/10.1109/tmi.2016.2593948}
	{\path{doi:10.1109/tmi.2016.2593948}}.
	
	\bibitem[Park, 2012]{Park2012}
	Park, H.~S. (2012).
	\newblock Asymptotic behavior of the kernel density estimator from a geometric
	viewpoint.
	\newblock {\em Communications in Statistics -- Simulation and Computation},
	41(19):3479--3496.
	\newblock \href {http://dx.doi.org/10.1080/03610926.2011.585009}
	{\path{doi:10.1080/03610926.2011.585009}}.
	
	\bibitem[Park, 2013]{Park2013}
	Park, H.~S. (2013).
	\newblock Comparison of relative efficiency of kernel density estimator with
	the exponential map.
	\newblock {\em Journal of the Korean Statistical Society}, 42(2):267--275.
	\newblock \href {http://dx.doi.org/10.1016/j.jkss.2012.08.007}
	{\path{doi:10.1016/j.jkss.2012.08.007}}.
	
	\bibitem[Peel et~al., 2001]{Peel2001}
	Peel, D., Whiten, W.~J., and McLachlan, G.~J. (2001).
	\newblock Fitting mixtures of {K}ent distributions to aid in joint set
	identification.
	\newblock {\em Journal of the American Statistical Association},
	96(453):56--63.
	\newblock \href {http://dx.doi.org/10.1198/016214501750332974}
	{\path{doi:10.1198/016214501750332974}}.
	
	\bibitem[Pelletier, 2005]{Pelletier2005}
	Pelletier, B. (2005).
	\newblock Kernel density estimation on {R}iemannian manifolds.
	\newblock {\em Statistics \& Probability Letters}, 73(3):297--304.
	\newblock \href {http://dx.doi.org/10.1016/j.spl.2005.04.004}
	{\path{doi:10.1016/j.spl.2005.04.004}}.
	
	\bibitem[Pennec, 2018]{Pennec2018}
	Pennec, X. (2018).
	\newblock Barycentric subspace analysis on manifolds.
	\newblock {\em The Annals of Statistics}, 46(6A):2711--2746.
	\newblock \href {http://dx.doi.org/10.1214/17-aos1636}
	{\path{doi:10.1214/17-aos1636}}.
	
	\bibitem[Pertsemlidis et~al., 2005]{Pertsemlidis2005}
	Pertsemlidis, A., Zelinka, J., Fondon, J.~W., Henderson, R.~K., and Otwinowski,
	Z. (2005).
	\newblock Bayesian statistical studies of the {R}amachandran distribution.
	\newblock {\em Statistical Applications in Genetics and Molecular Biology},
	4(1).
	\newblock \href {http://dx.doi.org/10.2202/1544-6115.1165}
	{\path{doi:10.2202/1544-6115.1165}}.
	
	\bibitem[Pewsey, 2000]{Pewsey2000}
	Pewsey, A. (2000).
	\newblock The wrapped skew-normal distribution on the circle.
	\newblock {\em Communications in Statistics -- Theory and Methods},
	29(11):2459--2472.
	\newblock \href {http://dx.doi.org/10.1080/03610920008832616}
	{\path{doi:10.1080/03610920008832616}}.
	
	\bibitem[Pewsey, 2002]{Pewsey2002a}
	Pewsey, A. (2002).
	\newblock Testing circular symmetry.
	\newblock {\em The Canadian Journal of Statistics}, 30(4):591--600.
	\newblock \href {http://dx.doi.org/10.2307/3316098}
	{\path{doi:10.2307/3316098}}.
	
	\bibitem[Pewsey, 2004a]{Pewsey2004a}
	Pewsey, A. (2004a).
	\newblock The large-sample joint distribution of key circular statistics.
	\newblock {\em Metrika}, 60(1):25--32.
	\newblock \href {http://dx.doi.org/10.1007/s001840300294}
	{\path{doi:10.1007/s001840300294}}.
	
	\bibitem[Pewsey, 2004b]{Pewsey2004}
	Pewsey, A. (2004b).
	\newblock Testing for circular reflective symmetry about a known median axis.
	\newblock {\em Journal of Applied Statistics}, 31(5):575--585.
	\newblock \href {http://dx.doi.org/10.1080/02664760410001681828}
	{\path{doi:10.1080/02664760410001681828}}.
	
	\bibitem[Pewsey, 2006]{Pewsey2006}
	Pewsey, A. (2006).
	\newblock Modelling asymmetrically distributed circular data using the wrapped
	skew-normal distribution.
	\newblock {\em Environmental and Ecological Statistics}, 13(3):257--269.
	\newblock \href {http://dx.doi.org/10.1007/s10651-005-0010-4}
	{\path{doi:10.1007/s10651-005-0010-4}}.
	
	\bibitem[Pewsey, 2008]{Pewsey2008}
	Pewsey, A. (2008).
	\newblock The wrapped stable family of distributions as a flexible model for
	circular data.
	\newblock {\em Computational Statistics \& Data Analysis}, 52(3):1516--1523.
	\newblock \href {http://dx.doi.org/10.1016/j.csda.2007.04.017}
	{\path{doi:10.1016/j.csda.2007.04.017}}.
	
	\bibitem[Pewsey, 2018]{Pewsey2018a}
	Pewsey, A. (2018).
	\newblock Applied directional statistics with {R}: an overview.
	\newblock In Ley, C. and Verdebout, T. (Eds.), {\em Applied Directional
		Statistics}, Chapman \& Hall/CRC Interdisciplinary Statistics Series, pp.
	277--290. CRC Press, Boca Raton.
	
	\bibitem[Pewsey and Jones, 2005]{Pewsey2005}
	Pewsey, A. and Jones, M.~C. (2005).
	\newblock Discrimination between the von {M}ises and wrapped normal
	distributions: just how big does the sample size have to be?
	\newblock {\em Statistics}, 39(2):81--89.
	\newblock \href {http://dx.doi.org/10.1080/02331880500031597}
	{\path{doi:10.1080/02331880500031597}}.
	
	\bibitem[Pewsey and Kato, 2016]{Pewsey2016}
	Pewsey, A. and Kato, S. (2016).
	\newblock Parametric bootstrap goodness-of-fit testing for {W}ehrly-{J}ohnson
	bivariate circular distributions.
	\newblock {\em Statistics and Computing}, 26(6):1307--1317.
	\newblock \href {http://dx.doi.org/10.1007/s11222-015-9605-2}
	{\path{doi:10.1007/s11222-015-9605-2}}.
	
	\bibitem[Pewsey et~al., 2007]{Pewsey2007}
	Pewsey, A., Lewis, T., and Jones, M.~C. (2007).
	\newblock The wrapped {$t$} family of circular distributions.
	\newblock {\em Australian \& New Zealand Journal of Statistics}, 49(1):79--91.
	\newblock \href {http://dx.doi.org/10.1111/j.1467-842x.2006.00465.x}
	{\path{doi:10.1111/j.1467-842x.2006.00465.x}}.
	
	\bibitem[Pewsey et~al., 2013]{Pewsey2013}
	Pewsey, A., Neuh\"auser, M., and Ruxton, G.~D. (2013).
	\newblock {\em Circular Statistics in {R}}.
	\newblock Oxford University Press, Oxford.
	
	\bibitem[Pham~Ngoc, 2019]{PhamNgoc2019}
	Pham~Ngoc, T.~M. (2019).
	\newblock Adaptive optimal kernel density estimation for directional data.
	\newblock {\em Journal of Multivariate Analysis}, 173:248--267.
	\newblock \href {http://dx.doi.org/10.1016/j.jmva.2019.02.009}
	{\path{doi:10.1016/j.jmva.2019.02.009}}.
	
	\bibitem[Pitt and Shephard, 1999]{Pitt1999}
	Pitt, M.~K. and Shephard, N. (1999).
	\newblock Filtering via simulation: auxiliary particle filters.
	\newblock {\em Journal of the American Statistical Association},
	94(446):590--599.
	\newblock \href {http://dx.doi.org/10.2307/2670179}
	{\path{doi:10.2307/2670179}}.
	
	\bibitem[Pizer et~al., 2013]{Pizer2013}
	Pizer, S.~M., Jung, S., Goswami, D., Vicory, J., Zhao, X., Chaudhuri, R.,
	Damon, J.~N., Huckemann, S., and Marron, J.~S. (2013).
	\newblock Nested sphere statistics of skeletal models.
	\newblock In Breu{\ss}, M., Bruckstein, A., and Maragos, P. (Eds.), {\em
		Innovations for Shape Analysis}, Mathematics and Visualization, pp. 93--115.
	Springer, Berlin.
	\newblock \href {http://dx.doi.org/10.1007/978-3-642-34141-0_5}
	{\path{doi:10.1007/978-3-642-34141-0_5}}.
	
	\bibitem[Polsen and Taylor, 2015]{Polsen2015}
	Polsen, O. and Taylor, C.~C. (2015).
	\newblock Parametric circular-circular regression and diagnostic analysis.
	\newblock In Dryden, I.~L. and Kent, J.~T. (Eds.), {\em Geometry Driven
		Statistics}, Wiley Series in Probability and Statistics, pp. 115--128. Wiley,
	Chichester.
	\newblock \href {http://dx.doi.org/10.1002/9781118866641.ch5}
	{\path{doi:10.1002/9781118866641.ch5}}.
	
	\bibitem[Porcu et~al., 2016]{Porcu2016}
	Porcu, E., Bevilacqua, M., and Genton, M.~G. (2016).
	\newblock Spatio-temporal covariance and cross-covariance functions of the
	great circle distance on a sphere.
	\newblock {\em Journal of the American Statistical Association},
	111(514):888--898.
	\newblock \href {http://dx.doi.org/10.1080/01621459.2015.1072541}
	{\path{doi:10.1080/01621459.2015.1072541}}.
	
	\bibitem[Porcu et~al., 2020]{Porcu2020}
	Porcu, E., Furrer, R., and Nychka, D. (2020).
	\newblock {30} years of space-time covariance functions.
	\newblock {\em WIREs Computational Statistics}, to appear:e1512.
	\newblock \href {http://dx.doi.org/10.1002/wics.1512}
	{\path{doi:10.1002/wics.1512}}.
	
	\bibitem[Presnell et~al., 1998]{Presnell1998}
	Presnell, B., Morrison, S.~P., and Littell, R.~C. (1998).
	\newblock Projected multivariate linear models for directional data.
	\newblock {\em Journal of the American Statistical Association},
	93(443):1068--1077.
	\newblock \href {http://dx.doi.org/10.2307/2669850}
	{\path{doi:10.2307/2669850}}.
	
	\bibitem[Pycke, 2007]{Pycke2007}
	Pycke, J.-R. (2007).
	\newblock A decomposition for invariant tests of uniformity on the sphere.
	\newblock {\em Proceedings of the American Mathematical Society},
	135(9):2983--2993.
	\newblock \href {http://dx.doi.org/10.1090/s0002-9939-07-08804-1}
	{\path{doi:10.1090/s0002-9939-07-08804-1}}.
	
	\bibitem[Pycke, 2010]{Pycke2010}
	Pycke, J.-R. (2010).
	\newblock Some tests for uniformity of circular distributions powerful against
	multimodal alternatives.
	\newblock {\em The Canadian Journal of Statistics}, 38(1):80--96.
	\newblock \href {http://dx.doi.org/10.1002/cjs.10048}
	{\path{doi:10.1002/cjs.10048}}.
	
	\bibitem[Qin et~al., 2011]{Qin2011}
	Qin, X., Zhang, J.-S., and Yan, X.-D. (2011).
	\newblock A nonparametric circular-linear multivariate regression model with a
	rule-of-thumb bandwidth selector.
	\newblock {\em Computers and Mathematics with Applications}, 62(8):3048--3055.
	\newblock \href {http://dx.doi.org/10.1016/j.camwa.2011.08.016}
	{\path{doi:10.1016/j.camwa.2011.08.016}}.
	
	\bibitem[Qiu et~al., 2015]{Qiu2015}
	Qiu, X., Wu, S., and Wu, H. (2015).
	\newblock A new information criterion based on {L}angevin mixture distribution
	for clustering circular data with application to time course genomic data.
	\newblock {\em Statistica Sinica}, 25(4):1459--1476.
	\newblock \href {http://dx.doi.org/10.5705/ss.2013.030}
	{\path{doi:10.5705/ss.2013.030}}.
	
	\bibitem[{R Core Team}, 2020]{R2020}
	{R Core Team} (2020).
	\newblock {\em R: A Language and Environment for Statistical Computing}.
	\newblock {R} Foundation for Statistical Computing.
	\newblock URL: \url{https://www.R-project.org/}.
	
	\bibitem[Rakhimberdiev et~al., 2017]{Rakhimberdiev2019}
	Rakhimberdiev, E., Saveliev, A., Piersma, T., and Karagicheva, J. (2017).
	\newblock {FLightR}: an {R} package for reconstructing animal paths from solar
	geolocation loggers.
	\newblock {\em Methods in Ecology and Evolution}, 8(11):1482--1487.
	\newblock \href {http://dx.doi.org/10.1111/2041-210X.12765}
	{\path{doi:10.1111/2041-210X.12765}}.
	
	\bibitem[Ranalli et~al., 2018]{Ranalli2018}
	Ranalli, M., Lagona, F., Picone, M., and Zambianchi, E. (2018).
	\newblock Segmentation of sea current fields by cylindrical hidden {M}arkov
	models: a composite likelihood approach.
	\newblock {\em Journal of the Royal Statistical Society, Series C (Applied
		Statistics)}, 67(3):575--598.
	\newblock \href {http://dx.doi.org/10.1111/rssc.12240}
	{\path{doi:10.1111/rssc.12240}}.
	
	\bibitem[Rayleigh, 1919]{Rayleigh1919}
	Rayleigh, {\relax Lord}. (1919).
	\newblock On the problem of random vibrations, and of random flights in one,
	two, or three dimensions.
	\newblock {\em The London, Edinburgh, and Dublin Philosophical Magazine and
		Journal of Science}, 37(220):321--347.
	\newblock \href {http://dx.doi.org/10.1080/14786440408635894}
	{\path{doi:10.1080/14786440408635894}}.
	
	\bibitem[Reed and Pewsey, 2009]{Reed2009}
	Reed, W.~J. and Pewsey, A. (2009).
	\newblock Two nested families of skew-symmetric circular distributions.
	\newblock {\em TEST}, 18(3):516--528.
	\newblock \href {http://dx.doi.org/10.1007/s11749-008-0111-0}
	{\path{doi:10.1007/s11749-008-0111-0}}.
	
	\bibitem[Riccardi et~al., 2009]{Riccardi2009}
	Riccardi, L., Nguyen, P.~H., and Stock, G. (2009).
	\newblock Free-energy landscape of {RNA} hairpins constructed via dihedral
	angle principal component analysis.
	\newblock {\em The Journal of Physical Chemistry B}, 113(52):16660--16668.
	\newblock \href {http://dx.doi.org/10.1021/jp9076036}
	{\path{doi:10.1021/jp9076036}}.
	
	\bibitem[Rivest, 1997]{Rivest1997}
	Rivest, L.-P. (1997).
	\newblock A decentred predictor for circular-circular regression.
	\newblock {\em Biometrika}, 84(3):717--726.
	\newblock \href {http://dx.doi.org/10.1093/biomet/84.3.717}
	{\path{doi:10.1093/biomet/84.3.717}}.
	
	\bibitem[Rivest, 1999]{Rivest1999}
	Rivest, L.-P. (1999).
	\newblock Some linear model techniques for analyzing small circle spherical
	data.
	\newblock {\em The Canadian Journal of Statistics}, 27(3):623--638.
	\newblock \href {http://dx.doi.org/10.2307/3316117}
	{\path{doi:10.2307/3316117}}.
	
	\bibitem[Rivest et~al., 2016]{Rivest2016}
	Rivest, L.-P., Duchesne, T., Nicosia, A., and Fortin, D. (2016).
	\newblock A general angular regression model for the analysis of data on animal
	movement in ecology.
	\newblock {\em Journal of the Royal Statistical Society, Series C (Applied
		Statistics)}, 65(3):445--463.
	\newblock \href {http://dx.doi.org/10.1111/rssc.12124}
	{\path{doi:10.1111/rssc.12124}}.
	
	\bibitem[Rivest and Kato, 2019]{Rivest2019}
	Rivest, L.-P. and Kato, S. (2019).
	\newblock A random-effects model for clustered circular data.
	\newblock {\em The Canadian Journal of Statistics}, 47(4):712--728.
	\newblock \href {http://dx.doi.org/10.1002/cjs.11520}
	{\path{doi:10.1002/cjs.11520}}.
	
	\bibitem[Rivest and Oualkacha, 2018]{Rivest2018}
	Rivest, L.-P. and Oualkacha, K. (2018).
	\newblock On modeling of {SE}(3) objects.
	\newblock In Ley, C. and Verdebout, T. (Eds.), {\em Applied Directional
		Statistics}, Chapman \& Hall/CRC Interdisciplinary Statistics Series, pp.
	111--127. CRC Press, Boca Raton.
	
	\bibitem[Rodgers et~al., 2014]{Rodgers2014}
	Rodgers, J.~L., Beasley, W.~H., and Schuelke, M. (2014).
	\newblock Graphical data analysis on the circle: wrap-around time series plots
	for (interrupted) time series designs.
	\newblock {\em Multivariate Behavioral Research}, 49(6):571--580.
	\newblock \href {http://dx.doi.org/10.1080/00273171.2014.946589}
	{\path{doi:10.1080/00273171.2014.946589}}.
	
	\bibitem[Rodr\'iguez et~al., 2020]{Rodriguez2020}
	Rodr\'iguez, C.~E., {N\'u\~{n}ez-Antonio}, G., and Escarela, G. (2020).
	\newblock A {B}ayesian mixture model for clustering circular data.
	\newblock {\em Computational Statistics \& Data Analysis}, 143:106842.
	\newblock \href {http://dx.doi.org/10.1016/j.csda.2019.106842}
	{\path{doi:10.1016/j.csda.2019.106842}}.
	
	\bibitem[Rodriguez-Lujan et~al., 2015]{RodriguezLujan2015}
	Rodriguez-Lujan, L., Bielza, C., and Larra\~{n}aga, P. (2015).
	\newblock Regularized multivariate von mises distribution.
	\newblock In Puerta, J.~M., G\'amez, J.~A., Dorronsoro, B., Barrenechea, E.,
	Troncoso, A., Baruque, B., and Galar, M. (Eds.), {\em Advances in Artificial
		Intelligence}, volume 9422 of {\em Lecture Notes in Computer Science}, pp.
	25--35, Cham. Springer.
	\newblock \href {http://dx.doi.org/10.1007/978-3-319-24598-0_3}
	{\path{doi:10.1007/978-3-319-24598-0_3}}.
	
	\bibitem[Rodriguez-Lujan et~al., 2017]{RodriguezLujan2017}
	Rodriguez-Lujan, L., Bielza, C., and Larra\~{n}aga, P. (2017).
	\newblock Frobenius norm regularization for the multivariate von {M}ises
	distribution.
	\newblock {\em International Journal of Intelligent Systems}, 32(2):153--176.
	\newblock \href {http://dx.doi.org/10.1002/int.21834}
	{\path{doi:10.1002/int.21834}}.
	
	\bibitem[Rosenthal et~al., 2014]{Rosenthal2014}
	Rosenthal, M., Wu, W., Klassen, E., and Srivastava, A. (2014).
	\newblock Spherical regression models using projective linear transformations.
	\newblock {\em Journal of the American Statistical Association},
	109(508):1615--1624.
	\newblock \href {http://dx.doi.org/10.1080/01621459.2014.892881}
	{\path{doi:10.1080/01621459.2014.892881}}.
	
	\bibitem[Rothman, 1972]{Rothman1972}
	Rothman, E.~D. (1972).
	\newblock Tests for uniformity of a circular distribution.
	\newblock {\em Sankhy\=a, Series A}, 34(1):23--32.
	
	\bibitem[Roy et~al., 2017]{Roy2017}
	Roy, A., Pal, A., and Garain, U. (2017).
	\newblock {JCLMM}: {A} finite mixture model for clustering of circular-linear
	data and its application to psoriatic plaque segmentation.
	\newblock {\em Pattern Recognition}, 66:160--173.
	\newblock \href {http://dx.doi.org/10.1016/j.patcog.2016.12.016}
	{\path{doi:10.1016/j.patcog.2016.12.016}}.
	
	\bibitem[Rueda et~al., 2016]{Rueda2016}
	Rueda, C., Fern\'andez, M.~A., Barrag\'an, S., Mardia, K.~V., and Peddada,
	S.~D. (2016).
	\newblock Circular piecewise regression with applications to cell-cycle data.
	\newblock {\em Biometrics}, 72(4):1266--1274.
	\newblock \href {http://dx.doi.org/10.1111/biom.12512}
	{\path{doi:10.1111/biom.12512}}.
	
	\bibitem[Rueda et~al., 2015]{Rueda2015}
	Rueda, C., Fern\'andez, M.~A., Barrag\'an, S., and Peddada, S.~D. (2015).
	\newblock Some advances in constrained inference for ordered circular
	parameters in oscillatory systems.
	\newblock In Dryden, I.~L. and Kent, J.~T. (Eds.), {\em Geometry Driven
		Statistics}, Wiley Series in Probability and Statistics, pp. 97--114. Wiley,
	Chichester.
	\newblock \href {http://dx.doi.org/10.1002/9781118866641.ch4}
	{\path{doi:10.1002/9781118866641.ch4}}.
	
	\bibitem[Rueda et~al., 2009]{Rueda2009}
	Rueda, C., Fern\'andez, M.~A., and Peddada, S.~D. (2009).
	\newblock Estimation of parameters subject to order restrictions on a circle
	with application to estimation of phase angles of cell cycle genes.
	\newblock {\em Journal of the American Statistical Association},
	104(485):338--347.
	\newblock \href {http://dx.doi.org/10.1198/jasa.2009.0120}
	{\path{doi:10.1198/jasa.2009.0120}}.
	
	\bibitem[Rumcheva and Presnell, 2017]{Rumcheva2017}
	Rumcheva, P. and Presnell, B. (2017).
	\newblock An improved test of equality of mean directions for the
	{L}angevin-von {M}ises-{F}isher distribution.
	\newblock {\em Australian \& New Zealand Journal of Statistics},
	59(1):119--135.
	\newblock \href {http://dx.doi.org/10.1111/anzs.12183}
	{\path{doi:10.1111/anzs.12183}}.
	
	\bibitem[Sadikon et~al., 2019]{Sadikon2019}
	Sadikon, N.~H., Ibrahim, A. I.~N., Mohamed, I., and Shimizu, K. (2019).
	\newblock A new test of discordancy in cylindrical data.
	\newblock {\em Communications in Statistics -- Simulation and Computation},
	48(8):2512--2522.
	\newblock \href {http://dx.doi.org/10.1080/03610918.2018.1458131}
	{\path{doi:10.1080/03610918.2018.1458131}}.
	
	\bibitem[Sahoo et~al., 2019]{Sahoo2019}
	Sahoo, I., Guinness, J., and Reich, B.~J. (2019).
	\newblock A test for isotropy on a sphere using spherical harmonic functions.
	\newblock {\em Statistica Sinica}, 29(3):1253--1276.
	\newblock \href {http://dx.doi.org/10.5705/ss.202017.0475}
	{\path{doi:10.5705/ss.202017.0475}}.
	
	\bibitem[Salah and Nadif, 2017]{Salah2017}
	Salah, A. and Nadif, M. (2017).
	\newblock Social regularized von {M}ises-{F}isher mixture model for item
	recommendation.
	\newblock {\em Data Mining and Knowledge Discovery}, 31(5, SI):1218--1241.
	\newblock \href {http://dx.doi.org/10.1007/s10618-017-0499-9}
	{\path{doi:10.1007/s10618-017-0499-9}}.
	
	\bibitem[Salah and Nadif, 2019]{Salah2019}
	Salah, A. and Nadif, M. (2019).
	\newblock Directional co-clustering.
	\newblock {\em Advances in Data Analysis and Classification}, 13(3):591--620.
	\newblock \href {http://dx.doi.org/10.1007/s11634-018-0323-4}
	{\path{doi:10.1007/s11634-018-0323-4}}.
	
	\bibitem[Sargsyan et~al., 2015]{Sargsyan2015}
	Sargsyan, K., Hua, Y.~H., and Lim, C. (2015).
	\newblock Clustangles: an open library for clustering angular data.
	\newblock {\em Journal of Chemical Information and Modeling}, 55(8):1517--1520.
	\newblock \href {http://dx.doi.org/10.1021/acs.jcim.5b00316}
	{\path{doi:10.1021/acs.jcim.5b00316}}.
	
	\bibitem[Sargsyan et~al., 2012]{Sargsyan2012}
	Sargsyan, K., Wright, J., and Lim, C. (2012).
	\newblock Geo{PCA}: a new tool for multivariate analysis of dihedral angles
	based on principal component geodesics.
	\newblock {\em Nucleic Acids Research}, 40(3):e25--e25.
	\newblock \href {http://dx.doi.org/10.1093/nar/gkv1000}
	{\path{doi:10.1093/nar/gkv1000}}.
	
	\bibitem[Sarma and Jammalamadaka, 1993]{Sarma1993}
	Sarma, Y.~R. and Jammalamadaka, S.~R. (1993).
	\newblock Circular regression.
	\newblock In Matsusita, K., Puri, M.~L., and Hayakawa, T. (Eds.), {\em
		Statistical Science and Data Analysis}, pp. 109--128, Utrecht. VSP.
	
	\bibitem[Sau and Rodriguez, 2018]{Sau2018}
	Sau, M.~F. and Rodriguez, D. (2018).
	\newblock Minimum distance method for directional data and outlier detection.
	\newblock {\em Advances in Data Analysis and Classification}, 12(3):587--603.
	\newblock \href {http://dx.doi.org/10.1007/s11634-017-0287-9}
	{\path{doi:10.1007/s11634-017-0287-9}}.
	
	\bibitem[Saw, 1983]{Saw1983}
	Saw, J.~G. (1983).
	\newblock Dependent unit vectors.
	\newblock {\em Biometrika}, 70(3):665--671.
	\newblock \href {http://dx.doi.org/10.1093/biomet/70.3.665}
	{\path{doi:10.1093/biomet/70.3.665}}.
	
	\bibitem[Scealy and Welsh, 2011]{Scealy2011}
	Scealy, J.~L. and Welsh, A.~H. (2011).
	\newblock Regression for compositional data by using distributions defined on
	the hypersphere.
	\newblock {\em Journal of the Royal Statistical Society, Series B (Statistical
		Methodology)}, 73(3):351--375.
	\newblock \href {http://dx.doi.org/10.1111/j.1467-9868.2010.00766.x}
	{\path{doi:10.1111/j.1467-9868.2010.00766.x}}.
	
	\bibitem[Scealy and Welsh, 2014a]{Scealy2014}
	Scealy, J.~L. and Welsh, A.~H. (2014a).
	\newblock Colours and cocktails: compositional data analysis: 2013 {L}ancaster
	lecture.
	\newblock {\em Australian \& New Zealand Journal of Statistics},
	56(2):145--169.
	\newblock \href {http://dx.doi.org/10.1111/anzs.12073}
	{\path{doi:10.1111/anzs.12073}}.
	
	\bibitem[Scealy and Welsh, 2014b]{Scealy2014a}
	Scealy, J.~L. and Welsh, A.~H. (2014b).
	\newblock Fitting {K}ent models to compositional data with small concentration.
	\newblock {\em Statistics and Computing}, 24(2):165--179.
	\newblock \href {http://dx.doi.org/10.1007/s11222-012-9361-5}
	{\path{doi:10.1007/s11222-012-9361-5}}.
	
	\bibitem[Scealy and Welsh, 2017]{Scealy2017}
	Scealy, J.~L. and Welsh, A.~H. (2017).
	\newblock A directional mixed effects model for compositional expenditure data.
	\newblock {\em Journal of the American Statistical Association},
	112(517):24--36.
	\newblock \href {http://dx.doi.org/10.1080/01621459.2016.1189336}
	{\path{doi:10.1080/01621459.2016.1189336}}.
	
	\bibitem[Scealy and Wood, 2019]{Scealy2019}
	Scealy, J.~L. and Wood, A. T.~A. (2019).
	\newblock Scaled von {M}ises--{F}isher distributions and regression models for
	paleomagnetic directional data.
	\newblock {\em Journal of the American Statistical Association},
	114(528):1547--1560.
	\newblock \href {http://dx.doi.org/10.1080/01621459.2019.1585249}
	{\path{doi:10.1080/01621459.2019.1585249}}.
	
	\bibitem[Schlather et~al., 2015]{Schlather2015}
	Schlather, M., Malinowski, A., Menck, P.~J., Oesting, M., and Strokorb, K.
	(2015).
	\newblock Analysis, simulation and prediction of multivariate random fields
	with package {RandomFields}.
	\newblock {\em Journal of Statistical Software}, 63(8):1--25.
	\newblock \href {http://dx.doi.org/10.18637/jss.v063.i08}
	{\path{doi:10.18637/jss.v063.i08}}.
	
	\bibitem[Schulz et~al., 2015]{Schulz2015}
	Schulz, J., Jung, S., Huckemann, S., Pierrynowski, M., Marron, J.~S., and
	Pizer, S.~M. (2015).
	\newblock Analysis of rotational deformations from directional data.
	\newblock {\em Journal of Computational and Graphical Statistics},
	24(2):539--560.
	\newblock \href {http://dx.doi.org/10.1080/10618600.2014.914947}
	{\path{doi:10.1080/10618600.2014.914947}}.
	
	\bibitem[Scott, 2011]{Scott2011}
	Scott, J.~G. (2011).
	\newblock Bayesian estimation of intensity surfaces on the sphere via needlet
	shrinkage and selection.
	\newblock {\em Bayesian Analysis}, 6(2):307--327.
	\newblock \href {http://dx.doi.org/10.1214/11-BA611}
	{\path{doi:10.1214/11-BA611}}.
	
	\bibitem[Self and Liang, 1987]{Self1987}
	Self, S.~G. and Liang, K.-Y. (1987).
	\newblock Asymptotic properties of maximum likelihood estimators and likelihood
	ratio tests under nonstandard conditions.
	\newblock {\em Journal of the American Statistical Association},
	82(398):605--610.
	\newblock \href {http://dx.doi.org/10.1080/01621459.1987.10478472}
	{\path{doi:10.1080/01621459.1987.10478472}}.
	
	\bibitem[SenGupta and Bhattacharya, 2015]{SenGupta2015}
	SenGupta, A. and Bhattacharya, S. (2015).
	\newblock Finite mixture-based {B}ayesian analysis of linear-circular models.
	\newblock {\em Environmental and Ecological Statistics}, 22(4):667--679.
	\newblock \href {http://dx.doi.org/10.1007/s10651-015-0325-8}
	{\path{doi:10.1007/s10651-015-0325-8}}.
	
	\bibitem[SenGupta and Laha, 2008]{SenGupta2008}
	SenGupta, A. and Laha, A.~K. (2008).
	\newblock A likelihood integrated method for exploratory graphical analysis of
	change point problem with directional data.
	\newblock {\em Communications in Statistics -- Theory and Methods},
	37(11-12):1783--1791.
	\newblock \href {http://dx.doi.org/10.1080/03610920701826401}
	{\path{doi:10.1080/03610920701826401}}.
	
	\bibitem[SenGupta and Pal, 2001]{SenGupta2001}
	SenGupta, A. and Pal, C. (2001).
	\newblock On optimal tests for isotropy against the symmetric wrapped
	stable-circular uniform mixture family.
	\newblock {\em Journal of Applied Statistics}, 28(1):129--143.
	\newblock \href {http://dx.doi.org/10.1080/02664760120011653}
	{\path{doi:10.1080/02664760120011653}}.
	
	\bibitem[SenGupta and Roy, 2005]{SenGupta2005}
	SenGupta, A. and Roy, S. (2005).
	\newblock A simple classification rule for directional data.
	\newblock In Balakrishnan, N., Nagaraja, H.~N., and Kannan, N. (Eds.), {\em
		Advances in Ranking and Selection, Multiple Comparisons, and Reliability},
	Statistics for Industry and Technology, pp. 81--90. Birkh\"auser, Boston.
	\newblock \href {http://dx.doi.org/10.1007/0-8176-4422-9_5}
	{\path{doi:10.1007/0-8176-4422-9_5}}.
	
	\bibitem[SenGupta and Ugwuowo, 2011]{SenGupta2011}
	SenGupta, A. and Ugwuowo, F.~I. (2011).
	\newblock A classification method for directional data with application to the
	human skull.
	\newblock {\em Communications in Statistics -- Theory and Methods},
	40(3):457--466.
	\newblock \href {http://dx.doi.org/10.1080/03610920903377807}
	{\path{doi:10.1080/03610920903377807}}.
	
	\bibitem[Shieh and Johnson, 2005]{Shieh2005}
	Shieh, G.~S. and Johnson, R.~A. (2005).
	\newblock Inference based on a bivariate distribution with von {M}ises
	marginals.
	\newblock {\em Annals of the Institute of Statistical Mathematics},
	57(4):789--802.
	\newblock \href {http://dx.doi.org/10.1007/bf02915439}
	{\path{doi:10.1007/bf02915439}}.
	
	\bibitem[Singh et~al., 2002]{Singh2002}
	Singh, H., Hnizdo, V., and Demchuk, E. (2002).
	\newblock Probabilistic model for two dependent circular variables.
	\newblock {\em Biometrika}, 89(3):719--723.
	\newblock \href {http://dx.doi.org/10.1093/biomet/89.3.719}
	{\path{doi:10.1093/biomet/89.3.719}}.
	
	\bibitem[Sinz et~al., 2018]{Berens2018}
	Sinz, F., Berens, B., Kuemmerer, M., and Wallis, T. (2018).
	\newblock {\em {PyCircStat}: Circular Statistics with {P}ython}.
	\newblock URL: \url{https://github.com/circstat/pycircstat}.
	
	\bibitem[Sittel et~al., 2017]{Sittel2017}
	Sittel, F., Filk, T., and Stock, G. (2017).
	\newblock Principal component analysis on a torus: theory and application to
	protein dynamics.
	\newblock {\em The Journal of Chemical Physics}, 147(24):244101.
	\newblock \href {http://dx.doi.org/10.1063/1.4998259}
	{\path{doi:10.1063/1.4998259}}.
	
	\bibitem[Sklar, 1959]{Sklar1959}
	Sklar, M. (1959).
	\newblock Fonctions de r\'epartition \`a {$n$} dimensions et leurs marges.
	\newblock {\em Publications de l'Institut de Statistique de l'Universit\'e de
		Paris}, 8:229--231.
	
	\bibitem[Small, 1987]{Small1987}
	Small, C.~G. (1987).
	\newblock Measures of centrality for multivariate and directional
	distributions.
	\newblock {\em The Canadian Journal of Statistics}, 15(1):31--39.
	\newblock \href {http://dx.doi.org/10.2307/3314859}
	{\path{doi:10.2307/3314859}}.
	
	\bibitem[Soetaert, 2019]{Soetaert2019}
	Soetaert, K. (2019).
	\newblock {\em {plot3D}: Plotting Multi-Dimensional Data}.
	\newblock {R} package version 1.3.
	\newblock URL: \url{https://CRAN.R-project.org/package=plot3D}.
	
	\bibitem[Sommer, 2013]{Sommer2013}
	Sommer, S. (2013).
	\newblock Horizontal dimensionality reduction and iterated frame bundle
	development.
	\newblock In Nielsen, F. and Barbaresco, F. (Eds.), {\em Geometric Science of
		Information}, volume 8085 of {\em Lecture Notes in Computer Science}, pp.
	76--83, Berlin. Springer.
	\newblock \href {http://dx.doi.org/10.1007/978-3-642-40020-9_7}
	{\path{doi:10.1007/978-3-642-40020-9_7}}.
	
	\bibitem[Sommer, 2019]{Sommer2019}
	Sommer, S. (2019).
	\newblock An infinitesimal probabilistic model for principal component analysis
	of manifold valued data.
	\newblock {\em Sankhy\=a, Series A}, 81(1):37--62.
	\newblock \href {http://dx.doi.org/10.1007/s13171-018-0139-5}
	{\path{doi:10.1007/s13171-018-0139-5}}.
	
	\bibitem[Sommer et~al., 2014]{Sommer2014}
	Sommer, S., Lauze, F., and Nielsen, M. (2014).
	\newblock Optimization over geodesics for exact principal geodesic analysis.
	\newblock {\em Advances in Computational Mathematics}, 40(2):283--313.
	\newblock \href {http://dx.doi.org/10.1007/s10444-013-9308-1}
	{\path{doi:10.1007/s10444-013-9308-1}}.
	
	\bibitem[Souden et~al., 2013]{Souden2013}
	Souden, M., Kinoshita, K., and Nakatani, T. (2013).
	\newblock An integration of source location cues for speech clustering in
	distributed microphone arrays.
	\newblock In {\em 2013 {IEEE} International Conference on Acoustics, Speech and
		Signal Processing}, pp. 111--115, New York. IEEE.
	\newblock \href {http://dx.doi.org/10.1109/icassp.2013.6637619}
	{\path{doi:10.1109/icassp.2013.6637619}}.
	
	\bibitem[Soukissian, 2014]{Soukissian2014}
	Soukissian, T.~H. (2014).
	\newblock Probabilistic modeling of directional and linear characteristics of
	wind and sea states.
	\newblock {\em Ocean Engineering}, 91:91--110.
	\newblock \href {http://dx.doi.org/10.1016/j.oceaneng.2014.08.018}
	{\path{doi:10.1016/j.oceaneng.2014.08.018}}.
	
	\bibitem[Sra, 2018]{Sra2018}
	Sra, S. (2018).
	\newblock Directional statistics in machine learning: a brief review.
	\newblock In Ley, C. and Verdebout, T. (Eds.), {\em Applied Directional
		Statistics}, Chapman \& Hall/CRC Interdisciplinary Statistics Series, pp.
	259--276. CRC Press, Boca Raton.
	
	\bibitem[Sra and Karp, 2013]{Sra2013}
	Sra, S. and Karp, D. (2013).
	\newblock The multivariate {W}atson distribution: maximum-likelihood estimation
	and other aspects.
	\newblock {\em Journal of Multivariate Analysis}, 114:256--269.
	\newblock \href {http://dx.doi.org/10.1016/j.jmva.2012.08.010}
	{\path{doi:10.1016/j.jmva.2012.08.010}}.
	
	\bibitem[Stephens, 1982]{Stephens1982}
	Stephens, M.~A. (1982).
	\newblock Use of the von {M}ises distribution to analyse continuous
	proportions.
	\newblock {\em Biometrika}, 69(1):197--203.
	\newblock \href {http://dx.doi.org/10.1093/biomet/69.1.197}
	{\path{doi:10.1093/biomet/69.1.197}}.
	
	\bibitem[Straub et~al., 2015]{Straub2015}
	Straub, J., Chang, J., Freifeld, O., and Fisher, J. W.~I. (2015).
	\newblock A {D}irichlet process mixture model for spherical data.
	\newblock In Lebanon, G. and Vishwanathan, S. V.~N. (Eds.), {\em Proceedings of
		the Eighteenth International Conference on Artificial Intelligence and
		Statistics}, volume~38 of {\em Proceedings of Machine Learning Research}, San
	Diego. PMLR.
	
	\bibitem[Su and Wu, 2011]{Su2011}
	Su, Y. and Wu, X.-K. (2011).
	\newblock Smooth test for uniformity on the surface of a unit sphere.
	\newblock In {\em 2011 International Conference on Machine Learning and
		Cybernetics}, pp. 867--872, New York. IEEE.
	\newblock \href {http://dx.doi.org/10.1109/icmlc.2011.6016757}
	{\path{doi:10.1109/icmlc.2011.6016757}}.
	
	\bibitem[Sun and Lockhart, 2019]{Sun2019}
	Sun, S.~Z. and Lockhart, R.~A. (2019).
	\newblock Bayesian optimality for {B}eran's class of tests of uniformity around
	the circle.
	\newblock {\em Journal of Statistical Planning and Inference}, 198:79--90.
	\newblock \href {http://dx.doi.org/10.1016/j.jspi.2018.03.006}
	{\path{doi:10.1016/j.jspi.2018.03.006}}.
	
	\bibitem[Taghia et~al., 2014]{Taghia2014}
	Taghia, J., Ma, Z., and Leijon, A. (2014).
	\newblock Bayesian estimation of the von-{M}ises {F}isher mixture model with
	variational inference.
	\newblock {\em IEEE Transactions on Pattern Analysis and Machine Intelligence},
	36(9):1701--1715.
	\newblock \href {http://dx.doi.org/10.1109/tpami.2014.2306426}
	{\path{doi:10.1109/tpami.2014.2306426}}.
	
	\bibitem[Taijeron et~al., 1994]{Taijeron1994}
	Taijeron, H.~J., Gibson, A.~G., and Chandler, C. (1994).
	\newblock Spline interpolation and smoothing on hyperspheres.
	\newblock {\em SIAM Journal on Scientific Computing}, 15(5):1111--1125.
	\newblock \href {http://dx.doi.org/10.1137/0915068}
	{\path{doi:10.1137/0915068}}.
	
	\bibitem[Takasu et~al., 2018]{Takasu2018}
	Takasu, Y., Yano, K., and Komaki, F. (2018).
	\newblock Scoring rules for statistical models on spheres.
	\newblock {\em Statistics \& Probability Letters}, 138:111--115.
	\newblock \href {http://dx.doi.org/10.1016/j.spl.2018.02.054}
	{\path{doi:10.1016/j.spl.2018.02.054}}.
	
	\bibitem[Tang et~al., 2009]{Tang2009}
	Tang, H., Chu, S.~M., and Huang, T.~S. (2009).
	\newblock Generative model-based speaker clustering via mixture of von
	{M}ises-{F}isher distributions.
	\newblock In {\em 2009 {IEEE} International Conference on Acoustics, Speech and
		Signal Processing}, pp. 4101--4104, New York. IEEE.
	\newblock \href {http://dx.doi.org/10.1109/icassp.2009.4960530}
	{\path{doi:10.1109/icassp.2009.4960530}}.
	
	\bibitem[Taniguchi et~al., 2020]{Taniguchi2020}
	Taniguchi, M., Kato, S., Ogata, H., and Pewsey, A. (2020).
	\newblock Models for circular data from time series spectra.
	\newblock {\em Journal of Time Series Analysis}, to appear.
	\newblock \href {http://dx.doi.org/10.1111/jtsa.12549}
	{\path{doi:10.1111/jtsa.12549}}.
	
	\bibitem[Taylor, 2008]{Taylor2008}
	Taylor, C.~C. (2008).
	\newblock Automatic bandwidth selection for circular density estimation.
	\newblock {\em Computational Statistics \& Data Analysis}, 52(7):3493--3500.
	\newblock \href {http://dx.doi.org/10.1016/j.csda.2007.11.003}
	{\path{doi:10.1016/j.csda.2007.11.003}}.
	
	\bibitem[Taylor et~al., 2018]{Taylor2018}
	Taylor, C.~C., Lafratta, G., and Fensore, S. (2018).
	\newblock {\em {nprotreg}: Nonparametric Rotations for Sphere-Sphere
		Regression}.
	\newblock {R} package version 1.0.1.
	\newblock URL: \url{https://CRAN.R-project.org/package=nprotreg}.
	
	\bibitem[Tipping and Bishop, 1999]{Tipping1999}
	Tipping, M.~E. and Bishop, C.~M. (1999).
	\newblock Probabilistic principal component analysis.
	\newblock {\em Journal of the Royal Statistical Society, Series B (Statistical
		Methodology)}, 61(3):611--622.
	\newblock \href {http://dx.doi.org/10.1111/1467-9868.00196}
	{\path{doi:10.1111/1467-9868.00196}}.
	
	\bibitem[Traa and Smaragdis, 2013]{Traa2013}
	Traa, J. and Smaragdis, P. (2013).
	\newblock A wrapped {K}alman filter for azimuthal speaker tracking.
	\newblock {\em IEEE Signal Processing Letters}, 20(12):1257--1260.
	\newblock \href {http://dx.doi.org/10.1109/lsp.2013.2287125}
	{\path{doi:10.1109/lsp.2013.2287125}}.
	
	\bibitem[Tsagris and Alenazi, 2019]{Tsagris2019}
	Tsagris, M. and Alenazi, A. (2019).
	\newblock Comparison of discriminant analysis methods on the sphere.
	\newblock {\em Communications in Statistics: Case Studies, Data Analysis and
		Applications}, 5(4):467--491.
	\newblock \href {http://dx.doi.org/10.1080/23737484.2019.1684854}
	{\path{doi:10.1080/23737484.2019.1684854}}.
	
	\bibitem[Tsagris et~al., 2020]{Tsagris2020}
	Tsagris, M., Athineou, G., Sajib, A., Amson, E., and Waldstein, M.~J. (2020).
	\newblock {\em {Directional}: Directional Statistics}.
	\newblock {R} package version 4.4.
	\newblock URL: \url{https://CRAN.R-project.org/package=Directional}.
	
	\bibitem[Tsai, 2009]{Tsai2009}
	Tsai, M.-T. (2009).
	\newblock Asymptotically efficient two-sample rank tests for modal directions
	on spheres.
	\newblock {\em Journal of Multivariate Analysis}, 100:445--458.
	\newblock \href {http://dx.doi.org/10.1016/j.jmva.2008.05.009}
	{\path{doi:10.1016/j.jmva.2008.05.009}}.
	
	\bibitem[Tsuruta and Sagae, 2017a]{Tsuruta2017a}
	Tsuruta, Y. and Sagae, M. (2017a).
	\newblock Asymptotic property of wrapped {C}auchy kernel density estimation on
	the circle.
	\newblock {\em Bulletin of Informatics and Cybernetics}, 49:1--10.
	\newblock \href {http://dx.doi.org/10.5109/2232318}
	{\path{doi:10.5109/2232318}}.
	
	\bibitem[Tsuruta and Sagae, 2017b]{Tsuruta2017}
	Tsuruta, Y. and Sagae, M. (2017b).
	\newblock Higher order kernel density estimation on the circle.
	\newblock {\em Statistics \& Probability Letters}, 131:46--50.
	\newblock \href {http://dx.doi.org/10.1016/j.spl.2017.08.003}
	{\path{doi:10.1016/j.spl.2017.08.003}}.
	
	\bibitem[Tsuruta and Sagae, 2018]{Tsuruta2018}
	Tsuruta, Y. and Sagae, M. (2018).
	\newblock Properties for circular nonparametric regressions by von {M}iese and
	wrapped {C}auchy kernels.
	\newblock {\em Bulletin of Informatics and Cybernetics}, 50:1--13.
	
	\bibitem[Tsuruta and Sagae, 2020]{Tsuruta2020}
	Tsuruta, Y. and Sagae, M. (2020).
	\newblock Theoretical properties of bandwidth selectors for kernel density
	estimation on the circle.
	\newblock {\em Annals of the Institute of Statistical Mathematics},
	72(2):511--530.
	\newblock \href {http://dx.doi.org/10.1007/s10463-018-0701-x}
	{\path{doi:10.1007/s10463-018-0701-x}}.
	
	\bibitem[Tung and Jammalamadaka, 2013]{Tung2013}
	Tung, D.~D. and Jammalamadaka, S.~R. (2013).
	\newblock On the {G}ini mean difference test for circular data.
	\newblock {\em Communications in Statistics -- Theory and Methods},
	42(11):1998--2008.
	\newblock \href {http://dx.doi.org/10.1080/03610926.2011.601947}
	{\path{doi:10.1080/03610926.2011.601947}}.
	
	\bibitem[Umbach and Jammalamadaka, 2009]{Umbach2009}
	Umbach, D. and Jammalamadaka, S.~R. (2009).
	\newblock Building asymmetry into circular distributions.
	\newblock {\em Statistics \& Probability Letters}, 79(5):659--663.
	\newblock \href {http://dx.doi.org/10.1016/j.spl.2008.10.022}
	{\path{doi:10.1016/j.spl.2008.10.022}}.
	
	\bibitem[van~der Vaart, 2000]{VanderVaart2000}
	van~der Vaart, A.~W. (2000).
	\newblock {\em Asymptotic Statistics}.
	\newblock Cambridge Series in Statistical and Probabilistic Mathematics.
	Cambridge University Press, Cambridge.
	\newblock \href {http://dx.doi.org/10.1017/CBO9780511802256}
	{\path{doi:10.1017/CBO9780511802256}}.
	
	\bibitem[Veeraraghavan et~al., 2009]{Veeraraghavan2009}
	Veeraraghavan, A., Srivastava, A., Roy-Chowdhury, A.~K., and Chellappa, R.
	(2009).
	\newblock Rate-invariant recognition of humans and their activities.
	\newblock {\em IEEE Transactions on Image Processing}, 18(6):1326--1339.
	\newblock \href {http://dx.doi.org/10.1109/TIP.2009.2017143}
	{\path{doi:10.1109/TIP.2009.2017143}}.
	
	\bibitem[Verdebout, 2015]{Verdebout2015}
	Verdebout, T. (2015).
	\newblock On some validity-robust tests for the homogeneity of concentrations
	on spheres.
	\newblock {\em Journal of Nonparametric Statistics}, 27(3):372--383.
	\newblock \href {http://dx.doi.org/10.1080/10485252.2015.1041945}
	{\path{doi:10.1080/10485252.2015.1041945}}.
	
	\bibitem[Verdebout, 2017]{Verdebout2017}
	Verdebout, T. (2017).
	\newblock On the efficiency of some rank-based test for the homogeneity of
	concentrations.
	\newblock {\em Journal of Statistical Planning and Inference}, 191:101--109.
	\newblock \href {http://dx.doi.org/10.1016/j.jspi.2017.05.009}
	{\path{doi:10.1016/j.jspi.2017.05.009}}.
	
	\bibitem[Vuollo and Holmstrom, 2018]{Vuollo2018}
	Vuollo, V. and Holmstrom, L. (2018).
	\newblock A scale space approach for exploring structure in spherical data.
	\newblock {\em Computational Statistics \& Data Analysis}, 125:57--69.
	\newblock \href {http://dx.doi.org/10.1016/j.csda.2018.03.014}
	{\path{doi:10.1016/j.csda.2018.03.014}}.
	
	\bibitem[Vuollo et~al., 2016]{Vuollo2016}
	Vuollo, V., Holmstr\"om, L., Aarnivala, H., Harila, V., Heikkinen, T.,
	Pirttiniemi, P., and Valkama, A.~M. (2016).
	\newblock Analyzing infant head flatness and asymmetry using kernel density
	estimation of directional surface data from a craniofacial 3{D} model.
	\newblock {\em Statistics in Medicine}, 35(26):4891--4904.
	\newblock \href {http://dx.doi.org/10.1002/sim.7032}
	{\path{doi:10.1002/sim.7032}}.
	
	\bibitem[Wang, 2013]{Wang2013}
	Wang, F. (2013).
	\newblock {\em Space and Space-Time Modeling of Directional Data}.
	\newblock PhD thesis, Duke University.
	
	\bibitem[Wang and Gelfand, 2013]{Wang2013a}
	Wang, F. and Gelfand, A.~E. (2013).
	\newblock Directional data analysis under the general projected normal
	distribution.
	\newblock {\em Statistical Methodology}, 10(1):113--127.
	\newblock \href {http://dx.doi.org/10.1016/j.stamet.2012.07.005}
	{\path{doi:10.1016/j.stamet.2012.07.005}}.
	
	\bibitem[Wang and Gelfand, 2014]{Wang2014}
	Wang, F. and Gelfand, A.~E. (2014).
	\newblock Modeling space and space-time directional data using projected
	{G}aussian processes.
	\newblock {\em Journal of the American Statistical Association},
	109(508):1565--1580.
	\newblock \href {http://dx.doi.org/10.1080/01621459.2014.934454}
	{\path{doi:10.1080/01621459.2014.934454}}.
	
	\bibitem[Wang et~al., 2015]{Wang2015}
	Wang, F., Gelfand, A.~E., and Jona-Lasinio, G. (2015).
	\newblock Joint spatio-temporal analysis of a linear and a directional
	variable: space-time modeling of wave heights and wave directions in the
	{A}driatic {S}ea.
	\newblock {\em Statistica Sinica}, 25(1):25--39.
	
	\bibitem[Wang et~al., 2004]{Wang2004}
	Wang, J., Boyer, J., and Genton, M.~G. (2004).
	\newblock A skew-symmetric representation of multivariate distributions.
	\newblock {\em Statistica Sinica}, 14(4):1259--1270.
	
	\bibitem[Wang and Shimizu, 2012]{Wang2012}
	Wang, M. and Shimizu, K. (2012).
	\newblock On applying {M}\"obius transformation to cardioid random variables.
	\newblock {\em Statistical Methodology}, 9(6):604--614.
	\newblock \href {http://dx.doi.org/10.1016/j.stamet.2012.04.001}
	{\path{doi:10.1016/j.stamet.2012.04.001}}.
	
	\bibitem[Wang and Wang, 2016]{Wang2016a}
	Wang, M. and Wang, D. (2016).
	\newblock {VMF-SNE}: embedding for spherical data.
	\newblock In {\em 2016 {IEEE} International Conference on Acoustics, Speech and
		Signal Processing ({ICASSP})}, pp. 2344--2348, New York. IEEE.
	\newblock \href {http://dx.doi.org/10.1109/icassp.2016.7472096}
	{\path{doi:10.1109/icassp.2016.7472096}}.
	
	\bibitem[Wang and Zhao, 2001]{Wang2001}
	Wang, X. and Zhao, L. (2001).
	\newblock Laws of the iterated logarithm for kernel estimator of density
	function of spherical data.
	\newblock {\em Journal of Systems Science and Mathematical Sciences},
	21(3):264--273.
	
	\bibitem[Wang et~al., 2000]{Wang2000a}
	Wang, X., Zhao, L., and Wu, Y. (2000).
	\newblock Distribution free laws of the iterated logarithm for kernel estimator
	of regression function based on directional data.
	\newblock {\em Chinese Annals of Mathematics. Series B}, 21(4):489--498.
	\newblock \href {http://dx.doi.org/10.1142/S0252959900000480}
	{\path{doi:10.1142/S0252959900000480}}.
	
	\bibitem[Wang, 2002]{Wang2002}
	Wang, X.~M. (2002).
	\newblock Exponential bounds of mean error for the kernel regression estimates
	with directional data.
	\newblock {\em Chinese Annals of Mathematics. Series A}, 23(1):55--62.
	
	\bibitem[Wang and Ma, 2000]{Wang2000b}
	Wang, X.~M. and Ma, L. (2000).
	\newblock Nearest neighbor estimator for density function of directional data.
	\newblock {\em Journal of Biomathematics}, 15(3):332--338.
	
	\bibitem[Wang and Zhao, 2003]{Wang2003}
	Wang, X.~M. and Zhao, L.~C. (2003).
	\newblock A law of logarithm for kernel density estimators with directional
	data.
	\newblock {\em Acta Mathematica Sinica, Chinese Series}, 46(5):865--874.
	
	\bibitem[Watamori and Jupp, 2005]{Watamori2005}
	Watamori, Y. and Jupp, P.~E. (2005).
	\newblock Improved likelihood ratio and score tests on concentration parameters
	of von {M}ises-{F}isher distributions.
	\newblock {\em Statistics \& Probability Letters}, 72(2):93--102.
	\newblock \href {http://dx.doi.org/10.1016/j.spl.2004.10.017}
	{\path{doi:10.1016/j.spl.2004.10.017}}.
	
	\bibitem[Watson, 1961]{Watson1961}
	Watson, G.~S. (1961).
	\newblock Goodness-of-fit tests on a circle.
	\newblock {\em Biometrika}, 48(1/2):109--114.
	\newblock \href {http://dx.doi.org/10.2307/2333135}
	{\path{doi:10.2307/2333135}}.
	
	\bibitem[Watson, 1983]{Watson1983}
	Watson, G.~S. (1983).
	\newblock {\em Statistics on Spheres}.
	\newblock University of Arkansas Lecture Notes in the Mathematical Sciences.
	Wiley, New York.
	
	\bibitem[Wehrly and Johnson, 1980]{Wehrly1980}
	Wehrly, T.~E. and Johnson, R.~A. (1980).
	\newblock Bivariate models for dependence of angular observations and a related
	{M}arkov process.
	\newblock {\em Biometrika}, 67(1):255--256.
	\newblock \href {http://dx.doi.org/10.1093/biomet/67.1.255}
	{\path{doi:10.1093/biomet/67.1.255}}.
	
	\bibitem[Wilson et~al., 2014]{Wilson2014}
	Wilson, R.~C., Hancock, E.~R., Pekalska, E., and Duin, R. P.~W. (2014).
	\newblock Spherical and hyperbolic embeddings of data.
	\newblock {\em IEEE Transactions on Pattern Analysis and Machine Intelligence},
	36(11):2255--2269.
	\newblock \href {http://dx.doi.org/10.1109/tpami.2014.2316836}
	{\path{doi:10.1109/tpami.2014.2316836}}.
	
	\bibitem[Wood, 2017]{Wood2017}
	Wood, S.~N. (2017).
	\newblock {\em Generalized Additive Models}.
	\newblock Chapman \& Hall/CRC Texts in Statistical Science Series. CRC Press,
	Boca Raton, second edition.
	\newblock \href {http://dx.doi.org/10.1201/9781315370279}
	{\path{doi:10.1201/9781315370279}}.
	
	\bibitem[Wouters et~al., 2009]{Wouters2009}
	Wouters, H., Thas, O., and Ottoy, J.-P. (2009).
	\newblock Data-driven smooth tests and a diagnostic tool for lack-of-fit for
	circular data.
	\newblock {\em Australian \& New Zealand Journal of Statistics},
	51(4):461--480.
	\newblock \href {http://dx.doi.org/10.1111/j.1467-842X.2009.00558.x}
	{\path{doi:10.1111/j.1467-842X.2009.00558.x}}.
	
	\bibitem[Xu and Wang, 2020]{Xu2020}
	Xu, D. and Wang, Y. (2020).
	\newblock Area-proportional visualization for circular data.
	\newblock {\em Journal of Computational and Graphical Statistics},
	29(2):351--357.
	\newblock \href {http://dx.doi.org/10.1080/10618600.2019.1654881}
	{\path{doi:10.1080/10618600.2019.1654881}}.
	
	\bibitem[Yamaji and Sato, 2011]{Yamaji2011}
	Yamaji, A. and Sato, K. (2011).
	\newblock Clustering of fracture orientations using a mixed {B}ingham
	distribution and its application to paleostress analysis from dike or vein
	orientations.
	\newblock {\em Journal of Structural Geology}, 33(7):1148--1157.
	\newblock \href {http://dx.doi.org/10.1016/j.jsg.2011.05.006}
	{\path{doi:10.1016/j.jsg.2011.05.006}}.
	
	\bibitem[Yang et~al., 2014]{Yang2014}
	Yang, M.-S., Chang-Chien, S.-J., and Kuo, H.-C. (2014).
	\newblock On mean shift clustering for directional data on a hypersphere.
	\newblock In Rutkowski, L., Korytkowski, M., Scherer, R., Tadeusiewicz, R.,
	Zadeh, L.~A., and Zurada, J.~M. (Eds.), {\em Artificial Intelligence and Soft
		Computing}, volume 8468 of {\em Lecture Notes in Compututer Scence}, pp.
	809--818, Cham. Springer.
	\newblock \href {http://dx.doi.org/10.1007/978-3-319-07176-3_70}
	{\path{doi:10.1007/978-3-319-07176-3_70}}.
	
	\bibitem[Yang and Pan, 1997]{Yang1997}
	Yang, M.-S. and Pan, J.-A. (1997).
	\newblock On fuzzy clustering of directional data.
	\newblock {\em Fuzzy Sets and Systems}, 91(3):319--326.
	\newblock \href {http://dx.doi.org/10.1016/s0165-0114(96)00157-1}
	{\path{doi:10.1016/s0165-0114(96)00157-1}}.
	
	\bibitem[Yeh et~al., 2013]{Yeh2013}
	Yeh, S.-Y., Harris, K. D.~M., and Jupp, P.~E. (2013).
	\newblock A drifting {M}arkov process on the circle, with physical
	applications.
	\newblock {\em Proceedings of the Royal Society A: Mathematical, Physical and
		Engineering Sciences}, 469(2156):20130092.
	\newblock \href {http://dx.doi.org/10.1098/rspa.2013.0092}
	{\path{doi:10.1098/rspa.2013.0092}}.
	
	\bibitem[You, 2020]{You2020}
	You, K. (2020).
	\newblock {\em {RiemBase}: Functions and {C}++ Header Files for Computation on
		Manifolds}.
	\newblock {R} package version 0.2.4.
	\newblock URL: \url{https://CRAN.R-project.org/package=RiemBase}.
	
	\bibitem[Zhan et~al., 2019]{Zhan2019}
	Zhan, X., Ma, T., Liu, S., and Shimizu, K. (2019).
	\newblock On circular correlation for data on the torus.
	\newblock {\em Statistical Papers}, 60(6):1827--1847.
	\newblock \href {http://dx.doi.org/10.1007/s00362-017-0897-5}
	{\path{doi:10.1007/s00362-017-0897-5}}.
	
	\bibitem[Zhang et~al., 2018]{Zhang2018}
	Zhang, L., Li, Q., Guo, Y., Yang, Z., and Zhang, L. (2018).
	\newblock An investigation of wind direction and speed in a featured wind farm
	using joint probability distribution methods.
	\newblock {\em Sustainability}, 10(12):4338.
	\newblock \href {http://dx.doi.org/10.3390/su10124338}
	{\path{doi:10.3390/su10124338}}.
	
	\bibitem[Zhang and Fletcher, 2013]{Zhang2013}
	Zhang, M. and Fletcher, T. (2013).
	\newblock Probabilistic principal geodesic analysis.
	\newblock In Burges, C. J.~C., Bottou, L., Welling, M., Ghahramani, Z., and
	Weinberger, K.~Q. (Eds.), {\em Advances in Neural Information Processing
		Systems 26}, pp. 1178--1186, Red Hook. Curran Associates.
	
	\bibitem[Zhang et~al., 2019]{Zhang2019}
	Zhang, Z., Klassen, E., and Srivastava, A. (2019).
	\newblock Robust comparison of kernel densities on spherical domains.
	\newblock {\em Sankhy\=a, Series A}, 81(1):144--171.
	\newblock \href {http://dx.doi.org/10.1007/s13171-018-0131-0}
	{\path{doi:10.1007/s13171-018-0131-0}}.
	
	\bibitem[Zhao and Wu, 2001]{Zhao2001}
	Zhao, L. and Wu, C. (2001).
	\newblock Central limit theorem for integrated squared error of kernel
	estimators of spherical density.
	\newblock {\em Science in China Series A: Mathematics}, 44(4):474--483.
	\newblock \href {http://dx.doi.org/10.1007/bf02881884}
	{\path{doi:10.1007/bf02881884}}.
	
	\bibitem[Zou et~al., 2007]{Zou2007}
	Zou, G., Hua, J., and Muzik, O. (2007).
	\newblock Non-rigid surface registration using spherical thin-plate splines.
	\newblock In Ayache, N., Ourselin, S., and Maeder, A. (Eds.), {\em Medical
		Image Computing and Computer-Assisted Intervention -- {MICCAI} 2007}, pp.
	367--374, Berlin. Springer.
	\newblock \href {http://dx.doi.org/10.1007/978-3-540-75757-3_45}
	{\path{doi:10.1007/978-3-540-75757-3_45}}.
	
	\bibitem[Zucchini et~al., 2016]{Zucchini2016}
	Zucchini, W., MacDonald, I.~L., and Langrock, R. (2016).
	\newblock {\em Hidden {M}arkov Models for Time Series}, volume 150 of {\em
		Monographs on Statistics and Applied Probability}.
	\newblock CRC Press, Boca Raton, second edition.
	\newblock \href {http://dx.doi.org/10.1201/b20790} {\path{doi:10.1201/b20790}}.
	
\end{thebibliography}

\fi

\end{document}